%% file: main.tex
\documentclass{NSE_style/nseJournal}

\usepackage{amssymb}
\usepackage{amsbsy,amsmath}
\usepackage{graphicx}
\usepackage[dvipsnames]{xcolor}
\usepackage{hyperref}
\usepackage[capitalise]{cleveref}
\usepackage[normalem]{ulem}  % for \sout
\usepackage{geometry}
\usepackage{subcaption}
\usepackage{cite}
\usepackage{esvect}
\usepackage{longtable}
\usepackage{booktabs}

\title{Producing a complete nuclear data library for adjoint Monte Carlo simulations}
\date{2026-05-13}

\newcommand{\tm}{\textsuperscript{\textregistered}}

\newcommand{\tripolir}{TRIPOLI-5\tm\,}

\begin{document}          % Do not remove

% Use the \addAuthor macro to add authors in the order they should appear. The second argument corresponds to
% the affiliation declared below.
% The corresponding author should be wrapped in \correspondingAuthor
\addAuthor{\correspondingAuthor{Paul Rovel}}{a,b}
% The corresponding author's email can be specified using \correspondingEmail
\correspondingEmail{paul.rovel@cea.fr}
\addAuthor{Coline Larmier}{a}
\addAuthor{Davide Mancusi}{a}
\addAuthor{Andrea Zoia}{a}

% Affiliations can be added in the order they should appear. For breaks in addresses, use either \\ or \tabularnewline
\addAffiliation{a}{Université Paris-Saclay, CEA,\\ Service d'études des réacteurs et de mathématiques appliquées\\ 91191 Gif-sur-Yvette, France}
\addAffiliation{b}{École nationale des ponts et chaussées, Institut Polytechnique de Paris,\\ 77455 Champs-sur-Marne, France}

% Add keywords to appear in Abstract in the order they should appear
\addKeyword{Monte Carlo}
\addKeyword{Adjoint}
\addKeyword{Transport}
\addKeyword{Neutron}
\addKeyword{Nuclear Data}

\titlePage

\begin{abstract}
   Radiation shielding applications related to reactor design typically involve situations where the source region (the core) is much larger than the detector region (a dosimeter). In such cases, the efficiency of Monte Carlo simulation might be significantly increased by solving the adjoint transport equation: adjoint particles are born from the detector, undergo adjoint (`reversed') flights and collisions, and accumulate their tallies in the source region. A key prerequisite to sample the adjoint collision events is the preparation of adjoint nuclear data. In this work, we propose a general method able to handle the diversity of nuclear reactions available in modern neutron data evaluations, and ultimately create a full adjoint nuclear data library. This is a stepping stone in view of implementing adjoint sampling schemes in \tripolir{}, the next-generation Monte Carlo code developed by CEA and ASNR. We validate this strategy based on a relevant continuous-energy benchmark configuration involving mixtures of heavy and light nuclides, and we compare our results to those obtained by standard forward Monte Carlo simulations.
\end{abstract}
 
\include{Introduction}

\include{Theory}

\include{Preparation}

\include{Discretization}

\include{Results}

\include{Conclusion}

\appendix
\crefalias{section}{appendix}

\renewcommand{\theequation}{\thesection.\arabic{equation}}

\include{Appendices}

\section*{Acknowledgements}
\tripolir{} is a registered trademark of CEA. We thank EDF and ASNR for partial financial support.

\newpage
\bibliographystyle{NSE_style/ans_js}                                                                           %custom ANS journal submission template bibliography style
\bibliography{biblio}

\end{document}

%% file: Introduction.tex
\section{Introduction}

The resolution of the adjoint Boltzmann equation by Monte Carlo methods has been explored since the early 1960s, initially in the framework of simple one-speed neutron transport problems \cite{maynard_application_1961}. These investigations have been mainly motivated by the reciprocity property: any detector response due to a particle source can equivalently be obtained by solving the direct or the adjoint transport equations, provided that the roles of source and detector are permuted \cite{duality}. The solution of the direct transport equation via Monte Carlo games involves sampling the histories of particles that are emitted from the source, undergo flights and collisions, and accumulate their tallies in the detector region \cite{lux_monte_2018}. In typical radiation shielding calculations where the source (e.g. reactor core or a spent fuel assembly) is larger that the detector (e.g. a dosimeter), sampling particle histories from the source will yield very few particles crossing the detector zone and contributing to the response, yielding a poor statistical convergence. Conversely, solving the adjoint transport equation by Monte Carlo games can be framed as sampling histories of fictitious adjoint particles (often called `adjunctons') that are emitted from the detector region, undergo adjoint flights and collisions, and accumulate their tallies in the source region \cite{de_matteis_phenomenological_1974}. In this situation, we expect a larger number of contributing histories, and better overall statistical convergence \cite{hoogenboom_adjoint_1977}.

The main difficulty of adjoint Monte Carlo games is the sampling of the adjoint collision \cite{hoogenboom_practical_1981}: although multi-group versions of adjoint simulation schemes have been implemented in production Monte Carlo codes such as MCNP \cite{mcnp63}, the full variety of reactions available in modern nuclear data libraries using the ENDF format \cite{endf_manual} has severely limited the development of adjoint transport routines in continuous-energy transport codes  \cite{grimstone1998extension,hoogenboom2003methodology,diop}.

To address these issues, in a previous work we presented a general framework able to handle the sampling of completely general, continuous-energy adjoint collisions \cite{rovel_general_2025}. Building upon the pioneering findings by Hoogenboom and others \cite{hoogenboom_adjoint_1977,matteis_new_1978}, our formalism uses adjoint cross sections and adjoint outgoing distribution laws to sample the adjoint collision events. In this paper, we set out to produce adjoint nuclear data from modern nuclear data libraries, based on the strategy outlined in Ref.~\cite{rovel_general_2025}. We will focus in particular on neutron reactions. The strategies proposed in this work can treat full modern neutron physics, including Thermal Scattering Laws (TSL) and Unresolved Resonance Range (URR). These results pave the way towards the implementation of adjoint transport capabilities in the next-generation Monte Carlo code \tripolir{}, jointly developed by CEA and ASNR \cite{t5_ref}.

This manuscript is organized as follows. First, in \cref{sec:theory} we will briefly recall the theory of adjoint transport and its solution using Monte Carlo games, along the lines presented in Ref.~\citenum{rovel_general_2025}. Then, in \cref{sec:prep_adj_data} we will detail the techniques used to prepare the corresponding adjoint cross sections and distribution laws, encompassing all reactions allowed in the ENDF format. In \cref{sec:disc_algo} we will focus on the discretization algorithms used extensively in the methods presented in \cref{sec:prep_adj_data}. Finally, in \cref{sec:results} we will test the produced adjoint nuclear data library using a simple benchmark configuration. Technical details and formulas concerning the thermal broadening of the elastic scattering kernel are provided in \cref{sec:proof_svt}.

%% file: Theory.tex
\section{Solving the adjoint transport equations by Monte Carlo games} 
\label{sec:theory}

We begin by briefly recalling the mathematical foundations of the adjoint transport equation and illustrating how it can be solved by Monte Carlo methods. Although the strategy considered here can be extended to other kinds of particle transport processes, we will mainly address the case of neutrons.

\subsection{Adjoint transport equations}

In fixed-source time-independent problems, neutrons emerge from the source $S$ and undergo successive flight and collision events. The collision density $\psi$ describes the expected number of collision events at point $P=(\mathbf{r},E,\mathbf{\Omega})$ in the phase space\footnote{Here $\mathbf{r}$ denotes the particle position, $E$ the particle energy, and $\mathbf{\Omega}$ the particle direction, as customary.}, and the emission density $\chi$ describes the expected number of emitted neutrons at $P$. The Boltzmann transport equation for the collision density reads
\begin{equation}
    \label{eq:boltzmann}
    \psi = \mathbb{T}\mathbb{C}\psi+ \mathbb{T}S,
\end{equation}
where $\mathbb{T}$ is the flight operator and $\mathbb{C}$ is the collision operator \cite{lux_monte_2018}. The sought response $R$ (typically a dose or a reaction rate) can be expressed in terms of the collision density $\psi$ as $R = \langle \psi,\eta_{\psi}\rangle$, where $\eta_\psi$ is the collision-based response function, and brackets denote integration over the phase space.

The adjoint emission density $\chi^\dagger$ corresponds to the neutron importance function, i.e.\ the expected contribution to the response function due to a neutron being emitted at point $P$ \cite{rovel_general_2025}. The quantity $\chi^\dagger(P)$ satisfies the adjoint Boltzmann equation
\begin{equation}
    \label{eq:boltzmann_adj}
    \chi^\dagger = \mathbb{T}^\dagger\mathbb{C}^\dagger\chi^\dagger+ \mathbb{T}^\dagger\eta_\psi,
\end{equation}
where $\mathbb{T}^\dagger$ is the adjoint flight operator, and $\mathbb{C}^\dagger$ the adjoint collision operator \cite{lux_monte_2018}. The elegant reciprocity theorem states that
\begin{equation}
\label{eq:reciprocity}
    R=\langle \psi,\eta_{\psi}\rangle=\langle S,\chi^\dagger \rangle ,
\end{equation}
which means that any response $R$ that can be obtained by solving the transport equation can be equivalently obtained by solving the adjoint transport equation \cite{hoogenboom_adjoint_1977,duality,rovel_general_2025}.

\subsection{Monte Carlo solution to the adjoint transport equation}

In the standard Monte Carlo approach to solve a direct transport problem, neutrons are sampled from the source $S$, successive flights and collisions are then sampled according to $\mathbb{T}$ and $\mathbb{C}$, respectively, and the tally for the response $R$ is accumulated at each collision event by evaluating $\eta_\psi$. \Cref{eq:reciprocity,eq:boltzmann_adj} state that $R$ can be also estimated by running an `adjoint' Monte Carlo game where adjoint particles (also called `adjunctons') are born from the response function $\eta_\psi$ (which becomes the adjoint source), undergo adjoint flight and collision events according to $\mathbb{T}^\dagger$ and $\mathbb{C}^\dagger$, respectively, and the tally for the response is accumulated at each adjoint collision event by evaluating $S$ (which becomes the adjoint response function) \cite{rovel_general_2025}. Several formalisms have been proposed in the literature for the implementation of adjoint Monte Carlo games: a non-exhaustive list can be found e.g. in References  \citenum{carter_coupled_1970,hoogenboom_adjoint_1977,matteis_new_1978,saracco_adjoint_2016}. In this work, we will specifically follow the strategy recently proposed in Ref.~\citenum{rovel_general_2025}.

\subsubsection{Adjoint flight events}

The flight kernel $T(P\to P')$ associated to the operator $\mathbb{T}$ describes the expected number of neutrons entering collision around point $P'$ for a neutron starting a flight at point $P$. It can be expressed as:
\begin{align}
    T(P\rightarrow P') =&\Sigma_t(\mathbf{r}',E') \exp\left(-\int_0 ^{|\mathbf{r}-\mathbf{r}'|}  \Sigma_t(\mathbf{r}+l\mathbf{\Omega},E)dl\right) \nonumber \\
    &\times\frac{\delta\left(\frac{\mathbf{r}'-\mathbf{r}}{|\mathbf{r}-\mathbf{r}'|} - \mathbf{\Omega} \right)}{|\mathbf{r}-\mathbf{r}'|^2} \delta \left(E-E' \right)  \delta(\mathbf{\Omega}-\mathbf{\Omega}'),
    \label{eq:decomposed_direct_T}
\end{align}
where $\Sigma_t(\mathbf{r},E)$ is the total macroscopic cross section at point $\mathbf{r}$ and energy $E$. Sampling from \cref{eq:decomposed_direct_T} is done by sampling the length $l$ along the chord defined by $\mathbf{r},\mathbf{\Omega}$. The length $l$ is usually determined either using surface tracking, or delta tracking (see Chap.~3 Sec.~I.B of Ref.~\citenum{lux_monte_2018}). The adjoint flight kernel can then be written as
\begin{align}
    &T^\dagger(P'\rightarrow P) = T(P\to P') \nonumber \\
    & =\Sigma_t(\mathbf{r}',E') \exp\left(-\int_0 ^{|\mathbf{r}-\mathbf{r}'|}  \Sigma_t(\mathbf{r}+l\mathbf{\Omega},E)dl\right) \times\frac{\delta\left(\frac{\mathbf{r}'-\mathbf{r}}{|\mathbf{r}-\mathbf{r}'|} - \mathbf{\Omega} \right)}{|\mathbf{r}-\mathbf{r}'|^2} \delta \left(E-E' \right)  \delta(\mathbf{\Omega}-\mathbf{\Omega'}) \nonumber\\
    &= \frac{\Sigma_t(\mathbf{r}',E')}{\Sigma_t(\mathbf{r},E)} \Sigma_t(\mathbf{r},E) \exp\left(-\int_0 ^{|\mathbf{r}'-\mathbf{r}|}  \Sigma_t(\mathbf{r}'-l\mathbf{\Omega}',E)dl\right) \times\frac{\delta\left(\frac{\mathbf{r}-\mathbf{r}'}{|\mathbf{r}'-\mathbf{r}|} +\mathbf{\Omega} \right)}{|\mathbf{r}'-\mathbf{r}|^2} \delta \left(E'-E \right)  \delta(\mathbf{\Omega}-\mathbf{\Omega'}) \nonumber\\
    &= \frac{\Sigma_t(\mathbf{r}',E')}{\Sigma_t(\mathbf{r},E)} T((\mathbf{r}',E',-\mathbf{\Omega}')\to (\mathbf{r},E,-\mathbf{\Omega})).
    \label{eq:adj_transport_kernel}
\end{align}
According to \cref{eq:adj_transport_kernel}, an adjoint flight event can be sampled by e.g.~drawing a flight from the direct kernel $T$ with starting point $P'$, in the direction $-\mathbf{\Omega'}$, and by applying a weight correction $\Sigma_t(\mathbf{r}',E')/\Sigma_t(\mathbf{r},E)$ once the point $P$ has been sampled. Observe that this sampling method is by no means unique, and other strategies could be conceived as well.

\subsubsection{Adjoint collision event}
\label{sec:adj_coll_event}

This work will mainly focus on the treatment of the collision events. The direct collision kernel $C(P\to P')$ associated to the operator $\mathbb{C}$ describes the expected number of neutrons exiting a collision around point $P'$ for a neutron entering the collision at point $P$. It is usually decomposed as
\begin{equation}
\label{eq:decomposed_direct_C}
    C(P\rightarrow P') = \sum_i \frac{N_i(\mathbf{r}) \sigma_{i,t}(E)}{\Sigma_t(\mathbf{r},E)} \times \sum_j \frac{\sigma_{i,j}(E)}{\sigma_{i,t}(E)} \times \nu_{i,j}(E)f_{i,j}(E\rightarrow E',\mathbf{\Omega}\cdot\mathbf{\Omega'}) \times \delta(\mathbf{r}'-\mathbf{r}),
\end{equation}
where $N_i$ denotes the concentration of nuclide $i$ in the mixture, $\sigma_{i,j}$ is the microscopic cross section of reaction $j$ on nuclide $i$, $\nu_{i,j}$ is the average multiplicity of reaction channel $\{i,j\}$, and $f_{i,j}$ is the distribution law of outgoing neutrons from channel $\{i,j\}$. Furthermore, $\sigma_{i,t} = \sum_j \sigma_{i,j}$ is the total microscopic cross section of nuclide $i$, and the total macroscopic cross section $\Sigma_t$ can be obtained via $\Sigma_t = \sum_i N_i \sigma_{i,t}$. In the Monte Carlo approach, the first term in the RHS of \cref{eq:decomposed_direct_C} (a normalized probability) is used to sample the nuclide $i$ on which the collision occurs; then, the second term (also a normalized probability) is used to sample the reaction channel $j$. The average multiplicity $\nu_{i,j}$ is taken into account by creating multiple outgoing neutrons, whose coordinates are sampled according to the normalized $f_{i,j}$, or by using the average multiplicity as a weight multiplier for a single emitted neutron with distribution $f_{i,j}$. The position $\mathbf{r}$ is left untouched.

The adjoint collision kernel $C^\dagger(P'\to P)$ associated to the operator $\mathbb{C}^\dagger$ can be dealt with using a similar strategy \cite{hoogenboom_adjoint_1977, rovel_general_2025}. Following Ref.~\citenum{matteis_new_1978}, we introduce arbitrary (positive) adjoint cross sections $\sigma^\dagger_{i,j}$, adjoint distribution laws $f_{i,j}^\dagger$ and adjoint multiplicities $\nu_{i,j}^\dagger$, and we decompose the adjoint collision kernel as:
\begin{equation}
\label{eq:decomposed_adjoint_C}
    C^\dagger(P'\rightarrow P) = \sum_i \frac{N_i(\mathbf{r}') \sigma^\dagger_{i,t}(P')}{\Sigma_t^\dagger(P')} \times \sum_j \frac{\sigma^\dagger_{i,j}(P')}{\sigma^\dagger_{i,t}(P')} \times \nu^\dagger_{i,j}(P',P)f_{i,j}^\dagger(P'\to P) \times \delta(\mathbf{r}-\mathbf{r}').
\end{equation}
The total microscopic adjoint cross section for nuclide $i$ is defined as
\begin{equation}
\label{eq:def_total_micro_adjoint}
\sigma_{i,t}^\dagger(P')=\sum_j \sigma_{i,j}^\dagger(P'),
\end{equation}
and the total macroscopic adjoint cross section as
\begin{equation}
    \label{eq:def_total_macro_adjoint}
    \Sigma_t^\dagger (P') = \sum_i N_i(\mathbf{r}')\sigma_{i,t}^\dagger (P'),
\end{equation}
by extension of the corresponding definitions used for the direct cross sections. Note that here all the arbitrary adjoint cross sections and distributions generally depend on the full phase-space coordinates $P$. To ensure the unbiasedness of the Monte Carlo game, we require $C^\dagger(P' \to P) = C(P\to P')$. The easiest way to enforce this condition is to impose
\begin{equation}\label{eq:unbiasedness_path}
    \frac{N_i(\mathbf{r}) \sigma^\dagger_{i,j}(P')}{\Sigma_t^\dagger(P')}\nu_{i,j}^\dagger(P',P)f_{i,j}^\dagger(P'\to P) = \frac{N_i(\mathbf{r}) \sigma_{i,j}(E)}{\Sigma_t(\mathbf{r},E)}\nu_{i,j}(E)f_{i,j}(E\rightarrow E',\mathbf{\Omega}\cdot\mathbf{\Omega'}),
\end{equation}
for each reaction channel $\{i,j\}$. \Cref{eq:unbiasedness_path} rules out adjoint cross sections $\sigma_{i,j}^\dagger$ and adjoint distribution laws $f^\dagger_{i,j}$ that are null where the direct cross section and distribution law are not. Excluding this case, \cref{eq:unbiasedness_path} can be achieved by using $\nu_{i,j}^\dagger$ as a weight correction for the adjoint particle, once the nuclide $i$, reaction $j$ and outgoing energies $E$ and direction $\mathbf{\Omega}$ are sampled from the distributions of the adjoint game, which yields:
\begin{equation}
    \label{eq:weight_corr}
    \nu_{i,j}^\dagger(P',P) =\frac{\Sigma^\dagger_t(P') \sigma_{i,j}(E)\nu_{i,j}(E)f_{i,j}(E\rightarrow E',\mathbf{\Omega}\cdot\mathbf{\Omega'})}{\Sigma_t(\mathbf{r},E)\sigma_{i,j}^\dagger(P')f_{i,j}^\dagger(P'\to P)}.
\end{equation}
In Ref.~\citenum{rovel_general_2025}, weight corrections occur at several distinct steps during the sampling of the adjoint collision event: \cref{eq:weight_corr} corresponds to a `global' weight correction, i.e.~the product of all the aforementioned corrections, and is therefore only applied once.

\subsubsection{Sources and detectors}

The source of an adjoint Monte Carlo calculation is the response function of the detector whose importance we want to assess. In principle, we should sample adjunctons from the detector response function $\eta_\psi(P)$, up to a normalization factor. In practice, we can sample from an arbitrary adjoint source $S^\dagger(P)$; in order to preserve the unbiasedness of the adjoint game, we apply then a weight correction $\eta_\psi(P)/S^\dagger(P)$, the only constraint being that $S^\dagger(P)$ needs to be non-zero where $\eta_\psi$ is also non-zero.

When we need to compute the average importance in a given region, we can use a detector counting particles in that region and divide by the corresponding phase space volume afterward. Using a collision counter (counting adjunctons after their adjoint flight) will yield the emitted importance $\chi^\dagger$, whereas using an emission counter (counting adjunctons before their adjoint flight) will yield collided importance $\psi^\dagger$. In order to compute a response $R$ using the reciprocity theorem, we need to use a collision counter with an estimator function $h(P)$ equal to the direct source $S(P)$, namely, $h(P)=S(P)$.

\subsection{Definition of the adjoint cross sections and distribution laws}
\label{sec:def_adj_data}

The adjoint cross sections $\sigma_{i,j}^\dagger$ and adjoint distribution laws $f_{i,j}^\dagger$ (which will be collectively be referred to as `adjoint nuclear data') introduced in \cref{sec:adj_coll_event} can be chosen arbitrarily: the corresponding adjoint Monte Carlo game will yield an unbiased result, provided that the weight correction (or adjoint multiplicity) $\nu_{i,j}^\dagger$ satisfies \cref{eq:weight_corr}. This degree of freedom has led to a broad spectrum of possible definitions for the adjoint data; see e.g.~the references \citenum{carter_coupled_1970,hoogenboom_adjoint_1977,matteis_new_1978,saracco_adjoint_2016}.

Contrary to the expected result of the adjoint simulation, which is unaffected by the arbitrary choice of adjoint nuclear data, the variance of the tally is highly sensitive to how $\sigma_{i,j}^\dagger$ and $f_{i,j}^\dagger$ are set. In our previous work \cite{rovel_general_2025}, we used the framework of zero-variance games to derive simple definitions for adjoint data enabling a significant reduction of the generated variance for adjoint Monte Carlo games. In particular, we have shown that, for a transport problem where the direct neutron flux $\varphi$ would show an energy dependence with a shape $g(E)$, then a convenient choice for the adjoint cross sections $\sigma_{i,j}^\dagger$ and the (normalized) adjoint distribution laws $f_{i,j}^\dagger$ would be
\begin{equation}
    \label{eq:adj_xs_flx_gen}
    \begin{aligned}
     \sigma_{i,j}^\dagger(E') &= \int g(E)\sigma_{i,j}(E)\nu_{i,j}(E)f_{i,j}(E\to E') dE \\
     f_{i,j}^\dagger(E'\to E,\mathbf{\Omega'}\cdot\mathbf{\Omega}) &= \frac{g(E)\sigma_{i,j}(E)\nu_{i,j}(E)f_{i,j}(E\to E',\mathbf{\Omega}\cdot\mathbf{\Omega'})}{\sigma_{i,j}^\dagger(E')}.
    \end{aligned}
\end{equation}
Here `convenient' means that the variance of the adjoint game would turn out to be reasonably small; for further details, see Ref.~\citenum{rovel_general_2025}. In \cref{eq:adj_xs_flx_gen}, we implicitly defined $f_{i,j}(E \to E')$ as the angularly integrated version of the original distribution law; we will apply the same convention for the adjoint distribution law $f^\dagger_{i,j}$. The definitions in \cref{eq:adj_xs_flx_gen} are quite versatile. For example, if the knowledge of the shape of the direct flux is missing, we can use \cref{eq:adj_xs_flx_gen} with $g(E)=1$ (corresponding to absence of `prior'), which corresponds to the `first set' of adjoint data definitions proposed by Hoogenboom in Refs.~\citenum{hoogenboom_adjoint_1977,hoogenboom_practical_1981}. Knowledge that the transport process is dominated by neutron slowing-down leads to assuming that the direct flux behaves roughly as $1/E$: plugging $g(E)=1/E$ into \cref{eq:adj_xs_flx_gen} corresponds to the `second set' of adjoint data definitions proposed by Hoogenboom in Refs.~\citenum{hoogenboom_adjoint_1977,hoogenboom_practical_1981}. Going further, we might add more details to the $g(E)$ function: for instance, we might want to add a Maxwellian shape at thermal energies or a fission spectrum at high energies. Finally, if we had the possibility of running a direct calculation beforehand, we could plug the actual shape of the flux in $g(E)$, in order to improve the quality of the adjoint game using the information content of the direct game; this approach has been thoroughly examined in Ref.~\citenum{carter_coupled_1970}.

Another noteworthy feature of the definitions in \cref{eq:adj_xs_flx_gen} is that the resulting adjoint cross sections $\sigma_{i,j}^\dagger$ only depend on energy $E'$, and the adjoint distributions laws $f_{i,j}^\dagger$ only depend on $E'$, $E$ and the deflection cosine $\mu_l:= \mathbf{\Omega}\cdot\mathbf{\Omega'}$. This makes the adjoint data similar to the direct nuclear data, and lends itself to easier sampling procedures.

In this work we will rely on the definitions in \cref{eq:adj_xs_flx_gen} for the adjoint data. However, a fundamental prerequisite for this approach consists in preparing the adjoint cross sections and adjoint distribution laws beforehand, so that they can be accessed and used rapidly on-the-fly during the adjoint game.

%% file: Preparation.tex
\section{Preparation of adjoint cross sections and distribution laws}
\label{sec:prep_adj_data}

\subsection{Requirements to prepare adjoint nuclear data}
\label{sec:prep_adj_data_requirements}
As stated in \cref{sec:def_adj_data}, we will use the definitions given in \cref{eq:adj_xs_flx_gen} to define the adjoint data that will be later used in our adjoint Monte Carlo calculations (See \cref{sec:results}). The implementation of the adjoint collision sampling process described in \cref{sec:adj_coll_event} requires access to the value of the adjoint cross sections $\sigma_{i,j}^\dagger$ at any arbitrary energy $E'$, which is mandatory to choose the reaction path $\{i,j\}$ once the collision site $P'$ is sampled. The total microscopic and total macroscopic adjoint cross sections can be obtained via \cref{eq:def_total_micro_adjoint,eq:def_total_macro_adjoint}. We then need to be able to sample the ingoing energy $E$ and deflection cosine $\mu_l$ at any arbitrary outgoing energy $E'$ from the adjoint distribution law $f_{i,j}^\dagger(E'\to E, \mu_l)$. Finally, once $E$ and $\mu_l$ are known, we need to be able to retrieve the adjoint density $f_{i,j}^\dagger(E'\to E,\mu_l)$ and the direct density $f_{i,j}(E\to E', \mu_l)$, or at least the ratio $f_{i,j}(E\to E', \mu_l)/f_{i,j}^\dagger(E'\to E,\mu_l)$, in order to compute the weight correction in \cref{eq:weight_corr}. The other terms occurring in \cref{eq:weight_corr} are customary accessible in standard (direct) Monte Carlo simulations. This whole set of quantities to be evaluated or sampled will be required at each adjoint collision; furthermore, since the values $E'$, $E$ and $\mu_l$ will likely change at each collision, the obtained results cannot be simply stored in memory. 

Numerically evaluating the terms of \cref{eq:adj_xs_flx_gen} at each adjoint collision would be too computationally expensive for the adjoint simulations to display better performances than direct simulations in real-world applications. In order to significantly reduce the execution time of the adjoint collision sampling, we thus need to pre-tabulate the values of $\sigma_{i,j}^\dagger$ and $f_{i,j}^\dagger(E'\to E, \mu_l)$, which allows these quantities to be readily accessed on-the-fly during the simulation. Furthermore, we also need to tabulate the cumulative density function of the distribution law $f_{i,j}^\dagger(E'\to E, \mu_l)$, to be able to sample from it efficiently. Therefore, a discretization scheme is required to store the tabulated values of $\sigma_{i,j}^\dagger$ and $f_{i,j}^\dagger$ at well-chosen points, combined with an interpolation scheme to retrieve the values in between stored data points. 

This discretization operation will inevitably cause small discrepancies between the stored  values and the exact values corresponding to the definitions in \cref{eq:adj_xs_flx_gen}. However, this will not bias the adjoint simulations, since the adjoint collision sampling procedure proposed in \cref{sec:adj_coll_event} is unbiased for any choice of $\sigma_{i,j}^\dagger$ and $f_{i,j}^\dagger$, as long as the weight correction in \cref{eq:weight_corr} is evaluated in a way that is consistent with the adopted adjoint data, i.e.~, provided that the discretized $f_{i,j}^\dagger$ used for the weight correction precisely corresponds to the probability density associated to the discretized cumulative distribution used for sampling. This fact was first emphasized by De Matteis and Simonini in Ref.~\citenum{matteis_new_1978}.

However, we still want the discretized data to be as close as possible to the definition given in \cref{eq:adj_xs_flx_gen}, so as to preserve the sought variance reduction effect.

\subsection{General strategy}
\label{sec:gen_strat}

To begin with, observe that the adjoint cross section $\sigma_{i,j}^\dagger$ in the definition \cref{eq:adj_xs_flx_gen} is used as normalization coefficient for the adjoint distribution law $f_{i,j}^\dagger$. In other words, introducing the non-normalized distribution law
\begin{equation}
    \tilde{f}^\dagger_{i,j}(E'\to E,\mu_l) : = \sigma_{i,j}(E)f_{i,j}(E\to E', \mu_l) \nu_{i,j}(E) g(E),
    \label{eq:def_non_norm_adj_distrib}
\end{equation}
and the associated non-normalized cumulative distribution in energy
\begin{equation}
\label{eq:def_non_norm_cum}
    \tilde{F}^\dagger_{i,j}(E',E) = \int _{E_\text{min}}^E \tilde{f}_{i,j}^\dagger(E'\to \tilde{E})d\tilde{E},
\end{equation}
we can retrieve the adjoint cross section using
\begin{equation}
    \label{eq:adj_xs_from_non_norm_cum}
    \sigma_{i,j}^\dagger(E') = \tilde{F}_{i,j}(E',E_\text{max})-\tilde{F}_{i,j}(E',E_\text{min}).
\end{equation}
Note that in \cref{eq:def_non_norm_cum,eq:adj_xs_from_non_norm_cum} we introduced global lower and upper bounds $E_\text{min}$ and $E_\text{max}$ on the energy range to avoid numerical integration from $-\infty$ to $\infty$. In order to sample the energy $E$ from $f_{i,j}^\dagger$ using the non-normalized cumulative, we can use the inverse transform method, sampling a uniform random number in a range $[\tilde{F}_{i,j}(E',E_\text{max}),\tilde{F}_{i,j}(E',E_\text{min})]$ instead of $[0,1]$. The density $f_{i,j}^\dagger$ can also be evaluated using
\begin{equation}
    f_{i,j}^\dagger(E\to E') = \frac{\tilde{f}_{i,j}^\dagger(E\to E')}{\sigma_{i,j}^\dagger(E')}.
\end{equation}
Storing the non-normalized adjoint distribution law $\tilde{f}^\dagger_{i,j}$ and its cumulative $\tilde{F}^\dagger_{i,j}$ will thus allow sampling from and evaluating the density of the adjoint distribution law $f_{i,j}^\dagger$, and evaluating the adjoint cross section $\sigma^\dagger_{i,j}$. The cumulative with respect to energy defined in \cref{eq:def_non_norm_cum} only allows sampling the energy $E$. One could similarly prepare the cumulative with respect to the angle to sample $\mu_l$. In practice, the sampling of the angular component of the collision events is strongly reaction-dependent, and will be detailed for each channel in the following. 

The strategy presented above needs to be specified for every type of direct distribution law that can be encountered in the nuclear data library. In view of an eventual implementation of adjoint transport routines in \tripolir, which relies on processed nuclear data in the form of ACE files \cite{conlin_compact_2019}, we will try to cover all distribution laws that can be met in ACE files in modern nuclear data libraries. For ease of identification, we will also provide for each the associated `MT' identifier corresponding to the ENDF format \cite{endf_manual}.

\subsection{Unidimensional and bidimensional data formats}
\label{sec:bidim_interp}

Before discussing the implementation details for all reaction channels, let us take some time to dwell on the specific procedure used to store and then read the adjoint nuclear data. For the needs of adjoint collision sampling, we will need to store unidimensional data, such as adjoint cross sections $\sigma_{i,j}^\dagger(E')$, which only depend on $E'$, and bidimensional data, such as adjoint distribution laws $f^\dagger_{i,j}(E'\to E)$, where to every energy $E'$ corresponds a distribution along $E$. 

We will illustrate the procedure of bidimensional data storage with the storage of a bidimensional non-normalized distribution law $\tilde{f}^\dagger(E'\to E)$ as defined in \cref{eq:def_non_norm_adj_distrib}. Since bidimensional distributions are composed of unidimensional distributions, this description will cover also the sampling of unidimensional non-normalized distributions. As mentioned, the specific treatment used to sample the angle $\mu_l$ will be provided later for each reaction. In this section, the $\{i,j\}$ dependence of the nuclear data will be dropped, for readability.

\begin{figure}[ht]
    \centering
    \includegraphics[width=.5\textwidth]{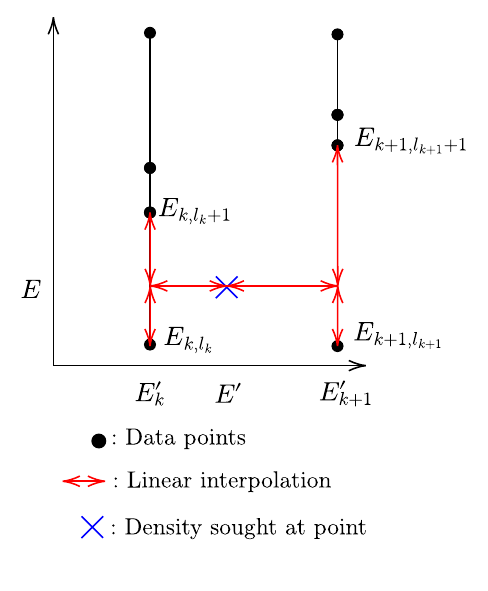}
    \caption{Bidimensional interpolation scheme for the storage of $\tilde{f}^\dagger(E'\to E)$. Data is first interpolated in $E$ in the two bracketing distributions in $E'_k$ and $E'_{k+1}$, then it is interpolated in $E'$ between the two previously obtained values to obtain the final value. See the main text for the detailed procedure. Black dots represent the positions of the discretization points in $(E',E)$ space; red arrows represent the successive interpolations and their directions (along $E$ or along $E'$), and the blue cross represents the point $(E',E)$ where the interpolated value is sought.}
    \label{fig:2D_interp}
\end{figure}

Using the definition in \cref{eq:def_non_norm_adj_distrib}, the non-normalized distribution law $\tilde{f}^\dagger$ is evaluated at a set of well-chosen incident energies $E'_k$, $k=1,\cdots,N$. For each $E'_k$, we evaluate $\tilde{f}^\dagger$ at a set of outgoing energies $E_{k,l}$, $l=1,\cdots,M$. Properly choosing the set of discretization points $E'_k$ and $E_{k,l}$ requires careful consideration, in order to ensure that interpolation is sufficiently precise, while avoiding prohibitive storage size. This will be achieved by using the generic bidimensional discretization algorithms described in \cref{sec:disc_algo}. The values of $\tilde{f}^\dagger$ at any arbitrary pair $(E',E)$ can be obtained by linearly interpolating twice. First, we find the two values $\{E'_k, E'_{k+1}\}$ that bracket $E'$ (i.e. such that $E'_k\le E' < E'_{k+1}$). This can be efficiently done by using binary search on the set of ordered $E'_{k}$. Both $E'_k$ and $E'_{k+1}$ correspond to functions of $E$: $\tilde{f}^\dagger(E'_k\to E)$ and $\tilde{f}^\dagger(E'_{k+1}\to E)$, respectively. We now need to interpolate the unidimensional functions along $E$. Again, we find $l_k$ and $l_{k+1}$ such that $\{E_{k,l_k},E_{k,l_k+1}\}$ and $\{E_{k+1,l_{k+1}},E_{k+1,l_{k+1}+1}\}$ bracket $E$. By linearly interpolating in $E$ we obtain the values $\tilde{f}^\dagger(E'_k\to E)$ and $\tilde{f}^\dagger(E'_{k+1}\to E)$. Finally, we obtain $\tilde{f}^\dagger (E' \to E)$ by linearly interpolating (with respect to $E'$) between these two values. This scheme is illustrated in \cref{fig:2D_interp}.

In order to sample from such bidimensional distributions, we numerically integrate the cumulative $\tilde{F}^\dagger$ at the same sets of points $E'_k$ and $E_{k,l}$. As the distribution $\tilde{f}$ at a fixed $E'_k$ is interpolated as a piecewise linear function in between values of $E_{k,l}$, numerical integration can be performed exactly by using the trapezoidal rule. The exact correspondence between the discretized cumulative and the discretized density is mandatory to ensure unbiasedness via the weight correction, as mentioned in \cref{sec:prep_adj_data_requirements}. Since the definition of $\tilde{f}^\dagger$ involves the arbitrary function $g(E)$, which can span several orders of magnitude as $E$ ranges from $E_\text{min}$ to $E_\text{max}$, the values of $\tilde{f}^\dagger$ can also span several orders of magnitude. When numerically integrating such a function along $E$, catastrophic loss of precision can occur if the small addition from $\tilde{f}$ at a point is negligible with respect to the accumulated value from $\tilde{f}$ at another point. To overcome this issue, it is sometimes preferable to conduct a reverse integration from $E_\text{max}$ to $E_\text{min}$, so that the smallest contributions of $\tilde{f}$ occur for values of $\tilde{F}$ closer to zero, which increases the precision of the floating point storage. Since the cumulative is non-normalized, the resulting shift in value will not influence the sampling of the distribution, and this is also true for the evaluation of the adjoint cross section $\sigma^\dagger$ using \cref{eq:adj_xs_from_non_norm_cum}. 

To sample from $\tilde{f}^\dagger$ at a given $E'$, we find $\{E'_k, E'_{k+1}\}$ that bracket $E'$. Then, we proceed by sampling $E$ from $\tilde{f}^\dagger(E'_k\to E)$ with a probability
\begin{equation}
    \label{eq:stoch_sampling}
    \mathbb{P}(k) = \frac{E'_{k+1}-E'}{E'_{k+1}-E'_{k}},
\end{equation}
and from $\tilde{f}^\dagger(E'_{k+1}\to E)$ with a probability $\mathbb{P}(k+1)= 1- \mathbb{P}(k)$. We will denote $m$ the index chosen at this step, which can take either $m=k$ or $m=k+1$. In order to sample from the unidimensional non-normalized density $\tilde{f}^\dagger(E'_m \to E)$, we use the inverse transform method, sampling a random number $\xi$ uniformly in $\left[\tilde{F}^\dagger(E'_m,E_{m,\text{min}}),\tilde{F}^\dagger(E'_m,E_{m,\text{max}})\right]$. When performing the inverse transform sampling, we need to take into account that, if the values of $\tilde{f}^\dagger$ are linearly interpolated in $E$, then the associated cumulative $\tilde{F}^\dagger$ is quadratically interpolated in $E$. It can easily be shown that the stochastic interpolation between $\tilde{f}^\dagger(E'_k\to E)$ and $\tilde{f}^\dagger(E'_{k+1}\to E)$ corresponds to sampling from a distribution that is exactly a linear interpolation (in $E'$) between the bracketing distributions. Therefore, the interpolated value of $\tilde{f}^\dagger$ correponds to the actual density we sampled from, and the procedure is unbiased.

Finally, in order to obtain the value of the adjoint cross section $\sigma_{i,j}^\dagger(E')$ at an energy $E'$, we can simply interpolate between the values $\sigma_{i,j}^\dagger(E'_k)$ and $\sigma_{i,j}^\dagger(E'_{k+1})$ obtained via \cref{eq:adj_xs_from_non_norm_cum} applied to distributions evaluated at $E'_k$ and $E'_{k+1}$.

\subsubsection{Distribution edges}

In some cases, the direct density $f(E\to E')$ has some specific boundaries that we want to preserve in the adjoint density $f^\dagger(E'\to E)$. For example, some reactions need to ensure that $E'\le E$, leading to a potentially sharp edge in the density $f$ at the line $E=E'$. 

\begin{figure}[hbtp]
    \centering
    \includegraphics[width=.7\textwidth]{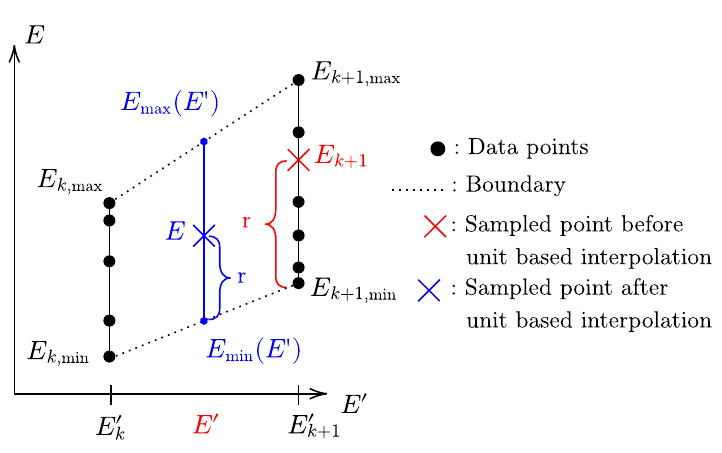}
    \caption{The bidimensional unit-based interpolation scheme to preserve the distribution edges. After sampling a preliminary value of $E_{k+1}$ in the distribution at $E'_{k+1}$ using stochastic interpolation, the distribution fraction $r$ is computed and used to interpolate the final value $E$ in between the bounds at $E'$. The distribution edges are interpolated linearly. See text for the detailed procedure. Black dots represent the positions of the discretization points in $(E',E)$ space, red quantities are relative to the stochastic sampling before unit-based interpolation, and the blue quantities are relative to the result after applying unit-based interpolation.}
    \label{fig:2D_Sampling_UBI}
\end{figure}

To preserve the boundary of the adjoint distribution, we use a procedure known as {\em unit-based interpolation} when sampling from it \cite{endf_manual}. This procedure is already widely used in existing direct distribution laws. For example, in commonly used Monte Carlo codes such as MCNP or \tripolir, unit-based interpolation is performed in the continuous tabular distribution (ENDF LAW 4), Kalbach distributions (ENDF LAW 44) and Kalbach tabular distributions (ENDF LAW 61); see e.g.~Section 2.4.3.5.4 of the MCNP user manual Ref.~\citenum{mcnp63}. After the stochastic sampling described above, we do not keep the sampled value of $E$ as is. We will denote $E_m$ the temporary $E$ value sampled with the previous technique from the distribution at $E'_m$. We evaluate the fraction $r$ within the bounds of the distribution at $E'_m$:
\begin{equation}
    r = \frac{E-E_{m,\text{min}}}{E_{m,\text{max}}-E_{m,\text{min}}}.
\end{equation}
The resulting value of $E$ will then be interpolated using $r$ between the bounds of the distribution at $E'$:
\begin{equation}
    E= rE_\text{max}(E')+(1-r)E_\text{min}(E'),
\end{equation}
where the bounds are usually obtained by linear interpolation of the bounds at the nearest distributions, i.e.
\begin{equation}
\label{eq:linear_bounds}
    \begin{aligned}
        E_\text{min}(E') &:= E_{k,\text{min}} + \left( E_{k+1,\text{min}}-E_{k,\text{min}}\right)\frac{E'-E_k}{E_{k+1}-E_{k}}\\
        E_\text{max}(E') &:= E_{k,\text{max}} + \left( E_{k+1,\text{max}}-E_{k,\text{max}}\right)\frac{E'-E_k}{E_{k+1}-E_{k}}.\\
    \end{aligned}
\end{equation}
The sampling process is summarized in \cref{fig:2D_Sampling_UBI}.

\begin{figure}[hbtp]
    \centering
    \includegraphics[width=.7\textwidth]{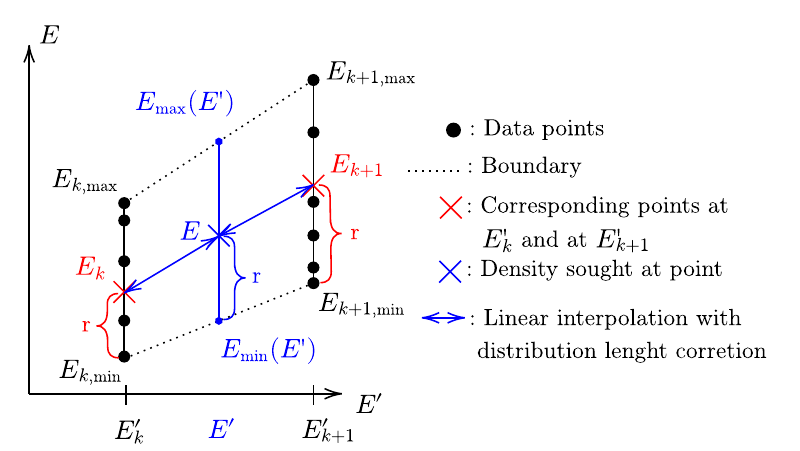}
    \caption{Unit-based interpolation scheme. See text for detailed procedure. Black dots represent the positions of the discretization points in $(E',E)$ space, the blue cross represents the point $E',E$ at which the density is to be evaluated, densities in the distribution at $E'_k$ and $E'_{k+1}$ are evaluated at $E_k$ and $E_{k+1}$ (red crosses), blue arrows represent interpolation in $E'$ with a correction factor for the relative lengths of the support of the distributions. }
    \label{fig:2D_Interp_UBI}
\end{figure}

Using this new interpolation technique in the sampling process, however, modifies the interpolation technique for the density that was presented above. Once the distributions at $E'_k$ and $E'_{k+1}$ have been found, we compute the bounds at $E'$ ($E_\text{min}(E')$ and $E_\text{max}(E')$) via \cref{eq:linear_bounds}. An inverse unit-based sampling is used to compute $E_{k}$, the corresponding value of $E$ from the distributions at $E'_{k}$, and $E_{k+1}$, the corresponding value of $E$ from the distributions at $E'_{k+1}$. The densities $f^\dagger(E'_k\to E_k)$ and $f^\dagger (E'_{k+1}\to E_{k+1})$ are then evaluated by linear interpolation. The density at $E'_k,E_k$ (respectively $E'_{k+1},E_{k+1}$) is then corrected to take into account that the unit-based interpolation changed the size of the support of the distribution. The correction factor is the ratio of the size of the support of the distribution at $E'_k$ (resp.~$E'_{k+1}$), that can be computed as $E_{k,\text{max}}-E_{k,\text{min}}$ (resp $E_{k+1,\text{max}}-E_{k+1,\text{min}}$), over the size of the support of the distribution at $E'$, that can be computed as $E_\text{max}(E')-E_\text{min}(E')$. Finally, the density $f^\dagger(E',E)$ is obtained as a linear interpolation (in $E'$) of the two corrected densities. This procedure is summarized in \cref{fig:2D_Interp_UBI}.

\subsection{Elastic scattering (MT2)}
\label{sec:el_scat}

The most common reaction met in neutron transport is elastic scattering (MT2). All the formulas regarding the kinematics of the elastic collision are derived in Appendix A of Ref.~\citenum{hoogenboom_adjoint_1977}. We consider first the case where thermal motion effects can be neglected, and we adopt the nucleus-at-rest hypothesis. For elastic scattering, deflection is usually characterized in terms of an angular distribution $g(E,\mu_{cm})$, which yields $\mu_{cm}$ in the center-of-mass frame for a given incident energy $E$ \cite{endf_manual}. The use of $\mu_{cm}$ is helpful in that the distribution always ranges from $\mu_{cm}=-1$ to $\mu_{cm}=1$. This would not be the case in laboratory frame for hydrogen, for example. The cosine $\mu_{cm}$ is related to the outgoing energy by a simple expression. Once $\mu_{cm}$ is sampled from $g(E,\mu_{cm})$, we retrieve the outgoing energy $E'$ from
\begin{equation}
\label{eq:el_scat_Ep_from_mu}
    E'= E\frac{A^2 + 2A\mu_{cm}+1}{(A+1)^2},
\end{equation}
where $A$ is the mass of the collided nuclide in units of neutron mass, and the deflection cosine $\mu_l$ in the laboratory frame from:
\begin{equation}
\label{eq:el_scat_mulab_from_mu}
    \mu_l = \frac{1+A\mu_{cm}}{\sqrt{A^2+2A\mu_{cm} +1}}.
\end{equation}

In order to construct the adjoint data via \cref{eq:adj_xs_flx_gen} and to apply weight correction via \cref{eq:weight_corr}, we also need to evaluate the energy distribution $f(E\to E')$ at an arbitrary point $E,E'$. For this purpose, given a pair $E,E'$ we begin by computing the corresponding $\mu_{cm}$ via:
\begin{equation}
    \label{eq:el_scat_mucm_from_E_Ep}
    \mu_{cm} = -1 +2 \frac{E'/E- \alpha}{1-\alpha},
\end{equation}
where $\alpha$ is customarily defined as $\alpha = (A-1)^2/(A+1)^2$. We compute then the density $g(E,\mu_{cm})$ via the bidimensional interpolation described in \cref{sec:bidim_interp}, and finally retrieve the sought density via
\begin{equation}
    f(E\rightarrow E')  = g(E,\mu_{cm}) \frac{d\mu_{cm}}{dE'}=\frac{2g(E,\mu_{cm})}{E(1-\alpha)}.
\end{equation}

In order to store the adjoint elastic collision reaction, we could simply store the non-normalized energy distribution $\tilde{f}^\dagger(E\to E')$ defined in \cref{eq:def_non_norm_adj_distrib} using the bidimensional storage procedure described in \cref{sec:bidim_interp}. This, however, would yield sharp energy edges along the lines $E'=E$ and $E'= \alpha E$. Storing the non-normalized angular distribution $\tilde{g}^\dagger(E',\mu_{cm})$ instead allows for `clean' boundaries at $\mu_{cm}=\pm1$. The quantity $\tilde{g}^\dagger(E',\mu_{cm})$ is obtained from $\tilde{f}^\dagger(E\to E')$ via:
\begin{equation}
    \label{eq:el_scat_adjoint_angular}
    \begin{aligned}
    \tilde{g}^\dagger(E',\mu_{cm})&=\tilde{f}^\dagger(E'\rightarrow E) \left|\frac{dE}{d\mu_{cm}}\right| \\
    &= \frac{2E^2A \tilde{f}^\dagger(E'\rightarrow E)}{E'(1+A)^2}.
    \end{aligned} 
\end{equation}
The adjoint non-normalized angular distribution $\tilde{g}^\dagger$ and its cumulative $\tilde{G}^\dagger$ are stored in a bidimensional data structure similarly to how $\tilde{f}^\dagger$ would have been stored. The adjoint cross section can similarly be retrieved via:
\begin{equation}
    \label{eq:el_scat_retrieve_adj_sigma}
    \sigma^\dagger(E') = \tilde{G}^\dagger(E',+1)-\tilde{G}^\dagger(E',-1).
\end{equation}

To sample the adjoint elastic collision event, a $\mu_c$ is first sampled at $E'$ via $\tilde{g}^\dagger(E',\mu_c)$. We then compute $E$ by reversing \cref{eq:el_scat_Ep_from_mu}:
\begin{equation}
    E = E'\frac{(A+1)^2}{A^2 + 2A\mu_{cm}+1},
\end{equation}
and compute $\mu_l$ via \cref{eq:el_scat_mulab_from_mu}. The value of $f^\dagger(E'\to E)$ for weight correction is obtained by inverting \cref{eq:el_scat_adjoint_angular}.

This technique works, but demands storing bidimensional data. This can lead to a memory burden if the stored data is highly heterogeneous and requires a lot of data points. Unfortunately, the definition of $\tilde{f}^\dagger$ (upon which $\tilde{g}^\dagger$ is based) includes the direct cross section $\sigma(E)$ (see \cref{eq:def_non_norm_adj_distrib}), which is highly heterogeneous in the resonance range for many heavy nuclei.

Fortunately, there exists an alternative, more efficient and resilient technique when the elastic scattering is only weakly non-isotropic, which is usually the case in the resonance region. In the case of isotropic elastic scattering, the distribution law reads
\begin{equation}
    f(E\to E')= 
    \begin{cases}
        \frac{1}{(1-\alpha)E}\quad &\text{if }E'\in[E'_\text{min}(E),E'_\text{max}(E)]\\
        0 \quad &\text{otherwise}
    \end{cases},
\end{equation}
with $E'_\text{min}(E)=\alpha E$ and $E'_\text{max}(E)=E$. In this case, the density depends on $E'$ only through the bounds of the distribution $E'_\text{min}(E)$ and  $E'_\text{max}(E)$. This allows us to store the adjoint distribution law as a unidimensional function of $E$, computing the bounds on-the-fly. We store the function:
\begin{equation}
    \tilde{h}^\dagger(E) := \frac{\sigma(E)\nu(E)g(E)}{(1-\alpha)E},
\end{equation}
(note that $\nu(E)=1$ for elastic scattering), and its cumulative $\tilde{H}^\dagger(E)$, again integrated via the trapezoidal rule in between the interpolation points. We can then easily retrieve the adjoint density via
\begin{equation}
    \label{eq:el_scat_retrieve_f_from_h}
    f^\dagger(E'\to E) = 
    \begin{cases}
        \tilde{h}^\dagger(E)/ \sigma^\dagger(E') \quad&\text{if }E\in[E_\text{min}(E'),E_\text{max}(E')]\\
        0 \quad &\text{otherwise}
    \end{cases},
\end{equation}
where $\sigma^\dagger(E')$ can be promptly computed as
\begin{equation}
    \label{eq:el_scat_retrieve_xs_from_H}
    \sigma^\dagger(E') = \tilde{H}^\dagger(E_\text{max}(E'))-\tilde{H}^\dagger(E_\text{min}(E')),
\end{equation}
with $E_\text{max}(E')=E'/\alpha$ and $E_\text{min}(E')=E'$. In order to sample from $f^\dagger(E'\to E)$, we proceed to an inverse transform sampling using the cumulative $\tilde{H}^\dagger$, drawing a random number uniformly in $[\tilde{H}^\dagger(E_\text{max}(E')),\tilde{H}^\dagger(E_\text{min}(E'))]$.

In practice, the adjoint elastic scattering data is stored in two parts: at lower energies, where the cross section is resonant and the angle distribution is typically close to isotropic, we use the isotropic technique; at higher energies, where the cross sections are usually smoother and the angle distribution is typically anisotropic, we store the adjoint distribution law as a bidimensional function. The pivot energy $E'_\text{pivot}$ that separates these regimes is chosen algorithmically by inspecting the angular distribution and the cross section and trying to find an energy higher than the resonances and lower than the anisotropic energy range; if the algorithm fails to do so, it prioritizes avoiding resonances. Using the isotropic technique in an energy domain where anisotropy occurs will not induce any bias, but might prove detrimental to the variance if the anisotropy is very strong.

\begin{figure}[htb]
    \centering
    \begin{subfigure}[t]{0.49\textwidth}
        \centering
        \includegraphics[trim={0 0.8cm 0 0.53cm},clip,width=\textwidth]{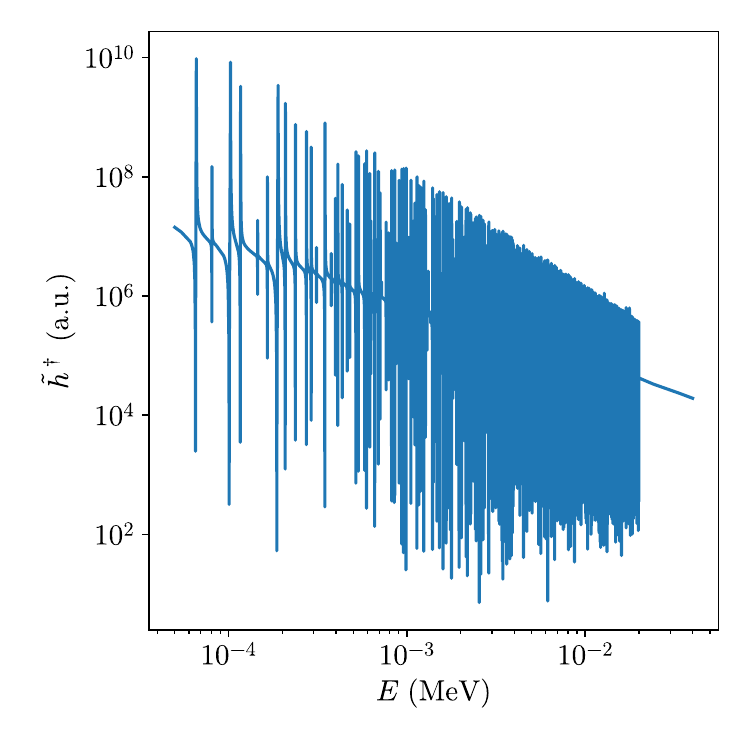}
    \end{subfigure}%
    ~ 
    \begin{subfigure}[t]{0.49\textwidth}
        \centering
        \includegraphics[trim={1cm 0.5cm 0 1.5cm},clip,width=\textwidth]{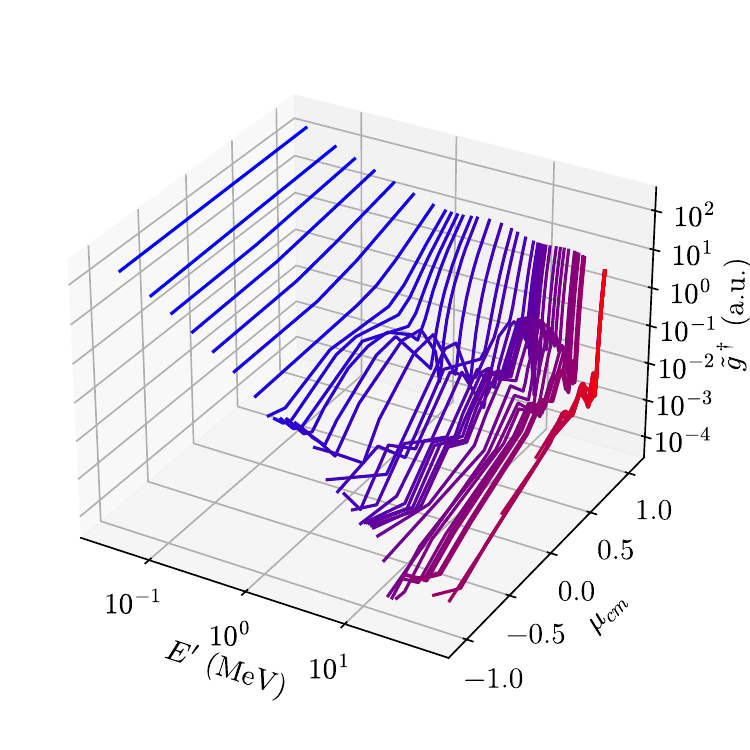}
    \end{subfigure}
    \caption{Adjoint distribution law for the elastic scattering of \textsuperscript{238}U prepared using definition from \cref{eq:adj_xs_flx_gen} with $g(E)=1/E$. \textbf{On the left}, the isotropic part of the distribution, stored as a unidimensional distribution $\tilde{h}^\dagger(E)$. This discretized distribution contains 14~965 data points. It ranges from $50\text{ eV}$ (which is the limit of the thermal treatment in this calculation) to $~40\text{ keV}$. It displays the resonances present in the cross section $\sigma(E)$, and the global $1/E$ shape from $g(E)$. \textbf{On the right}, the anisotropic part of the distribution, stored as a bidimensional adjoint angular distribution $\tilde{g}^\dagger (E',\mu_{cm})$. This discretized distribution contains 735 data points, split into 57 unidimensional distributions, each with an average of 13 data points. It ranges from $E'_\text{pivot}=39\text{ keV}$ to $E_\text{max}= 50\text{ MeV}$. The unidimensional distributions are colored with a gradient corresponding to the index $k$ of their corresponding entering energy $E'_k$. The distribution contains the anisotropy present at high energies, as well as the $1/E$ shape from $g(E)$.}
    \label{fig:distrib_U8_elscat}
\end{figure}

\begin{figure}[htb]
    \centering
    \begin{subfigure}[t]{0.49\textwidth}
        \centering
        \includegraphics[trim={0 0.8cm 0 0.53cm},clip,width=\textwidth]{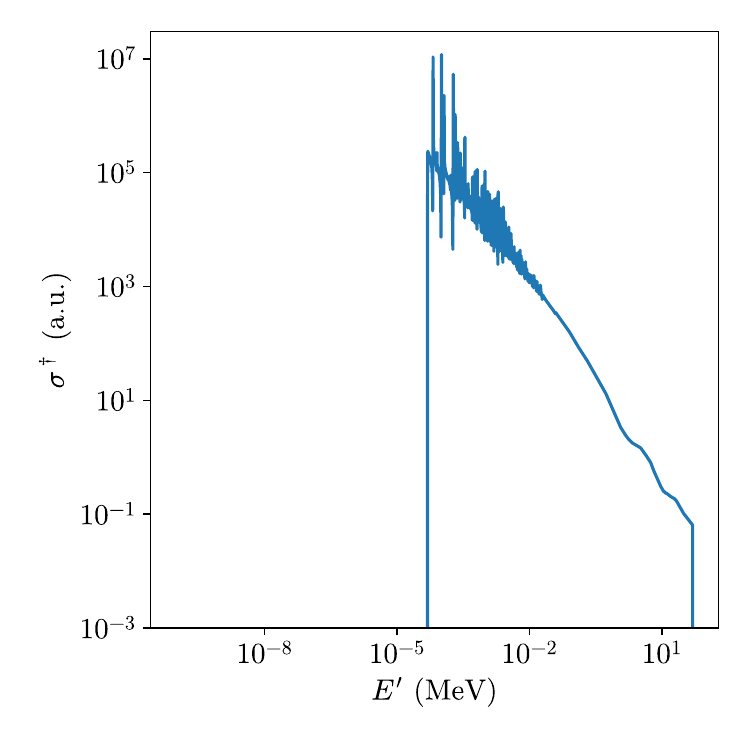}
    \end{subfigure}%
    ~ 
    \begin{subfigure}[t]{0.49\textwidth}
        \centering
        \includegraphics[trim={0 0.8cm 0 0.53cm},clip,width=\textwidth]{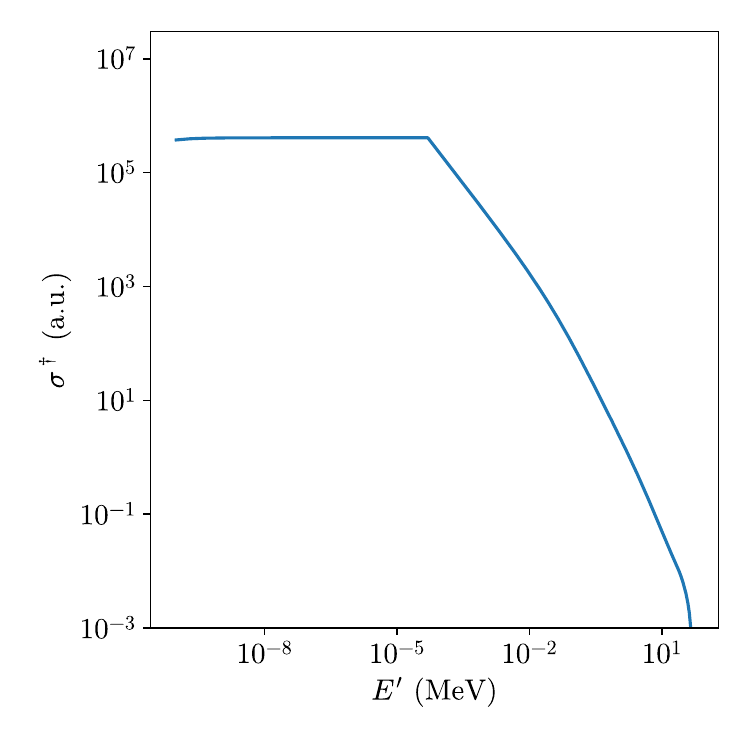}
    \end{subfigure}
    \caption{Adjoint cross sections $\sigma^\dagger(E')$ for elastic scattering of \textsuperscript{238}U (left) and \textsuperscript{1}H (right). As the formula for the adjoint cross section \cref{eq:adj_xs_flx_gen} effectively acts as a rolling average over the range of accessible energies, the adjoint cross section features smoothed resonances. The shape of $g(E)=1/E$ is visible in both cross sections. Even if the non-thermal elastic scattering reaction is turned off below $50 \text{ keV}$, the adjoint cross section of \textsuperscript{1}H is constant below that energy as adjoint elastic scattering on $A\approx1$ can always go to $E>50 \text{ keV}$. }
    \label{fig:adj_xs_U_H_elscat}
\end{figure}

Examples of adjoint distribution laws and adjoint cross sections produced for the elastic scattering reaction on \textsuperscript{238}U and \textsuperscript{1}H are provided in \cref{fig:distrib_U8_elscat,fig:adj_xs_U_H_elscat}.

\subsection{Inelastic scattering on discrete level (MT51-90)}

Inelastic scattering on discrete energy levels (MT51-90) will be treated along the same lines as for the case of elastic scattering presented in \cref{sec:el_scat}. All the formulas regarding the kinematics of the inelastic collision on a discrete level are derived in Appendix A of Ref.~\citenum{hoogenboom_adjoint_1977}. Similarly to elastic scattering, the direct angular distribution is again provided through an angular distribution $g(E,\mu_{cm})$. A relation between the deflection cosine and the outgoing energy exists, although the formula is different from the case of elastic scattering. Once $\mu_{cm}$ is sampled, the outgoing energy $E'$ is determined from
\begin{equation}
\label{eq:inel_scat_Ep_from_mu}
    E' = E \frac{A^2(1-\varepsilon/E)+2A\mu_{cm}\sqrt{1-\varepsilon/E}+1}{(A+1)^2} ,
\end{equation}
where $\varepsilon = Q(A+1)/A$, with $Q$ the discrete energy level. Note that if $Q=0$, the inelastic scattering turns into elastic scattering and \cref{eq:inel_scat_Ep_from_mu} turns into \cref{eq:el_scat_Ep_from_mu}. The deflection cosine $\mu_l$ in the laboratory frame is obtained from
\begin{equation}
\label{eq:inel_scat_mulab_from_mu}
    \mu_l = \frac{A\mu_{cm}\sqrt{1-\varepsilon/E}+1}{\sqrt{A^2(1-\varepsilon/E)+2A\mu_{cm}\sqrt{1-\varepsilon/E}+1}}.
\end{equation}

In order to evaluate the energy distribution $f(E\to E')$ at an arbitrary point $E,E'$, we begin by computing the corresponding $\mu_{cm}$ for a given pair $E,E'$ via:
\begin{equation}
    \mu_{cm} = \frac {2} {(1-\alpha)\sqrt{1-\epsilon/E}}\frac{E'}{E}-\frac 1 {2A\sqrt{1-\epsilon/E}} - \frac{A}{2}\sqrt{1-\epsilon/E}.
\end{equation}
Then, we compute the density $g(E,\mu_{cm})$, and finally retrieve the sought density via
\begin{equation}
    f(E\rightarrow E')  = g(E,\mu_{cm}) \frac{d\mu_{cm}}{dE'}=\frac{2g(E,\mu_{cm})}{E(1-\alpha)\sqrt{1-\varepsilon/E}}.
\end{equation}

In order to store the adjoint distribution, we also use a non-normalized angular distribution $\tilde{g}^\dagger(E',\mu_{cm})$. Note that here the relation between $E$, $E'$ and $\mu_{cm}$ is not as straightforward as in the case of elastic scattering, and for energies $E'$ lower than\footnote{Note that in some very rare cases (e.g. the $(n,n2)$ -MT52- reaction of \textsuperscript{186}Re\textsuperscript{*} in the ENDF-B/VIII.0 nuclear data library) the discrete energy level $Q$ can be negative. The only effect of this is the limit value $\varepsilon/(A+1)^2$ that is replaced with $|\varepsilon|A^2/(A+1)^2$. All the other equations and sampling procedures presented in this section remain correct.} $\varepsilon/(A+1)^{2}$ there exist two values of $E$ that can reach $E'$ with a deflection of $\mu_{cm}$. An alternative strategy would be to store the adjoint distribution using $\mu_l$ instead of $\mu_{cm}$, since using $\mu_l$ the pair $E',\mu_l$ determines a unique $E$. A thorough treatment is provided in the appendices of Hoogenboom's PhD thesis \cite{hoogenboom_adjoint_1977}. In practice, we ensure that the limiting value $\varepsilon/(A+1)^{2}$ always lies in the isotropic portion of the distribution (see the choice of $E'_\text{pivot}$ at the end of this section), which circumvents this problem. The non-normalized adjoint angular distribution $\tilde{g}^\dagger(E',\mu_{cm})$ is obtained from $\tilde{f}^\dagger(E'\rightarrow E)$ using
\begin{equation}
    \label{eq:inel_scat_adjoint_angular}
    \begin{aligned}
    \tilde{g}^\dagger(E',\mu_{cm})&=\tilde{f}^\dagger(E'\rightarrow E) \left|\frac{dE}{d\mu_{cm}}\right| \\
    &= 
 \tilde{f}^\dagger(E'\rightarrow E)\frac{2AE^2(1-\varepsilon/E)}{A\mu_{cm}\varepsilon+(A^2\varepsilon+(A+1)^2E')\sqrt{1-\varepsilon/E}}.
    \end{aligned} 
\end{equation}
Again, we retrieve the adjoint cross section via \cref{eq:el_scat_retrieve_adj_sigma}.

In order to sample the adjoint collision event, a deflection cosine $\mu_c$ is sampled at $E'$ via $\tilde{g}^\dagger(E',\mu_c)$. We can then compute $E$ using
\begin{equation}
    \label{eq:inel_scat_E_from_Ep_mu_c}
    \begin{aligned}
        E &= \frac{QA(A+1)}{A^2-C^2},\quad\text{with}\\
        C &= \frac{\sqrt{\mu_{cm}^2+(1+\gamma E')(A^2\gamma E'-1) }-\mu_{cm}}{1+\gamma E'} \quad \text{and}\\
        \gamma &=  \frac{1+A}{QA},
    \end{aligned}
\end{equation}
and compute $\mu_l$ using \cref{eq:inel_scat_mulab_from_mu}. Both \cref{eq:inel_scat_adjoint_angular,eq:inel_scat_E_from_Ep_mu_c} are only valid for $E'>\varepsilon/(A+1)^{2}$, due to the aforementioned issue. The evaluation of $f^\dagger(E'\to E)$ required for weight correction is obtained by inverting \cref{eq:inel_scat_adjoint_angular}.

For the same reasons as for elastic scattering, in the case where the inelastic scattering is weakly non-isotropic, we can store the distribution law in a unidimensional function of $E$. If the inelastic scattering is isotropic, the distribution law reads
\begin{equation}
    f(E\to E')= 
    \begin{cases}
        \frac{1}{(1-\alpha)E\sqrt{1-\varepsilon/E}}\quad &\text{if }E'\in[E'_\text{min}(E),E'_\text{max}(E)]\\
        0 \quad &\text{otherwise}
    \end{cases},
\end{equation}
where the bounds are 
\begin{equation}
    \label{eq:inel_scat_bounds_Ep}
    \begin{aligned}
        E'_{min} &= E \left( \frac{A\sqrt{1-\varepsilon/E}-1}{A+1} \right)^2 \\
        E'_{max} &= E\left( \frac{A\sqrt{1-\varepsilon/E}+1}{A+1} \right)^2 ,
    \end{aligned}
\end{equation}
which can be obtained by evaluating \cref{eq:inel_scat_Ep_from_mu} at the extremal values $\mu_{cm}= \pm1$. We store the function
\begin{equation}
    \tilde{h}^\dagger(E) := \frac{\sigma(E)\nu(E)g(E)}{(1-\alpha)E\sqrt{1-\varepsilon/E}},
\end{equation}
(once again, note that $\nu(E)=1$ for inelastic scattering), and its cumulative $\tilde{H}^\dagger(E)$. We can then easily retrieve the adjoint density via \cref{eq:el_scat_retrieve_f_from_h}, and $\sigma^\dagger(E')$ can also easily be evaluated using \cref{eq:el_scat_retrieve_xs_from_H}. The formula to obtain the bounds of $E$ knowing $E'$ cannot be obtained from \cref{eq:inel_scat_E_from_Ep_mu_c} by taking $\mu_{cm}=\pm1$, since \cref{eq:inel_scat_E_from_Ep_mu_c} is not valid on the whole energy range. Therefore, we have to use the relation derived by Hoogenboom in Ref.~\citenum{hoogenboom_adjoint_1977}
\begin{equation}
    \label{eq:inel_scat_E_from_Ep_and_mu_l}
    \begin{aligned}
         E = \bigg[&(A+1)^2(A^2+2\mu_l^2-1)E' + A^2(A^2-1)\varepsilon \\&-2\mu_l(A+1)\sqrt{(A+1)^2(A^2+\mu_l^2-1)E'^2+A^2(A^2-1)\varepsilon E'}\bigg] /(A^2-1)^2,
    \end{aligned}
\end{equation}
which allows determining $E$ from $E'$ and $\mu_l$ over the whole range. We can then obtain the bound $E_\text{max}(E)$ (resp. $E_\text{min}(E)$) by evaluating \cref{eq:inel_scat_E_from_Ep_and_mu_l} at $\mu_l=-1$ (resp. $\mu_l=1$).

The adjoint inelastic scattering is stored in two parts: at lower energies, we use the isotropic technique; at higher energies, where the angle distribution is generally non-isotropic, we store the adjoint distribution law as a bidimensional function. The pivot energy $E'_\text{pivot}$ between the two regimes is chosen algorithmically by inspecting the angular distribution and the cross section. We ensure that the pivot energy $E'_\text{pivot}$ is kept above $\varepsilon/(A+1)^{2}$, so that \cref{eq:inel_scat_adjoint_angular,eq:inel_scat_E_from_Ep_mu_c} remain valid.

\begin{figure}[htb]
    \centering
    \begin{subfigure}[t]{0.49\textwidth}
        \centering
        \includegraphics[trim={0 0.8cm 0 0.53cm},clip,width=\textwidth]{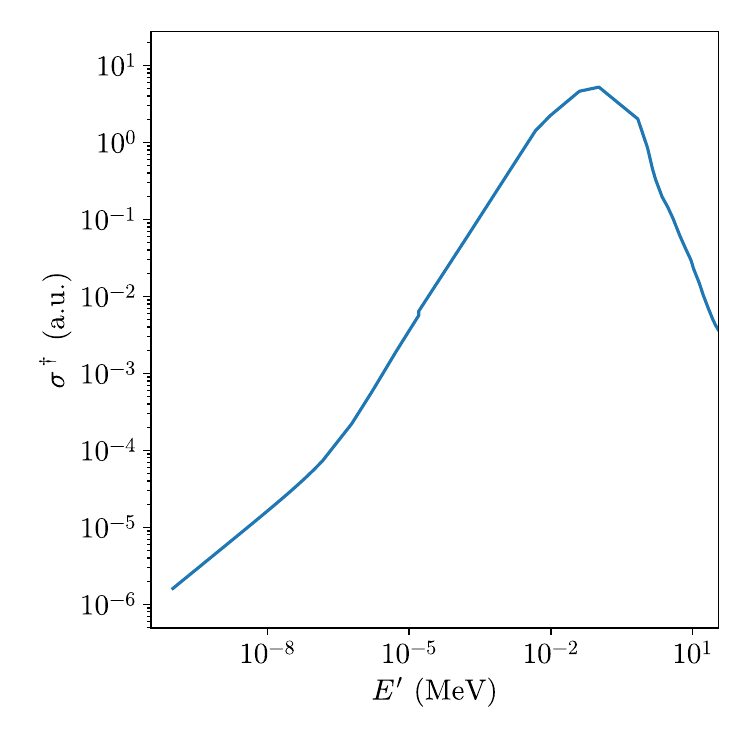}
    \end{subfigure}%
    ~ 
    \begin{subfigure}[t]{0.49\textwidth}
        \centering
        \includegraphics[trim={1cm 0.5cm 0 1.5cm},clip,width=\textwidth]{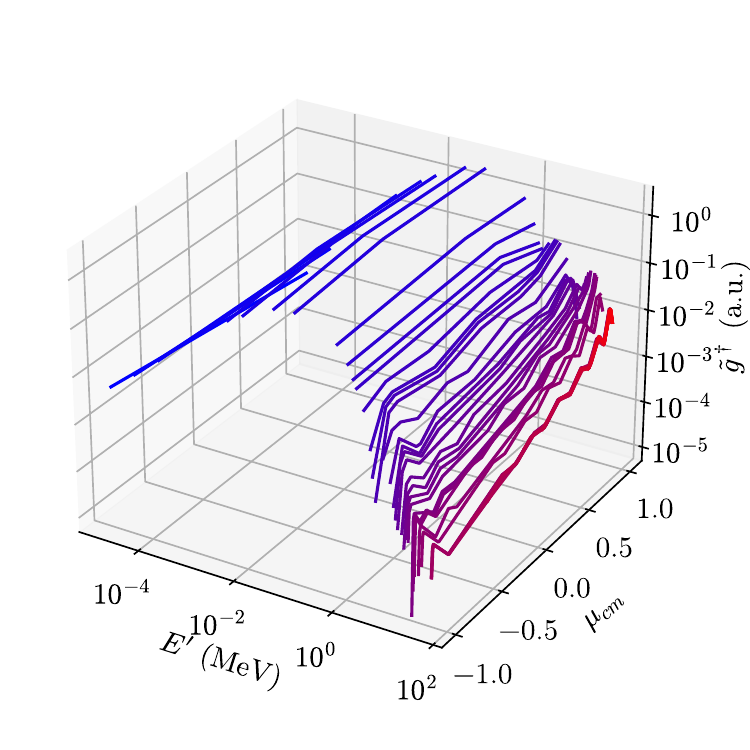}
    \end{subfigure}
    \caption{Adjoint nuclear data for the inelastic scattering on the first discrete level ($Q=45\text{ keV}$) of \textsuperscript{238}U prepared using definitions from \cref{eq:adj_xs_flx_gen} with $g(E)=1/E$. \textbf{On the left}, the adjoint cross section $\sigma^\dagger(E')$. At energies above the level energy, the adjoint cross section features the $1/E$ shape. \textbf{On the right}, the anisotropic part of the adjoint distribution law $f^\dagger$ stored as a bidimensional non-normalized angular distribution law $\tilde{g}^\dagger$. This discretized distribution contains 452 data points, split into 51 unidimensional distributions, each with an average of 8 data points. It ranges from $E'_\text{pivot}=16\text{ eV}$ to $E_\text{max}= 50\text{ MeV}$. The unidimensional distributions are colored with a gradient corresponding to the index $k$ of their corresponding $E'_k$ entering energy. As the pivot energy $E'_\text{pivot}$ was selected at a very low energy (corresponding to the hard-coded limit of $20 \varepsilon(A+1)^{-2}$) the isotropic part of the distribution is very small and not shown here.
    }
    \label{fig:U8_inel_scat}
\end{figure}

An example of adjoint data produced using this technique is shown in \cref{fig:U8_inel_scat}.

\subsection{Fission (MT18)}

The fission reaction (MT18) can be described in several ways in ACE files: it is provided as a Maxwellian fission spectrum, a Watt spectrum, or a generic energy-to-energy law \cite{conlin_compact_2019}; the angular distribution is sampled isotropically. In rare other cases (e.g. \textsuperscript{232}Th in ENDF-B/VIII.0) it can be described by other more precise laws such as Kalbach Tabular laws (ENDF LAW 61), in these cases the adjoint sampling procedure is described in \cref{sec:misc_freegas}.

In the case of a Maxwellian fission spectrum (ENDF LAW 7), a function $T_\text{fission}(E)$, tabulated as a function of $E$, stores the effective temperature $T$ to feed the Maxwellian distribution for a neutron inducing fission at energy $E$. Furthermore, a restriction energy $U$ is assigned to define a proper upper limit for the outgoing energy $E'< E-U$. The outgoing distribution density is then
\begin{equation}
    f(E\to E')= \frac{\sqrt{E'}}{I(E)}\text{exp}\left[-\frac{E'}{T_\text{fission}(E)}\right],
\end{equation}
with the normalisation constant
\begin{equation}
    I(E) = T_\text{fission}^{3/2}\frac{\sqrt{\pi}}{2}\text{erf}\left(\sqrt{\frac{E-U}{T_\text{fission}(E)}}\right)- \sqrt{\frac{E-U}{T_\text{fission}(E)}}\text{exp}\left( -\frac{E-U}{T_\text{fission}(E)} \right),
\end{equation}
which can be used to evaluate $f(E\to E')$ on-the-fly for the weight correction (see \cref{eq:weight_corr}).

In the case of a Watt spectrum (ENDF LAW 11), two functions $a(E)$ and $b(E)$, tabulated as a function of $E$, are used alongside the same restriction energy $U$ as presented above. The outgoing distribution density is then
\begin{equation}
    f(E\to E') = \frac{\text{exp}\left(- \frac{E'}{a(E)}\right)}{I(E)}\text{sinh}\left(\sqrt{b E'}\right),
\end{equation}
with the normalisation constant
\begin{equation}
    \begin{aligned}
        I(E) =& \sqrt{\frac{\pi a^3(E)b(E)}{16}}e^\frac{a(e)b(e)}{4}\Bigg[ \text{erf}\left(\sqrt{\frac{E-U}{a(E)}}-\sqrt{\frac{a(E)b(E)}{4}}\right)\\&+\text{erf}\left(\sqrt{\frac{E-U}{a(E)}}+ \sqrt{\frac{a(E)b(E)}{4}}\right) \Bigg] 
        - a(E) e^{-\frac{E-U}{a(E)}}\text{sinh}\sqrt{b(E)(E-U)}
    \end{aligned},
\end{equation}
which can be used to evaluate $f(E\to E')$ on-the-fly for the weight correction (see \cref{eq:weight_corr}).

Finally, in the case of a generic energy-to-energy law (ENDF LAW 4), the distribution $f(E\to E')$ is stored as 2D tabulated data using the storage format presented in \cref{sec:bidim_interp}. Evaluating the density can be done using the techniques presented therein.

Storing the adjoint distribution law could be done using a generic energy-to-energy law, as discussed in \cref{sec:bidim_interp}. However, in most cases, there exists a much more efficient and resilient technique based only on unidimensional data, which leverages the fact that the fission emission spectrum denoted $\chi_f(E')$ is mostly independent of the incident neutron energy $E$. When stored using Maxwellian fission spectra or Watt spectra presented above, the restriction energy $U$ is often negative around $-20\text{ MeV}$, ensuring that the spectrum produced is not restricted depending on the entering energy. The adjoint distribution law from \cref{eq:adj_xs_flx_gen} becomes then
\begin{equation}
    \label{eq:fiss_adj_distb_law}
    f^\dagger(E'\to E)= \frac{\sigma(E)\nu(E)g(E)\chi_f(E')}{\int \sigma(\tilde{E})\nu(\tilde{E})g(\tilde{E})\chi_f(E')d\tilde{E}} = \frac{\sigma(E)\nu(E)g(E)}{\int \sigma(\tilde{E})\nu(\tilde{E})g(\tilde{E})d\tilde{E}},
\end{equation}
which only depends on $E$ and can be stored in a unidimensional data distribution. The adjoint cross section can be obtained by computing the normalisation constant of \cref{eq:fiss_adj_distb_law}, namely $c_\chi:=\int \sigma(\tilde{E})\nu(\tilde{E})g(\tilde{E})d\tilde{E}$, which yields
\begin{equation}
    \sigma^\dagger(E')= c_\chi \chi_f(E').
\end{equation}
As always, the weight correction takes into account the true direct distribution: the approximation of $\chi_f$ being independent of $E$ does not lead to a bias.

Since the actual fission spectrum $\chi_f$ might generally depend on the incident energy $E$, to produce the adjoint data we need to choose a specific spectrum $\chi_f$ at a given energy $E$ among those available in the direct nuclear data library. A good strategy is to choose the spectrum that is non-zero over the maximal energy range : this avoids setting an adjoint cross section to zero in an energy range where it would actually be possible for a fission neutron to be emitted, which would create bias. For Maxwellian and Watt distributions, this constraint is satisfied by choosing the highest energy $E$, since the upper bound is $E-U$. For the generic energy-to-energy law, we choose the spectrum $\chi_f$ spanning the largest energy range among all the available spectra (at different entering energies $E_k$).

In rare cases, (e.g. \textsuperscript{266}U in ENDF-B/VIII.0) the fission reaction is not described as a single reaction (MT18) but as 1\textsuperscript{st}, 2\textsuperscript{nd}, 3\textsuperscript{rd} and 4\textsuperscript{th} chance fission reaction (corresponding to MT19, MT20, MT21 and MT38). In the cases of $n^\text{th}$ chance fission with $n>1$, the reactions often use Watt or Maxwellian spectra with a positive restriction energy $U$. This effectively voids the approximation of emission spectrum independent of the entering energy. Therefore, when $U$ is above an arbitrary threshold of $- 10\text{ MeV}$, the spectrum is considered dependent on the entering energy and the adjoint reaction is sampled using a generic energy-to-energy law, similarly to the sampling of the evaporation spectrum law (ENDF LAW 9) described in \cref{sec:misc_freegas}.

Regarding the deflection cosine $\mu_l$, in the cases treated in this section, it is sampled isotropically, which is also how we sample it in the adjoint collision. Since the adjoint and direct angular distribution are equal, no weight correction is needed here.

Fission reactions often also include delayed neutron energy spectra for the different precursor families. This means that the reaction comes with several outgoing distributions for all the families (each associated with a probability of being produced \cite{conlin_compact_2019}). For the adjoint distribution, each of these out-products is treated as an independent fission reaction that will be associated with its own adjoint fission reaction (the probability of sampling the out-product is taken into account in the multiplicity of the reaction). Delayed neutron energy spectra are also weakly dependent on the incident energy $E$, and the method proposed above is valid.

\begin{figure}[htb]
    \centering
    \begin{subfigure}[t]{0.49\textwidth}
        \centering
        \includegraphics[trim={0 0.8cm 0 0.53cm},clip,width=\textwidth]{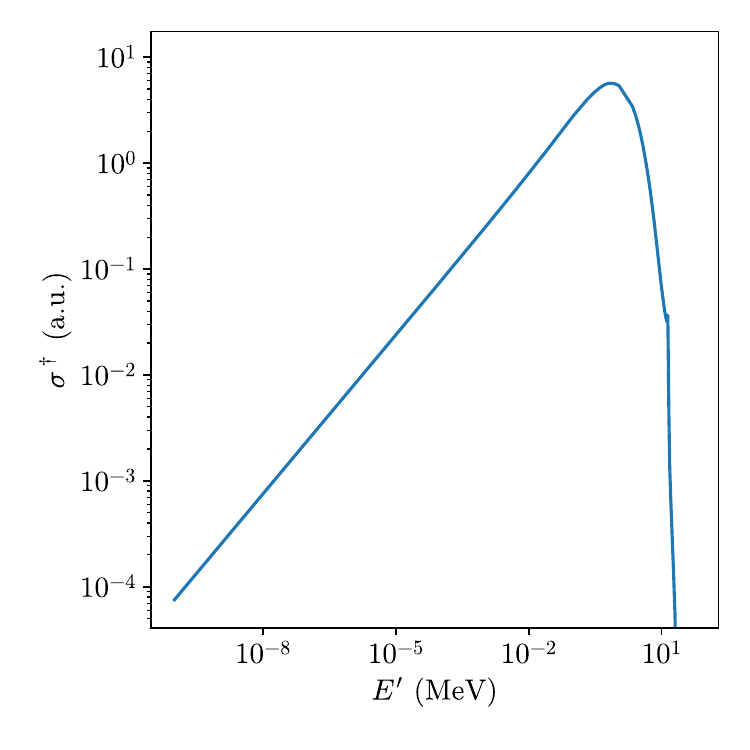}
    \end{subfigure}%
    ~ 
    \begin{subfigure}[t]{0.49\textwidth}
        \centering
        \includegraphics[trim={0 0.8cm 0 0.53cm},clip,width=\textwidth]{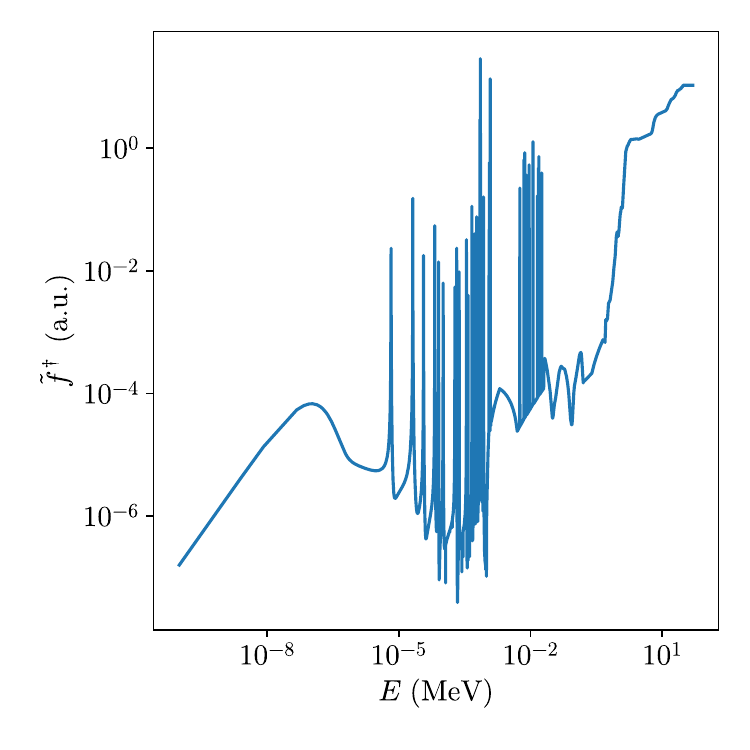}
    \end{subfigure}
    \caption{Adjoint nuclear data of the fission reaction of \textsuperscript{238}U. \textbf{On the left}, the adjoint cross section $\sigma^\dagger(E')$ has the shape of the direct fission spectrum $\chi_f(E')$. \textbf{On the right}, the non-normalized adjoint distribution law $\tilde{f}^\dagger$ has the shape of the direct fission cross section $\sigma_f(E)$. This distribution is stored as a unidimensional distribution and contains 1~748 data points.}
    \label{fig:adj_U_fiss}
\end{figure}

An example of adjoint nuclear data produced by the technique described above for the fission reaction on \textsuperscript{238}U is presented in \cref{fig:adj_U_fiss}.

\subsection{Other miscellaneous reactions}
\label{sec:misc_freegas}

Reactions not leading to neutron production are assiminated to capture and do not have a counterpart for adjoint transport \cite{rovel_general_2025}.

In addition to elastic, discrete inelastic and fission reactions, a plethora of other neutron-emitting reactions are available in modern nuclear data libraries, such as continuum inelastic scattering (MT91), $(n,xn)$ reactions (MT16, MT17 and MT37), $n^\text{th}$ chance fission (with $n>1$) (MT20, MT21 and MT38), or the $(z,\text{anything})$ (MT5) reaction.

These reactions are described using specific distribution laws, such as evaporation spectrum (ENDF LAW 9), N-body (ENDF LAW 66), generic energy-to-energy (ENDF LAW 4), Kalbach (ENDF LAW 44) and Kalbach tabular (ENDF LAW 61) laws \cite{conlin_compact_2019, endf_manual}. These distribution laws can be assigned in the laboratory frame (this is usually the case for the evaporation spectrum and generic energy-to-energy laws), or in the center-of-mass frame (this is more often the case for the N-body, Kalbach and Kalbach tabular laws).

\subsubsection{Laboratory frame}
\label{sec:lab_frame}

The evaporation spectrum uses a tabulated temperature function $T(E)$. Then, the outgoing distribution law is
\begin{equation}
    f(E\to E')= \frac{E'}{I(E)}\text{exp}\left[-\frac{E'}{T(E)}\right],
\end{equation}
with the normalization constant
\begin{equation}
    I(E) = T^2(E)\left[1-e^{\frac{E-U}{T(E)}}\left(1+\frac{E-U}{T(E)}\right)\right],
\end{equation}
which can be used to obtain the weight correction. The generic energy-to-energy distribution follows the bidimensional data storage procedure detailed in \cref{sec:bidim_interp}, and its density can be evaluated following the procedure described therein.

\begin{figure}[hbtp]
    \centering
    \includegraphics[width=.8\textwidth]{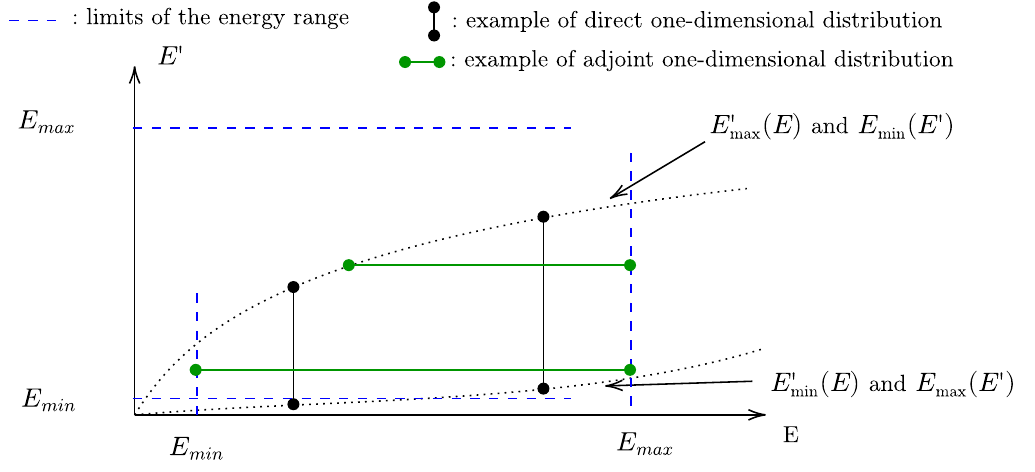}
    \caption{Procedure to obtain the bounds of the adjoint distribution. The distribution is shown in $(E,E')$ space. Blue dotted lines represent the global energy bounds of the problem; black lines represent some of the direct unidimensional distributions; green lines represent some of the adjoint unidimensional distributions.  The bounds of the direct distributions, that need to be increasing, are inverted to obtain the bounds of the adjoint distribution. The global bounds of the energy range are also taken into account.}
    \label{fig:Adjoint_bounds_E_to_E}
\end{figure}

In order to invert these distributions given in the laboratory frame, one follows the generic adjoint energy-to-energy discretization procedure provided in \cref{sec:bidim_interp}. The bounds $E_\text{max}(E')$ and $E_\text{min}(E')$ required for the adjoint distribution law are obtained as follows. We first retrieve the bounds of the direct distribution. For the evaporation spectrum, the upper bound is $E'_\text{max}(E)=E-U$ and the lower bound is $E'_\text{min}=0$. For the generic energy-to-energy law, we find the bounds by linearly interpolating between the unidimensional distributions stored within the bidimensional distribution; see \cref{sec:bidim_interp} for more details about the bounds of a bidimensional distribution. An important detail is that we need the direct bounds to be increasing functions of $E$ to be able to invert them into adjoint bounds as a function of $E'$. Although this condition is usually fulfilled by distributions given in ACE files, as a sanity check we additionally modify the bounds to make them increasing functions if they are not for specific points. To avoid the adjoint distribution being zero where the direct distribution is non-zero, we need the modified upper bound to be larger than or equal to the original upper bound, and the modified lower bound to be smaller than or equal to the original lower bound. Once the increasing boundaries have been obtained, we obtain the adjoint bounds by using the inverse of the direct bounds functions, namely
\begin{equation}
    \begin{aligned}
        E_\text{min}(E')&=(E'_\text{max})^{-1}(E')\\
        E_\text{max}(E')&=(E'_\text{min})^{-1}(E')
    \end{aligned}.
\end{equation}
We further truncate these bounds so that they remain in between the energy bounds of our problem. This procedure is summarized in \cref{fig:Adjoint_bounds_E_to_E}.

\subsubsection{Center-of-mass frame}

Inverting the distributions given in the center-of-mass frame is more intricate. Both the weight correction formula in \cref{eq:weight_corr} and the adjoint data definition in \cref{eq:adj_xs_flx_gen} need to access to the distribution law in the laboratory frame. The rules used to convert from $E'_{cm},\mu_{cm}$ to $E',\mu_l$ are given by
\begin{equation}
    \begin{aligned}
        E'&= E'_{cm}+\frac{E}{(A+1)^2} + 2 \mu_{cm}\frac{\sqrt{E'_{cm}E}}{A+1}\\
        \mu_{l}&=\sqrt{\frac{E'_{cm}}{E'}}\mu_{cm}+\frac{1}{A+1}\sqrt{\frac{E}{E'}}.
    \end{aligned}
    \label{eq:convert_cm_lab}
\end{equation}
In these formulas, the incident energy $E$ and the outgoing energy $E'$ (without subscript) are always expressed in the laboratory frame (consistently with the notation throughout the paper).

The rules used to convert backward from the laboratory frame to the center-of-mass frame are
\begin{equation}
    \begin{aligned}
        E'_{cm}&= E'+\frac{E}{(A+1)^2} - 2 \mu_{l}\frac{\sqrt{E'E}}{A+1}\\
        \mu_{cm}&=\sqrt{\frac{E'}{E'_{cm}}}\mu_{l}-\frac{1}{A+1}\sqrt{\frac{E}{E'_{cm}}}.
    \end{aligned}
    \label{eq:convert_lab_cm}
\end{equation}
To obtain the densities in the laboratory frame at a given point $(E',\mu_l)$, we first use \cref{eq:convert_lab_cm} to get the corresponding $(E'_{cm},\mu_{cm})$, and we obtain the density in the center-of-mass frame at this point. Finally, we multiply by the Jacobian of the transformation in \cref{eq:convert_cm_lab}, and we obtain
\begin{equation}
    \label{eq:convert_densities_cm_lab}
    f(E\to E',\mu_l) = f(E\to E'_{cm},\mu_{cm})\sqrt{\frac{E'}{E'_{cm}}}.
\end{equation}

The technique using non-normalized distributions to store generic energy-to-energy distributions described in \cref{sec:gen_strat,sec:bidim_interp} requires the access to the energy-to-energy direct distribution law $f(E\to E')$ already integrated in $\mu_l$. However, the formulas \cref{eq:convert_cm_lab,eq:convert_lab_cm,eq:convert_densities_cm_lab} demand both the energy and deflection cosine to compute the change of frame. We therefore produce non-integrated densities $f(E\to E',\mu_l)$, and we integrate them numerically between $\mu_l=-1$ and $\mu_l=1$ to obtain an approximation of $f(E\to E')$ that we will use to produce the adjoint generic energy-to-energy distribution $\tilde{f}^\dagger(E'\to E)$. This integration step is only needed during the preparation of the adjoint distribution, and can be carried out with arbitrary accuracy, since no bias will be induced. Using the trapezoidal rule, we integrate in $\mu_l \in [-1,1]$ with 3 points if $A>10$, and 5 points if $A\le 10$. This allows for fast integration, since this procedure must be applied at every point $(E,E')$ that we want to discretize.

\Cref{eq:convert_densities_cm_lab} includes a $\sqrt{E'/E_{cm}}$ term that will create singularities (points where the density in the laboratory frame goes to infinity) when $E_{cm}=0$. These singularities disappear when integrated in $\mu_l$. However, since we integrate numerically, using a small number of samples, the singularity can lead to `infinities' in the discretization\footnote{We might argue that the probability of landing on a point $(E,E',\mu_l)$ leading to $E'_{cm}=0$ and to a diverging value is null. However, numerical evidence shows that the discretization algorithm selecting the position of the discretization points $(E,E')$ tends to be `attracted' by those singularities, and will try to discretize the adjoint distribution thereby.}. Since the singularities occur for $E'_{cm}=0$, corresponding to $E'=E/(A+1)^2$ and $\mu_l=1$, we can avoid these pathological cases by computing a relative distance to the singularity
\begin{equation}
    d_\text{r.s.}(E,E') = \frac{\left|E'-\frac{E}{(A+1)^2}\right|}{E'},
\end{equation}
for every point $(E,E')$. Then, if the current point is close to the singularity, instead of integrating over the segment $\mu_l \in [-1,1]$, we integrate over $\mu_l \in [-1,0.98]$, which solves the problem.

In order to obtain the energy bounds required for the adjoint distribution by following the procedure given in \cref{sec:lab_frame}, we first need to obtain the bounds of the direct distribution in the laboratory frame. For this purpose, we need to convert the energy bounds available in the direct nuclear data in the center-of-mass frame to bounds in the laboratory frame. Inspecting the energy the change of frame in \cref{eq:convert_cm_lab}, since all the energy terms are positive, $E'_\text{max}(E)$ is achieved for $\mu_{cm}=1$ and $E'_{cm}=E'_{cm,\text{max}}(E)$. We obtain
\begin{equation}
    E'_\text{max}(E) = \left(\sqrt{E'_{cm,\text{max}}(E)}+\frac{\sqrt{E}}{A+1} \right)^2.
    \label{eq:max_bound_lab_from_cm}
\end{equation}
The minimum bound will be reached for $\mu_{cm}=-1$: \cref{eq:convert_cm_lab} yields
\begin{equation}
    E' = \left(\sqrt{E'_{cm}}-\frac{\sqrt{E}}{A+1}\right)^2,
\end{equation}
and the minimum value is
\begin{equation}
    E'_\text{min}(E) =
    \begin{cases}
        \left(\sqrt{E'_{cm,\text{min}}(E)}-\frac{\sqrt{E}}{A+1}\right)^2 & \text{if } E'_{cm,\text{min}}(E)>E/(A+1)^2\\
        \left(\frac{\sqrt{E}}{A+1}-\sqrt{E'_{cm,\text{max}}(E)}\right)^2 & \text{if } E'_{cm,\text{max}}(E)<E/(A+1)^2\\
        0 &\text{otherwise}\\
        
    \end{cases}.
    \label{eq:min_bound_lab_from_cm}
\end{equation}
Since the procedure in \cref{sec:lab_frame} requires increasing functions for the energy bounds, to simplify the procedure we can simply resort to using $E'_\text{min}(E)=0$ everywhere.

The adjoint distribution $\tilde{f}^\dagger(E\to E')$ contains no angular information. To avoid storing the angular distribution for every pair $(E,E')$, we simply skip this step and sample the angular component isotropically. This does not follow the definition in \cref{eq:adj_xs_flx_gen}, and might generate a large variance, especially if the reactions are strongly non-isotropic. However, we observe that the adjoint reactions belonging to this class occur less often than elastic and discrete inelastic scattering, or fission, for most configurations of interest. The impact of the choice concerning the angular distribution is thus expected to be minor on the overall variance of the adjoint simulation.

The last task is to specify, for each reaction type, how its direct density can be obtained in the center-of-mass frame (the conversion to laboratory frame is done via \cref{eq:convert_densities_cm_lab}), and how its energy boundaries can be obtained.

The N-body phase space distribution is parametrized by the integer $n$ that can take the values $\{ 3,4,5\}$ \cite{endf_manual}. The corresponding density in the center-of-mass frame is:
\begin{equation}
    f(E\to E'_{cm})= C_n \sqrt{E'_{cm}}\left(E'_{cm,\text{max}}(E)-E'_{cm} \right)^{\frac{3n}{2}-4},
\end{equation}
where $E'_{cm,\text{max}}(E)$ is defined as 
\begin{equation}
    \label{eq:n-body_bound}
    E'_{cm,\text{max}}= \frac{A-1}{A}\left(\frac{A}{A+1}E+Q\right),
\end{equation}
with $Q$ the discrete energy emitted by the reaction. The normalization constants are
\begin{equation}
    \begin{aligned}
        C_3 &= \frac{4}{\pi (E'_{cm,\text{max}}(E))^2}\\
        C_4 &= \frac{105}{32 (E'_{cm,\text{max}}(E))^{7/2}}\\
        C_5 &= \frac{256}{14\pi (E'_{cm,\text{max}}(E))^5}.\\
    \end{aligned}
\end{equation}
The energy bounds are $E'_{cm,\text{max}}$ (given in \cref{eq:n-body_bound}) and $E'_{cm,\text{min}}=0$.

Kalbach distributions (ENDF LAW 44) and Kalbach tabular distributions (ENDF LAW 61) are correlated energy-angle distributions that have an energy-to-energy tabulated distribution using the formalism described in \cref{sec:bidim_interp}, but an angular distribution is associated to every value $E'_{k,l}$. When sampling from such a distribution, we follow the procedure described in \cref{sec:bidim_interp} to sample the energy, and sample the deflection cosine in the angular distribution closest to the sampled point. In Kalbach distributions, the angular distribution is described by 
\begin{equation}
    \label{eq:kalbach_distrib}
    g_{k,l}(\mu)= \frac{a_{k,l}}{2\text{sinh}(a_{k,l})}\big[\text{cosh}( a_{k,l}\mu)+r_{k,l} \text{sinh}(a_{k,l} \mu)\big],
\end{equation}
with $a_{k,l}$ and $r_{k,l}$ some coefficients stored at every point $E'_{k,l}$ \cite{endf_manual}. In Kalbach tabular distributions, the angular distributions are simply unidimensional tabulated distributions \cite{endf_manual}. To obtain the densities from these distributions, we mainly follow the procedure from \cref{sec:bidim_interp} for the energy distribution. Once the energy densities have been obtained for the $k^{th}$ and $k+1^{th}$ distributions, we determine the angular distribution to use for the $k^{th}$ and $k+1^{th}$ distributions (by finding the closest tabulated $E'_{k,l}$ and $E'_{k+1,l}$ values), and we multiply the energy densities by their corresponding angular distribution to get the $k^{th}$ and $k+1^{th}$ energy-angle densities. We can then proceed with the final interpolation, as was done in \cref{sec:bidim_interp} to get the final energy-angle density. The energy bounds can be found using only the energy part of the distributions, similarly to what was done above (\cref{eq:max_bound_lab_from_cm,eq:min_bound_lab_from_cm}).

\begin{figure}[htb]
    \centering
    \begin{subfigure}[t]{0.49\textwidth}
        \centering
        \includegraphics[trim={0 0.8cm 0 0.53cm},clip,width=\textwidth]{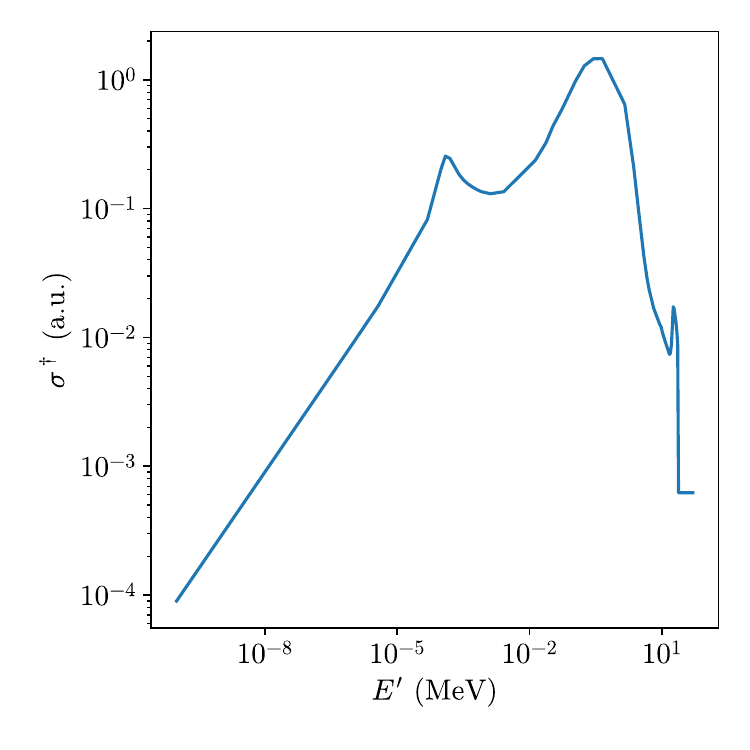}
    \end{subfigure}%
    ~ 
    \begin{subfigure}[t]{0.49\textwidth}
        \centering
        \includegraphics[trim={1cm 0.8cm 0cm 0.53cm},clip,width=\textwidth]{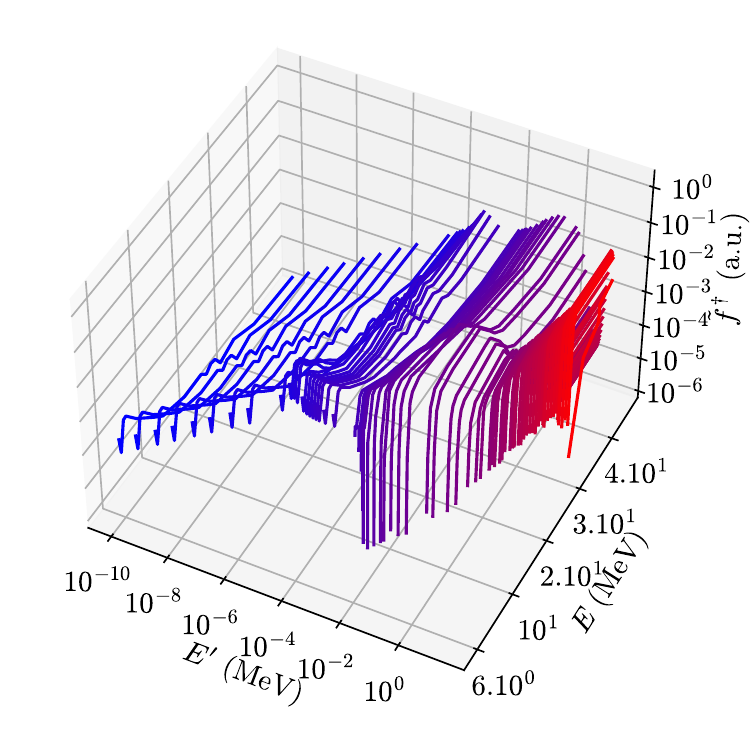}
    \end{subfigure}
    \caption{Adjoint nuclear data of the $(n,2n)$ reaction of \textsuperscript{238}U. \textbf{On the left}, the adjoint cross section $\sigma^\dagger(E')$. \textbf{On the right}, the non-normalized adjoint distribution law $\tilde{f}^\dagger$. This discretized distribution contains 1020 data points, split into 73 unidimensional distributions, each with an average of 13 data points.}
    \label{fig:adj_U_n2n}
\end{figure}

An example of adjoint nuclear data produced by these procedures for the $(n,2n)$ reaction on \textsuperscript{238}U is presented in \cref{fig:adj_U_n2n}.

\subsection{Thermal broadening of the scattering kernel}
\label{sec:svt_reac}

All reactions discussed so far assumed the target nuclide to be at rest. However, at low neutron energies, we have to take into account the thermal motion of the collided nuclide. For practical reasons, thermal motion corrections are only applied to elastic scattering (MT2). In the absence of data concerning the thermal scattering laws (TSL) for a given nuclide, by default the most commonly used model to describe neutron-nuclide kinematics accounting for thermal motion is the `Sampling the Velocity of the Target nucleus' (SVT) approximation. This consists in supposing that the velocity of the target nucleus obeys a Maxwellian distribution at temperature $T$. The SVT model does not take into account any specific chemical or crystallographic bond effect, and is thus preferentially replaced (at least in the lower part of the energy range) by TSL data, when available in the nuclear data library (see Sec.~\ref{sec:tsl}), which explicitly models the solid-state effects in the neutron-matter interactions.

\subsubsection{Direct sampling}
\label{sec:svt_direct_sampling}

We will denote $\mathbf{v}$ the velocity of the neutron,  $\mathbf{V}$ the velocity of the target nuclide, and $\mu_t$ the cosine of the angle between $\mathbf{v}$ and $\mathbf{V}$. Non-bold quantities will refer to the norm of the corresponding vector. By assumption, $V$ follows a Maxwellian density at temperature $T$, namely
\begin{equation}
    \mathcal{M}_T(V) = \frac{4}{\sqrt{\pi}}\beta^{3/2}V^2e^{-\beta V^2},
\end{equation}
with
\begin{equation}
    \beta = \frac{A}{2k_B T},
\end{equation}
where $k_B$ is the Boltzmann constant. We express the speed in units such that the mass of the neutron is $m_n=1$, whence $E=v^2/2$, which will simplify the following equations. Taking into account the relative probability that a collision occurs, for a collision with a neutron at speed $v$ the distribution $\mathbb{P}_v(V,\mu_t)$ of the nucleus speed reads
\begin{equation}
    \mathbb{P}_v(V,\mu_t)dV d\mu_t = C^{-1}(v)\times \frac{2\beta^{3/2}}{v\sqrt{\pi}} \times v_r \sigma_{0}(v_r) V^2e^{-\beta V^2} dV d\mu_t,
    \label{eq:full_target_speed_distrib}
\end{equation}
with $\sigma_0$ the cross section at $0\text{ }K$ and
\begin{equation}
    v_r = \sqrt{v^2+V^2-2\mu_t v V}
\end{equation}
the relative velocity between the neutron and the target nuclide \cite{bell_nuclear_1970}. The normalization constant is
\begin{equation}
    C(v) := \sigma_T(v)= \frac{2\beta^{3/2}}{v\sqrt{\pi}} \times\int_{-1}^1\int_0^\infty v_r \sigma_{0}(v_r) V^2e^{-\beta V^2} dV d\mu_t,
    \label{eq:full_target_speed_distrib_renorm}
\end{equation}
with $\sigma_T(v)$ the Doppler-broadened cross section at speed $v$. Once the velocity of the nuclide $\mathbf{V}$ has been sampled from \cref{eq:full_target_speed_distrib}, one can easily change frame to the center-of-mass frame, sample the collision using the center-of-mass angular distribution, and go back to the laboratory frame to compute the final speed of the neutron $\mathbf{v}'$.

Sampling from \cref{eq:full_target_speed_distrib} can be difficult as it involves the cross section $\sigma_0$. In the constant cross section approximation (CXS), $\sigma_0$ is supposed to be constant around the speed $v$, which transforms \cref{eq:full_target_speed_distrib} into
\begin{equation}
    \label{eq:target_speed_distrib}
    \tilde{\mathbb{P}}_v(V,\mu_t)dV d\mu_t = \tilde{C}^{-1}(v) \times \frac{2\beta^{3/2}}{v\sqrt{\pi}} \times v_r  V^2e^{-\beta V^2} dV d\mu_t ,
\end{equation}
with the normalization constant
\begin{equation}
    \label{eq:target_speed_distrib_renorm}
    \tilde{C}(v)=  \frac{\sigma_T(v)}{\sigma_0(v)}=\left(1+\frac{1}{2a^2(v)}\right)\erf(a(v))+\frac{e^{-a^2(v)}}{a(v)\sqrt{\pi}},
\end{equation}
and $a(v)$ a dimensionless neutron speed defined as
\begin{equation}
    a(v) = \sqrt{\frac{AE}{k_BT}} = \sqrt{\frac{A v^2}{2k_BT}}.
\end{equation}

If the elastic cross section of the considered nuclide is resonant at thermal temperatures, the CXS approximation is known to fail; see e.g.~Ref.~\citenum{zoia_doppler_2013,rothenstein_neutron_1996} for a thorough discussion. In this case, the Doppler Broadening Rejection Correction (DBRC) strategy should be used instead \cite{zoia_doppler_2013,rothenstein_neutron_1996}. The DBRC algorithm allows us to sample from the original \cref{eq:full_target_speed_distrib}; this is achieved by sampling from \cref{eq:target_speed_distrib} and applying a rejection to take into account the missing dependence on $\sigma_0$. The values sampled from \cref{eq:target_speed_distrib} are accepted with probability $\sigma_0(v_r)/\sigma_{0,\text{max}}(v)$, where $\sigma_{0,\text{max}}(v)$ is the maximum value of the $0\; K$ cross section in an interval around $v$. Strictly speaking, the maximum cross section value should be computed over the entire speed range, since it is in principle possible to sample a nuclide speed that would lead to arbitrary values of $v_r$. However, since these events are unlikely, it has been proposed to consider speeds $v_r$ such that $a(v_r)\in[a(v)-4,a(v)+4]$ when searching for the maximum cross section \cite{zoia_doppler_2013}.

\subsubsection{Obtaining the direct density}

In order to be able to apply the weight correction in \cref{eq:weight_corr}, we need to access the direct distribution law $f(E\to E',\mu_l)$. However, the sampling method described in \cref{sec:svt_direct_sampling} does not provide us with an easy-to-obtain density. In the specific case where we used the CXS approximation and the elastic scattering is perfectly isotropic (which is very often the case at thermal energies), this law takes the form
\begin{equation}
    \label{eq:kernel_svt_iso}
    f(E \to E', \mu_l) = \left(\frac{A+1}{A}\right)^2\frac{\beta^{1/2}\tilde{C}^{-1}(v)}{2\pi^{1/2}q}\sqrt{\frac{E'}{E}}\text{exp}\left(-\frac{\beta}{q^2}\left(E'-E+\frac{q^2}{2A}\right)^2\right),
\end{equation}
as shown in \cref{sec:proof_svt}, where $q$ is the momentum transfer
\begin{equation}
    q := ||\mathbf{v'}-\mathbf{v}||= \sqrt{2\left(E'+E-2\mu_l\sqrt{EE'}\right)}.
    \label{eq:svt_defq}
\end{equation}
\Cref{eq:kernel_svt_iso} has a simple expression and can be evaluated on-the-fly during the simulation.

In the general case, we can show (see \cref{sec:proof_svt}) that the law can be expressed as
\begin{equation}
    \label{eq:kernel_dbrc}
    \begin{aligned}
        f(E\to E',\mu_l) &= \left(\frac{A+1}{A}\right)^2\frac{\beta^{1/2}C^{-1}(v)}{\pi^{1/2}q}\sqrt{\frac{E'}{E}}\text{exp}\left(-\frac{\beta}{q^2}(E'-E+\frac{q^2}{2A})^2\right)\\
        &\times \int_{x=-\infty}^\infty  \int_{y=-\infty}^\infty g(\mu_{cm})  \sigma_{0}(v_r)\frac{\beta}{\pi}e^{-\beta(x^2+y^2) }dxdy,
    \end{aligned}
\end{equation}
where $g(\mu_{cm})$ is the angular scattering distribution in the center-of-mass frame. The quantities $\mu_{cm}$ and $v_r$ occurring in the integral in \cref{eq:kernel_dbrc} are functions of $x$ and $y$ that can be computed as follows. First we create a (three-dimensional) vector $\mathbf{v}$ of norm $v$, such as:
\begin{equation}
    \mathbf{v} = v \begin{bmatrix} 1\\0\\0 \end{bmatrix}.
\end{equation}
Then, we construct a vector $\mathbf{v'}$ of norm $v'$ whose cosine of deflection with respect to $\mathbf{v}$ is $\mu_l$; for example:
\begin{equation}
    \mathbf{v}' = v' \begin{bmatrix} \mu_l\\\sqrt{1-\mu_l^2}\\0 \end{bmatrix}.
\end{equation}
We define $\mathbf{q}:=\mathbf{v}'-\mathbf{v}$, with $\mathbf{n}= \mathbf{q}/||\mathbf{q}||$ the direction of $\mathbf{q}$. We construct the normalized vectors $\mathbf{e}_1$ and $\mathbf{e}_2$ that form an orthonormal basis together with $\mathbf{n}$. We construct the vector $\mathbf{V}$:
\begin{equation}
    \mathbf{V} = \left( \frac{A+1}{2A}q+\mathbf{v}\cdot\mathbf{n}\right)\mathbf{n}+x\mathbf{e}_1+ y\mathbf{e}_2.
\end{equation}
We can now compute the relative speed $\mathbf{v_r}= \mathbf{v}-\mathbf{V}$, the center-of-mass speed $\boldsymbol{\mathcal{V}}=(\mathbf{v}+A\mathbf{V})/(A+1)$, the neutron speed in the center-of-mass frame $\mathbf{v}_{cm}=\mathbf{v}-\boldsymbol{\mathcal{V}}$, and the outgoing neutron speed in the center-of-mass frame $\mathbf{v}'_{cm}=\mathbf{v}'-\boldsymbol{\mathcal{V}}$. Finally we can compute the deflection cosine in the center-of-mass frame as 
\begin{equation}
    \mu_{cm}= \frac{\mathbf{v}_{cm}\cdot\mathbf{v}'_{cm}}{v_{cm}v'_{cm}}.
\end{equation}

The general case formula presented in \cref{eq:kernel_dbrc} cannot be evaluated on-the-fly, since accurately evaluating the integral occurring therein is computationally prohibitive. However, in order for the weight correction procedure presented in \cref{sec:adj_coll_event} to ensure the unbiasedness of the adjoint Monte Carlo game, we do not need the weight correction to follow \cref{eq:weight_corr} precisely: we only require the weight correction to obey \cref{eq:weight_corr} \emph{on average}. Since the direct distribution density $f(E\to E',\mu_l)$ appears in the numerator of the weight correction, the minima requirement is that its evaluation is unbiased. Therefore, we can estimate the integral in \cref{eq:kernel_dbrc} using a Monte Carlo method. Although the convergence of this method is quite crude, the unbiasedness of the estimation is enough to ensure that, on average, the weight correction procedure remains unbiased. To speed-up the evaluation of \cref{eq:kernel_dbrc}, we estimate the following integral
\begin{equation}
    \label{eq:kernel_dbrc_MC}
    \int_{x=-\infty}^\infty  \int_{y=-\infty}^\infty g(\mu_{cm})  \sigma_{0}(v_r)\frac{\beta}{\pi}e^{-\beta(x^2+y^2) }dxdy
\end{equation}
by sampling $(x,y)$ only once. For each Monte Carlo sampling, we draw $x$ and $y$ independently from a Gaussian density with null average and standard deviation $(2\beta)^{-1/2}$; given the sampled pair $(x,y)$, we then undergo the evaluation procedure of $\mu_{cm}$ and $v_r$ described above, which yields the estimate of  $g(\mu_{cm})\sigma_0(v_r)$. The main advantage of this method is that, when the elastic scattering is close to isotropic and the associated cross section is nearly constant (which is often the case according to our experience), the variance of the integral estimation will be very small. In the limiting case of isotropic scattering with constant cross section, the estimation is actually exact. 

In order to evaluate \cref{eq:kernel_dbrc}, we also have to know the normalization constant $C^{-1}(v)$ defined in \cref{eq:full_target_speed_distrib_renorm}. If we use the CXS approximation, the we simply have $C(v)=\sigma_0(v)\tilde{C}(v)$. In the general case, however, \cref{eq:full_target_speed_distrib_renorm} involves an integral that would be expensive to compute on-the-fly. Observe that $C(v)$ corresponds to the Doppler-broadened cross section $\sigma_{T}(v)$: if we had an accurate estimate of this quantity, we could estimate $C(v)$. Nonetheless, this strategy is not very satisfactory, since it requires a very precise evaluation of $\sigma_{T}(v)$ at exactly the right temperature, to minimize the bias. We will therefore resort again to a Monte Carlo evaluation of the integral in \cref{eq:full_target_speed_distrib_renorm}. Contrary to the previous case, here we need an unbiased estimation of the inverse of the integral, since the term appearing in \cref{eq:kernel_dbrc} corresponds to the inverse of $C(v)$. To solve this problem, it is interesting to observe what is the role of the constant $C(v)$ in the sampling of the velocity of the target (using DBRC) with the distribution given in \cref{eq:full_target_speed_distrib}. Looking at the procedure described in \cref{sec:svt_direct_sampling}, we first sample in \cref{eq:target_speed_distrib}, whose normalization constant $\tilde{C}(v)$ is known analytically from  \cref{eq:target_speed_distrib_renorm}. The difficulty of accessing $C(v)$ emerges once we perform the rejection correction step with the factor $\sigma_0(v_r)/\sigma_{0,\text{max}}(v)$. This issue is thus an instance of a general class of problems related to estimating the density of a direct sampling based on rejection.

Let us consider a sampling technique that consists in drawing values from an unknown density $f(x)$ by first sampling from a known distribution $g(x)$, and then performing rejection using the acceptance function $a(x)$. The full density to be sampled from is 
\begin{equation}
    f(x) = \frac{g(x)a(x)}{\int g(y)a(y)dy}.
\end{equation}
Although we can evaluate $g(x)a(x)$ easily, the normalisation constant $(\int g(y)a(y)dy)^{-1}$ is not known. We denote $N_s$ the random variable that counts the number of times we have to go through the sampling procedure before acceptance ($N_s=1$ if the value is accepted right away and increases for each rejection). The probability for a single sampling to be rejected is:
\begin{equation}
    \mathbb{P}(\text{rejection})= \int g(x)(1-a(x))dx.
\end{equation}
Therefore, the expected value of $N_s$ can be expressed as
    \begin{align}
        \mathbb{E}(N_s) &= \sum_{n_s=1}^\infty \mathbb{P}(N_s\ge n_s)\nonumber\\
        &=  \sum_{n_r=0}^\infty \mathbb{P}(\text{rejection)}^{n_r}\nonumber\\
        &= \frac{1}{1-\mathbb{P}(\text{rejection})}\nonumber\\
        &= \frac{1}{1-\int g(x)(1-a(x))dx}\nonumber\\
        &= \frac{1}{\int g(x)a(x)dx},
    \end{align}
where we indexed by $n_s$ the number of samples, and by $n_r$ the number of rejections. Therefore, to obtain an unbiased estimator of the normalisation constant of $g(x)a(x)$, we can simply perform the rejection procedure and count the total number $N_s$ of trials. Since $N_s$ follows a geometric law of parameter $\int g(x)a(x)dx$, we can obtain its relative variance as:
\begin{equation}
    \mathbb{V}_r(N_s):=\frac{\mathbb{V}(N_s)}{\mathbb{E}(N_s)^2} = \mathbb{P}(\text{rejection}).
\end{equation}
Thus, the estimation improves by reducing rejection. Note that this whole procedure is generalized to the evaluation of any reciprocal of integrals by T.~E.~Booth in Ref.~\citenum{booth_unbiased_2007}.

We can now focus again on the problem of estimating $C^{-1}(v)$ for the DBRC procedure. Using a similar argument, we can record the number $N_s$ of trials required to accept the target speed when using the CXS approximation for the sapling and $\sigma_0(v_r)/\sigma_{0,\text{max}}$ as acceptance probability. This gives us an unbiased estimate of
\begin{equation}
    \mathbb{E}(N_s) = \left[ \int \frac{\sigma_0(v_r)}{\sigma_\text{max}}  \tilde{C}^{-1}(v) \frac{1}{v} v_r \frac{\beta^{3/2}}{\pi^{3/2}}e^{-\beta V^2}d\mathbf{V} \right]^{-1}.
\end{equation}
We can therefore obtain an unbiased estimate of $C^{-1}(v)$ via
\begin{align}
    C^{-1}(\mathbf{v})&=\left[ \frac{1}{v}\int v_r \sigma_{0}(v_r)\frac{\beta^{3/2}}{\pi^{3/2}}e^{-\beta V^2}d\mathbf{V}\right]^{-1}\nonumber\\
&= \mathbb{E}(N_s) / \left( \sigma_{0,\text{max}}(v)\tilde{C}(v)\right).
\end{align}
If the scattering cross section at 0 K is close to constant, the rejection function $\sigma_0(v_r)/\sigma_{0,\text{max}}(v)$ is close to 1, and the variance of our estimator will be very small.

Finally, observe that the product of the two Monte Carlo estimations performed to evaluate the direct distribution law will be an unbiased estimator of \cref{eq:kernel_dbrc}, provided that the two sampling procedures are independent.

\subsubsection{Adjoint distribution law}

To sample from the adjoint distribution law of the elastic scattering at thermal energies, we will use a generic energy-to-energy law as described in \cref{sec:bidim_interp}, and then sample the deflection cosine $\mu_l$ using an ad hoc technique. To simplify the construction of the adjoint data, we will produce it as if the considered reaction were perfectly isotropic, and using the CXS approximation. The true angular distribution and real cross section will be taken into account by the weight correction procedure. To follow the storage procedure in \cref{sec:bidim_interp}, we need to be able to access the energy part of the direct distribution law $f(E\to E')$. In the case of isotropic scattering and CXS approximation, we have the following expression (see \cref{sec:proof_svt}):
\begin{equation}
    \begin{aligned}
        f(E\to E') =& \tilde{C}^{-1}(E)\frac{\eta^2}{2E}\Bigg[ 
        \text{exp}\left(\frac{E-E'}{k_BT}\right)\times\left(\text{erf}\left(\frac{\eta\sqrt{E}-\rho\sqrt{E'}}{\sqrt{k_B T}}\right) \pm\text{erf}\left(\frac{\eta\sqrt{E}+\rho\sqrt{E'}}{\sqrt{k_B T}}\right)\right) \\
         &+ \text{erf}\left( \frac{\eta\sqrt{E'}-\rho\sqrt{E}}{\sqrt{k_B T}} \right)
        \mp \text{erf} \left( \frac{\eta\sqrt{E'}+\rho\sqrt{E}}{\sqrt{k_B T}} \right)
\Bigg],
    \end{aligned}
    \label{eq:svt_energy_to_energy_kernel}
\end{equation}
where 
\begin{equation}
    \eta = \frac{A+1}{2\sqrt{A}},
\end{equation}

\begin{equation}
    \rho = \frac{A-1}{2\sqrt{A}},
\end{equation}
and we use the upper signs when $ E<E'$, and lower signs when $E>E'$. Numerical problems can emerge when evaluating \cref{eq:svt_energy_to_energy_kernel}. When the energy loss $E-E'$ is very high compared to the thermal energy $k_BT$, the numerical evaluation exponential term can overflow, while at the same time the two error functions vanish. To avoid this problem, when $(E-E')/(k_BT)>20$ we use an asymptotic approximation for the error function, namely
\begin{equation}
    \text{erf}(x) \approx 1- e^{-x^2}/\sqrt{\pi}\times \left( \frac{1}{x}-\frac{1}{2x^3}+\frac{3}{4x^5}-\frac{15}{8x^7}\right),
    \label{eq:approx_erf}
\end{equation}
for $x\gg1$; see for example Eq.~\href{https://dlmf.nist.gov/7.12#i}{(7.12.1)} of NIST's Digital Library of Mathematical Functions \cite{NIST:DLMF}. Using this approximation in the two first error functions in \cref{eq:svt_energy_to_energy_kernel}, the unit terms cancel out, and the remaining exponential term in \cref{eq:approx_erf} cancels the problematic exponential term in \cref{eq:svt_energy_to_energy_kernel}. Conversely, when $E'\gg E$, the two last error functions in \cref{eq:svt_energy_to_energy_kernel} cancel out, leading to catastrophic cancellation. In order to avoid this problem, when $E'>E$ we express the difference of error functions as a difference of complementary error functions using 
\begin{equation}
    \text{erfc}(x) = 1- \text{erf}(x),
\end{equation}
which avoids the loss of precision in the numerical evaluation.

The bounds of the adjoint energy-to-energy law theoretically span the full energy range. Indeed, as the speed of the target nuclide is theoretically unbounded, the neutron can gain and lose arbitrary amounts of energy. However, setting the boundary at the limits of the energy range can cause some problems of interpolation. Near the upper limit of $400\times k_BT$, the Doppler-broadened elastic scattering reaction converges towards an immobile-nucleus elastic scattering, which has sharp edges at $E'=E$ and $E'=\alpha E$. Furthermore, for heavy nuclides, the $(1-\alpha)E$ `width' of the distribution is very narrow compared to the full energy range. This makes the correct interpolation of the energy-to-energy law difficult. To avoid this problem, we define moving boundaries that `follow' the $E=E'$ line at high energies. As the bidimensional data uses unit-based interpolation, this makes the interpolation easier. Strictly speaking, these boundaries make the adjoint distribution null where the direct distribution is not (which generates a bias). Therefore, we have to make sure that the part of the distribution that we are truncating is negligible. To assess this effect, we assume that the kinetic energy of the target will be below an arbitrary value $E_\text{target,max}$. We can then compute the corresponding speed $V_\text{max}= \sqrt{2E_\text{target,max}/A}$. At a given neutron speed $v\gg V_\text{max}$, we assume that the maximum attainable speed is then $v'_\text{max}=v+2V_\text{max}$ (corresponding to a collision with $A=\infty$), and the minimum is $v'_\text{min}=\sqrt{\alpha}v-2AV_\text{max}/(A+1)$. This gives the following boundaries
\begin{equation}
    \label{eq:svt_boundaries}
    \begin{aligned}
        E_\text{min}(E')&=\frac{1}{2}\left(\sqrt{2E'}-2V_\text{max} \right)^2\\
        E_\text{max}(E')&=\frac{1}{2\alpha}\left(\sqrt{2E'}+\frac{2A}{A+1}V_\text{max} \right)^2.
    \end{aligned}
\end{equation}
In our Monte Carlo code we take $E_\text{target,max}=20k_BT$. This value has been chosen arbitrarily; the probability that the target nuclide has an energy exceeding this value is about $10^{-8}$, leading to a bias that we consider acceptable. 

\begin{figure}[htb]
    \centering
    \begin{subfigure}[t]{0.49\textwidth}
        \centering
        \includegraphics[trim={0 0.8cm 0 0.53cm},clip,width=\textwidth]{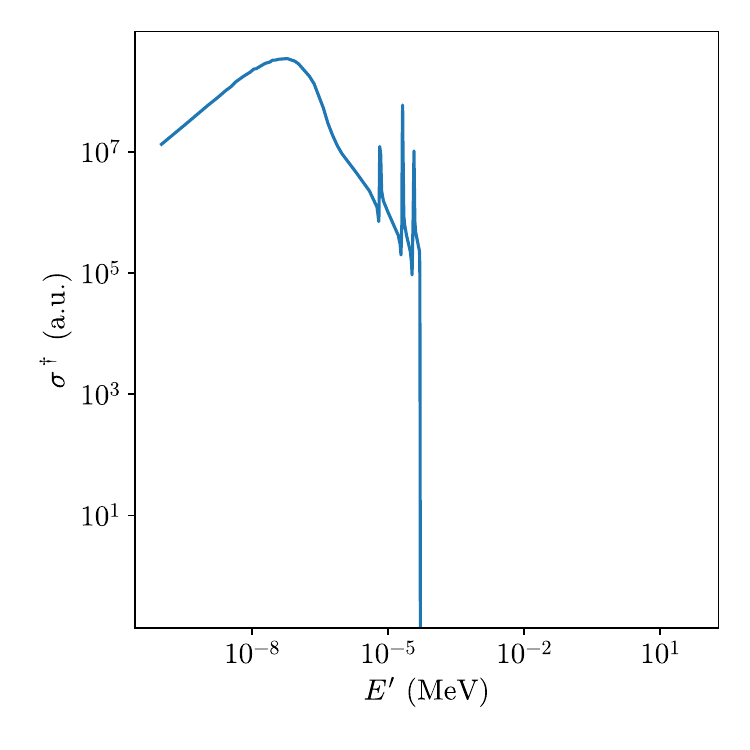}
    \end{subfigure}%
    ~ 
    \begin{subfigure}[t]{0.49\textwidth}
        \centering
        \includegraphics[trim={1cm 0.8cm 0cm 0.53cm},clip,width=\textwidth]{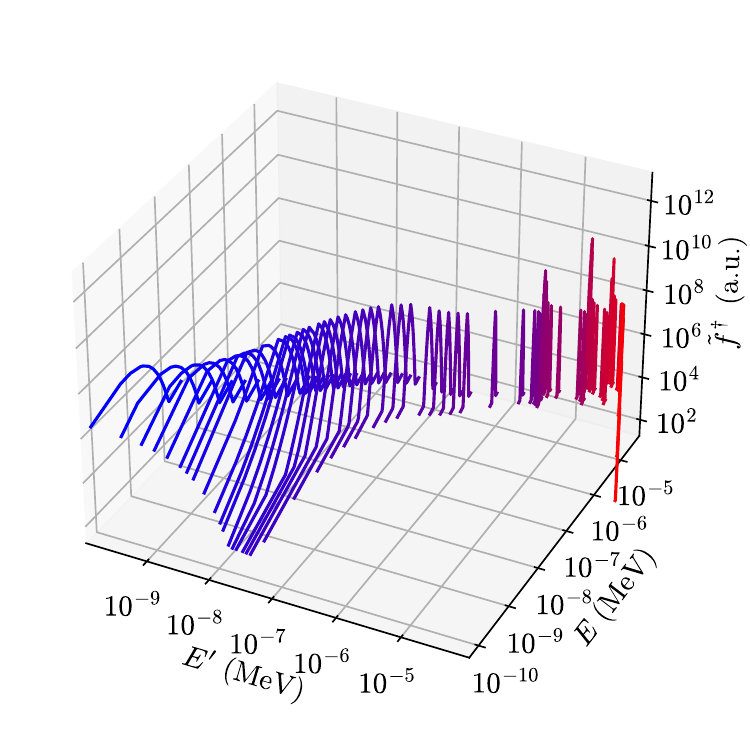}
    \end{subfigure}
    \caption{Adjoint nuclear data of the thermal part (SVT approximation) of the elastic scattering reaction of \textsuperscript{238}U. Data produced using \cref{eq:adj_xs_flx_gen} with a $g(E)$ featuring a $1/E$ neutron slowdown shape and a thermal Maxwellian at $k_BT=0.1\text{ eV}$ ($\approx1200\text{ K}$). In this example, no TSL reaction is used at low energy, and the code defaults to using the SVT approximation over the whole thermal range. \textbf{On the left}, the adjoint cross section $\sigma^\dagger(E')$. It features a thermal part (generated by the $g(E)$ function) and a neutron slowdown part. The first resonances of \textsuperscript{238}U reach into the $400 \times k_BT$ energy limit and are visible. \textbf{On the right}, the non-normalized adjoint distribution law $\tilde{f}^\dagger$. This discretized distribution contains 1673 data points, split into 77 unidimensional distributions, each with an average of 22 data points. The first resonances of \textsuperscript{238}U are also visible in the distribution law.}
    \label{fig:adj_U_svt}
\end{figure}

An example of adjoint energy-to-energy distribution is shown in \cref{fig:adj_U_svt}.

Once the energy-to-energy sampling is done, the last step concerns the sampling of the deflection cosine $\mu_l$. The angular distribution $g(E,E',\mu_l)$ for Doppler-broadened scattering at fixed energies $E,E'$ is not trivial to approximate. At thermal energies, it is roughly isotropic, but at higher energies (typically around $400 \times k_BT$) it converges towards the target-at-rest angular distribution, which is deterministic (see \cref{eq:el_scat_mucm_from_E_Ep}). We therefore choose to sample the adjoint angular distribution exactly like the direct angular distribution (using again the CXS and isotropic hypotheses). The angular distribution is then:
\begin{equation}
    \label{eq:svt_angular_distrib}
    g(E,E',\mu_l)= \frac{f(E\to E',\mu_l)}{f(E\to E')},
\end{equation}
where the numerator is computed using \cref{eq:kernel_svt_iso} and the denominator using \cref{eq:svt_energy_to_energy_kernel}. To sample from \cref{eq:svt_angular_distrib}, the only term depending on the deflection cosine $\mu_l$ in \cref{eq:kernel_svt_iso} is $q$, via the relation \cref{eq:svt_defq}. We can then sample directly $k:=q^2/E$ from a density
\begin{equation}
    \label{eq:svt_densityk}
    f(k)\propto \frac{1}{\sqrt{k}}\text{exp}\left( -\frac{1}{2}\left(\frac{a^2(E-E')^2}{E^2k}+\frac{a^2k}{4A^2}\right)\right),
\end{equation}
within the bounds $k\in[2(1-\sqrt{E'/E})^2,2(1+\sqrt{E'/E})^2]$, and then retrieve $\mu_l$ via the relation:
\begin{equation}
    \mu_l = \frac{1}{2}\left(1+\frac{E'}{E}-\frac{k}{2} \right)\sqrt{\frac{E}{E'}}.
\end{equation}
The density in \cref{eq:svt_densityk} is actually a Generalized Inverse Gaussian (GIG), which can be expressed under the form
\begin{equation}
    f(x) \propto x^{\lambda -1}\text{exp}\left(-\frac{1}{2}\left(\frac{\chi}{2x}+\psi x\right)\right),
\end{equation}
with the parameters $\lambda =0.5$, $\chi=a^2(E-E')^2/E^2$ and $\psi=a^2/(4A^2)$. We can then sample from this distribution using the efficient sampling technique proposed in Ref.~\citenum{hormann_generating_2014}, followed by a rejection if $k$ does not fall within the bounds. In the rare cases where the bounds fall really close to each other, which makes the rejection very inefficient, we sample instead from a uniform distribution within the bounds, and perform a rejection using \cref{eq:svt_densityk}. Useful information regarding the shape of GIG distribution laws that can be used to determine the maximum value of \cref{eq:svt_densityk} are  found e.g.~in Ref.~\citenum{hormann_generating_2014}. 

\subsection{Thermal scattering laws (TSL)}
\label{sec:tsl}

Thermal scattering laws (TSL) are used to describe the elastic scattering (MT2) with a nuclide taking into account solid-state effects (molecular bonds, phonon spectra, etc.) due to the medium. These laws are given directly in the laboratory frame, using no intermediary sampling, which makes their adjoint sampling easier than the SVT or DBRC procedures presented in \cref{sec:svt_reac}.

There exist two types of TSL: elastic laws that preserve the neutron energy ($E=E'$), and inelastic laws that can change the neutron energy. The adjoint law of an elastic TSL law is then the same as the direct law (we simply have to sample $\mu_l$ like in the direct distribution). The adjoint cross section is 
\begin{equation}
    \sigma^\dagger(E')=g(E')\sigma(E').
\end{equation}
Since the direct and adjoint distributions are the same, the factor $f(E\to E',\mu_l)/f^\dagger (E \to E',\mu_l)$ in the weight correction \cref{eq:weight_corr} is equal to unity.

Inelastic thermal reactions can be described by several formats (depending on the flags used in the ENDF format). We will focus here on the more modern `continuous' energy distribution, since it is more precise and easier to sample backward for our goals. Continuous inelastic TSL distributions are provided as energy-to-energy tabulated bidimensional data, coupled with an angular distribution for each $(E,E')$, similarly to what was done for Kalbach and Kalbach Tabular distributions (see \cref{sec:misc_freegas}). However, contrary to all the energy-to-energy distributions previously discussed, the interpolation between the different distributions is done differently. For a set of entering energies $E_k$, a tabulated unidimensional distribution along $E'$ is stored at points $E'_{k,l}$. To sample $f(E\to E',\mu_l)$, instead of performing stochastic interpolation between the distributions at $E_k$ and $E_{k+1}$, we always use $m$ so that $E_m$ is the discretized value closest to $E$. We then sample $E'_m$ from the unidimensional distribution at $E_m$, and sample the deflection cosine $\mu_l$ in the angular distribution corresponding to the closest $E'$ value. Finally, contrary to the previous examples, unit-based interpolation is not performed on the sampled $E'_m$ to obtain $E'$. The following interpolation is used:
\begin{itemize}
    \item if $E'_m>0.5 E'_{m,\text{max}}$ we use 
    \begin{equation}
        E' = E'_m + (E-E_m)
        \label{eq:tsl_translation_interp}
    \end{equation}
    \item if $E'_m\le0.5 E'_{m,\text{max}}$ we use
    \begin{equation}
        E' =  E'_m \times (2E - E_m)/E_m.
        \label{eq:tsl_scaled_interp}
    \end{equation}
\end{itemize}
This interpolation technique is used since the TSL distribution displays a spike at $E=E'$ (when $k_BT\ll E$) that needs to be properly interpolated between two consecutive $E_k$ distributions. The first part of the interpolation presented above (see \cref{eq:tsl_translation_interp}) effectively translates the distribution along the $E=E'$ line, preserving the spike. To prevent some portions of the distribution from landing at $E'<0$ when translating along $E=E'$, for low $E'$ energies, the distribution is `scaled' using the second part of the interpolation (see \cref{eq:tsl_scaled_interp}).

\begin{figure}[hbtp]
    \centering
    \includegraphics[width=.8\textwidth]{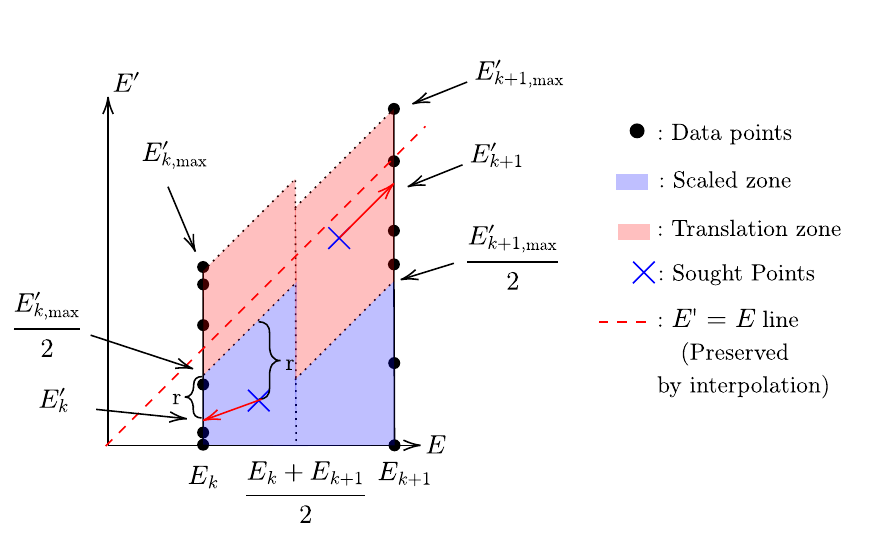}
    \caption{Procedure to obtain the density with the interpolation of TSL laws. One finds the closest distribution $E_m$, then computes $E'_m$ depending on the `interpolation zone' (translated or scaled) using \cref{eq:tsl_translation_inv,eq:tsl_scaled_inv}, and corrects the density using \cref{eq:tsl_density_correction}. }
    \label{fig:Interp2D_TSL}
\end{figure}

The energy density corresponding to such a sampling procedure is the following. At a given point $(E,E')$, we compute the corresponding closest energy $E_m$. If $E'>E-E_{m,\text{max}}/2$, then the $(E,E')$ point is in the `translated' part of the distribution, and $E'_m$ is computed as 
\begin{equation}
    E'_m = E' + E_m-E.
    \label{eq:tsl_translation_inv}
\end{equation}
Otherwise, the point is in the `scaled' part of the distribution, and $E'_m$ is computed as
\begin{equation}
    E'_m = E' \times \frac{E_m}{2E-E_m}.
    \label{eq:tsl_scaled_inv}
\end{equation}
The energy density is then computed at $E'_m$ in the unidimensional distribution at $E_m$. If $E'$ is in the translated part of the distribution, the value is used as-is, otherwise the density is corrected for the scaling factor, and 
\begin{equation}
    f(E\to E')= f(E_m\to E'_m) \times \frac{E_m}{2E-E_m}.
    \label{eq:tsl_density_correction}
\end{equation}
The interpolation procedure is summarized in \cref{fig:Interp2D_TSL}.

Obtaining the angular distribution density would be more complicated, since the sampling procedure involves drawing discrete $\mu_l$ values and then doing some `smearing' over the value to avoid neutron spikes at certain angles. To overcome this problem, we only prepare an adjoint distribution for the energy law (using the procedure in \cref{sec:bidim_interp}), and sample the angular part exactly like the direct distribution: we sample from the angular part of the direct distribution corresponding to the point $(E,E')$ sampled by the adjoint energy distribution. Therefore, we only use the ratio of the energy distribution laws $f(E\to E')/f^\dagger(E'\to E)$ in the weight correction \cref{eq:weight_corr}.

\begin{figure}[htb]
    \centering
    \begin{subfigure}[t]{0.49\textwidth}
        \centering
        \includegraphics[trim={0 0.8cm 0 0.53cm},clip,width=\textwidth]{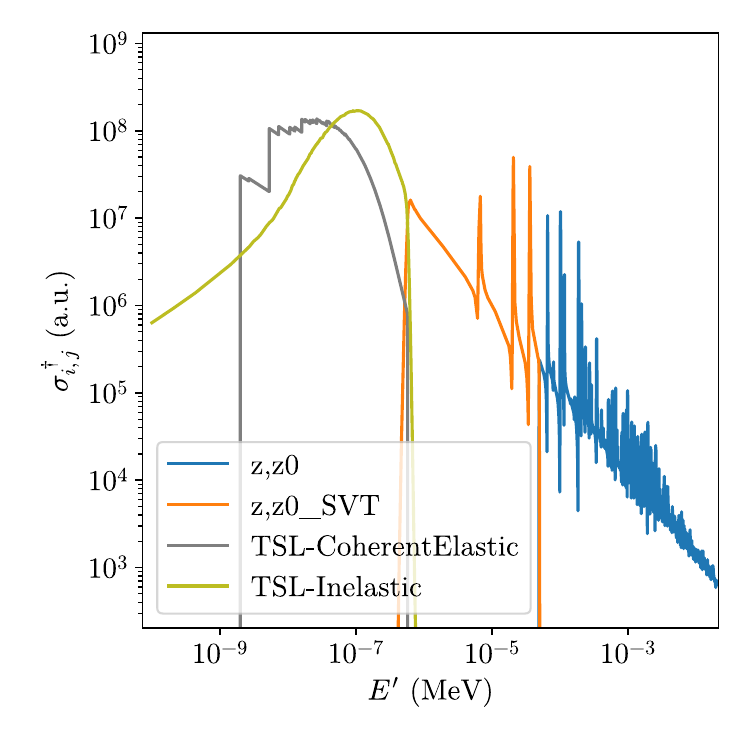}
    \end{subfigure}%
    ~ 
    \begin{subfigure}[t]{0.49\textwidth}
        \centering
        \includegraphics[trim={1cm 0.8cm 0cm 0.53cm},clip,width=\textwidth]{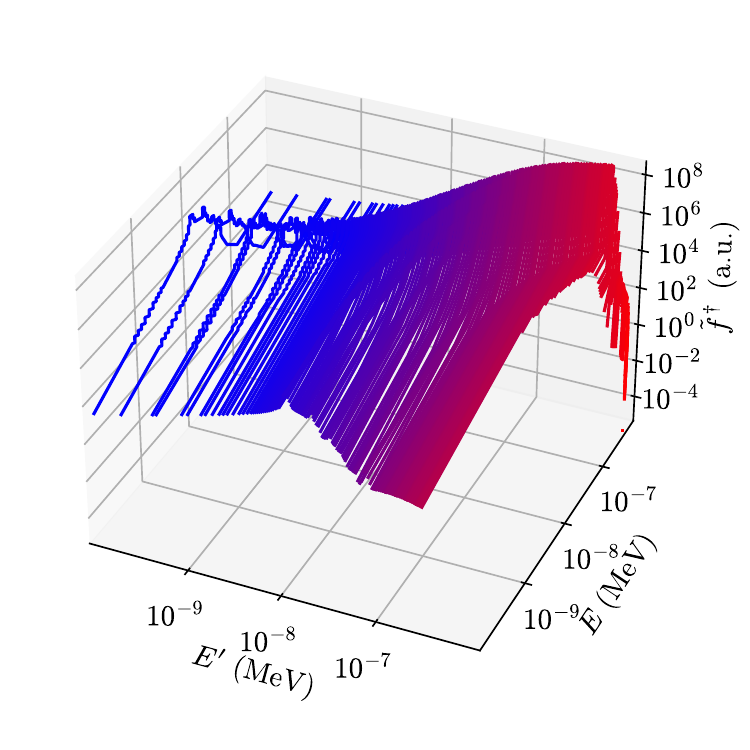}
    \end{subfigure}
    \caption{Adjoint nuclear data of the TSL of \textsuperscript{238}U (in UO\textsubscript{2} at $1200\text{ K}$). Data produced using \cref{eq:adj_xs_flx_gen} with a $g(E)$ featuring a $1/E$ neutron slowdown shape and a thermal Maxwellian shape at $k_BT=0.1\text{ eV}$ ($\approx1200\text{ K}$). \textbf{On the left}, the adjoint cross section $\sigma^\dagger(E')$ of the different TSL (elastic and inelastic) and also the truncated version of the SVT and free-gas elastic scattering. The thermal Maxwellian shape and the resonances are visible. The Bragg cross section of the coherent elastic TSL law are also visible in the adjoint cross section. \textbf{On the right}, the non-normalized adjoint distribution law $\tilde{f}^\dagger$ of the inelastic TSL reaction. This discretized distribution contains 6826 data points, split into 207 unidimensional distributions, each with an average of 33 data points. The jagged distributions at low $E'$ are due to the interpolation method used for the direct energy-to-energy distribution (see \cref{fig:Interp2D_TSL}), which creates small discontinuities. These discontinuities are then included in the adjoint distribution law $\tilde{f}^\dagger$.}
    \label{fig:adj_U_tsl}
\end{figure}

An example of the adjoint data produced using this technique is shown in \cref{fig:adj_U_tsl}. In the left part of \cref{fig:adj_U_tsl} we display the different reactions representing the elastic scattering (MT2). In a direct simulation, the TSL are used in the energy range where they are available; then, below $400 \times k_BT$, the SVT approximation is used, and finally, the target-at-rest model is used. Those energy bounds are not used for adjoint data. Since a target-at-rest elastic scattering can send neutrons below the $400 \times k_BT$ threshold, the adjoint target-at-rest elastic reaction needs to be present below the threshold. The same holds true for the SVT reaction with respect to the TSL energy range. When producing the adjoint reaction, we need to ensure that the cross section of the corresponding direct reaction is set to null wherever the energy is outside the energy range for which the reaction is accessible. This means that, if a nuclide is present in two media, one with TSL data and one without, we need to prepare two SVT adjoint reaction data for each medium: with the direct cross section truncated only for the medium that uses TSL.

\subsection{Unresolved Resonance Range (URR)}

The treatment of the unresolved resonance range (URR) is required to entirely cover the energy domain typically considered for neutron transport problems. In the unresolved resonance energy range the energy of the incoming particle is not enough to determine the corresponding value of the cross section for each nuclide $i$, a state variable (denoted $s_i$) is sampled uniformly in $[0,1]$, then the value of the cross section is read in a `probability table' corresponding to the energy $E$ that converts the state $s_i$ into the actual cross section value. Three reactions are usually represented by this model: elastic scattering, absorption, and fission. The details of how the URR probability tables are stored and interpolated can vary depending on the method used. Details can be found in Ref.~\citenum{conlin_compact_2019,sublet_calendf-2010_2011}, but are not needed for the adjoint URR implementation, and will not be provided here.

To understand how the adjoint implementation of the URR works, let us first recall that the specific state $s_i$ is only used to determine the cross section $\sigma_{i,j}$: it is used during the particle flight to determine the total macroscopic cross section $\Sigma_t$, and during the collision event to choose the reaction path $\{i,j\}$; however, it is not required to sample the distribution law $f_{i,j}(E\to E',\mu_l)$. Conversely, during the adjoint collision event, the state $s_i$ is not necessary to sample the adjoint distribution law $f_{i,j}^\dagger(E'\to E, \mu_l)$, and is only useful to compute the direct cross sections $\sigma_{i,j}(E,s_i)$ and $\Sigma_t(\mathbf{r},E,(s_i)_{i\in I})$ needed to compute the weight correction \cref{eq:weight_corr}. Therefore, we can prepare the adjoint distribution law $f_{i,j}^\dagger$ using the continuous cross section (as if there were no URR) using \cref{eq:adj_xs_flx_gen}, and sample from the adjoint distribution law normally. If the adjuncton lands at an energy $E$ within the URR energy range, then we need to sample all the states $s_i$ necessary to evaluate the cross sections in the current medium. The sampling of the URR states $s_i$ can be done uniformly in $[0,1]$, similarly to the direct sampling. Also, similarly to the direct transport, if more URR states prove to be necessary during an adjoint flight, we can sample them according to the same law. This requires no weight correction, since we are sampling the state $s_i$ from the same law as in direct transport.

\begin{figure}[hbtp]
    \centering
    \includegraphics[trim={0 0.8cm 0 0.53cm},clip,width=.49\textwidth]{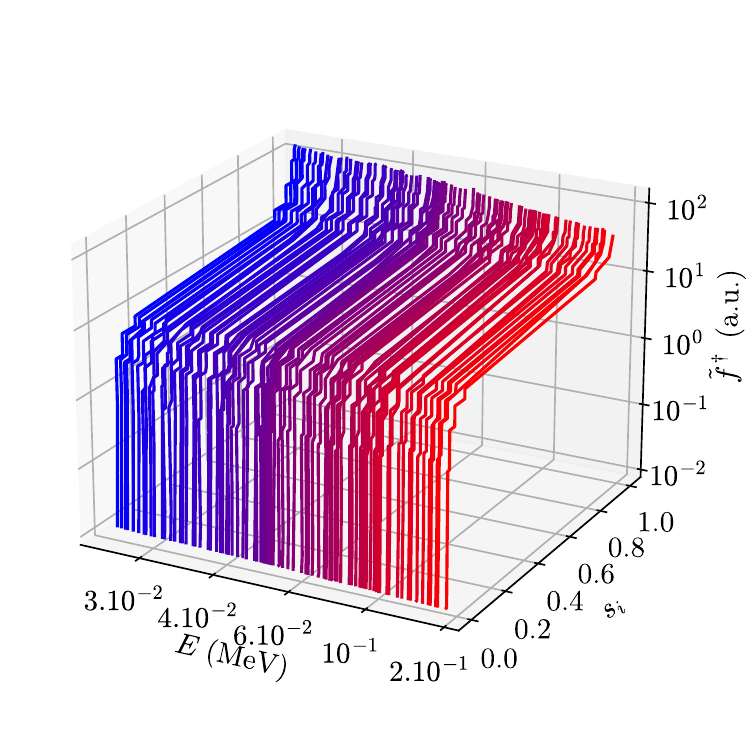}
    \caption{Adjoint URR state distribution of the elastic scattering of \textsuperscript{238}U. This discretized distribution contains 1462 data points, split into 81 unidimensional distributions, each with an average of 18 data points. The `step' nature of the adjoint distribution comes from the interpolation procedure of the URR that performs no interpolation between the cross sections values in the probability table.}
    \label{fig:adj_urr_U}
\end{figure}

In the specific case of the state $s_i$ corresponding to the reaction path $\{i,j\}$ that the adjuncton took to land at energy $E$, sampling $s_i$ might not be ideal. In the spirit of the definition of the adjoint distribution law in \cref{eq:adj_xs_flx_gen} that sends particles to energy $E$ proportionally to the cross section at that energy $\sigma_{i,j}(E)$, we would like to sample the state $s_i$ proportionally to the cross section at $(E,s_i)$. This leads to an adjoint state distribution:
\begin{equation}
    f_{i,j}^\dagger(E,s_i) = \frac{\sigma_{i,j}(E,s_i)}{\int_0^1\sigma_{i,j}(E,\tilde{s}_i)d\tilde{s}_i},
    \label{eq:adj_state_distrib}
\end{equation}
which needs to be compensated for by a weight correction $1/f_{i,j}^\dagger(E,s_i)$ to ensure unbiasedness. Note that, although the state $s_i$ is shared for all reactions of a given nuclide $i$, we propose a different distribution to sample $s_i$ for each reaction $\{i,j\}$. To store this quantity, we again use a non-normalized version 
\begin{equation}
    \tilde{f}_{i,j}^\dagger(E,s_i)= \sigma_{i,j}(E,s_i),
\end{equation}
stored using the bidimensional data treatment technique presented in \cref{sec:bidim_interp}. An example of adjoint URR state distribution is shown in \cref{fig:adj_urr_U}.

%% file: Discretization.tex
\section{Discretization algorithms}
\label{sec:disc_algo}

\subsection{Requirements of the discretized data}

Throughout the preparation of the adjoint cross sections and adjoint distribution laws in \cref{sec:prep_adj_data}, we required the storage of discretized uni and bidimensional functions. In \cref{sec:bidim_interp}, we proposed a technique to interpolate these functions (and distributions) in between the discretization points. We will now discuss the choice of the positions of the discretization points using uni and bidimensional algorithms.

Throughout this section, we will denote $f_o$ the original function that needs to be discretized, and $f_d$ the function that has been discretized and reconstructed through interpolation in between the data points. When considering unidimensional functions, the variable of these functions will be called $x$, and the discretization space will be an interval $I_x$. When considering bidimensional functions, the variables of these functions will be called $x$ and $w$. Following the interpolation procedure detailed in \cref{sec:bidim_interp}, $f_d$ will be composed of an ensemble of unidimensional functions along $x$, for an ensemble of $w$ values. The variable $w$ will span an interval $I_w$, and for a given $w$ the variable $x$ will span an interval $I_x(w)$; the interval $I_x(w)$ is delimited by boundary functions or edge functions denoted $x_\text{min}(w)$ and $x_\text{max}(w)$. The domain over which we want to discretize the original function $f_o$ is denoted $\mathcal{O}$. Following the notation of \cref{sec:bidim_interp}, the dicretized $w$ points will be denoted $w_k$, and the discretized $x$ points in the unidimensional function at $w_k$ will be denoted $x_{k,l}$; when simply referring to a unidimensional function, the index will be denoted $x_k$. The discretization position in two dimensions will be denoted $p_{k,l}=(w_k,x_{k,l})$; in one dimension, we will use $p_l= (x_l)$. The ensemble of all discretization positions will be denoted $\mathcal{D}$.

First, let us state our goals for the discretization procedure:
\begin{itemize}
    \item We want our discretized function $f_d$ to resemble closely the original function $f_0$; i.e., we want enough discretization points at relevant positions so that $f_d$ remains within a given tolerance from $f_o$. The tolerance can be an absolute distance, a relative distance, or a more complicated indicator. Tolerance will be detailed in \cref{sec:notion_of_tolerance}.
    \item We want the number of discretization points in $\mathcal{D}$ to be small. This ensures that the discretized data does not use too much memory, but also keeps the access to the data fast.
    \item We want our discretization algorithm to run at an acceptable speed.
\end{itemize}
These objectives will be made `quantitative' through the analysis carried out in this section. 

Bearing these broad objectives in mind, we can already rule out a naive strategy that would consist in storing data points at regularly placed intervals in $\mathcal{O}$, either in logarithmic scale if the variable ranges several orders of magnitude (like for the energy $E$, which usually ranges from $10^{-6}\text{ eV}$ to $20\text{ MeV}$), or in linear scale otherwise (like for $\mu_l$, which lies in $[-1,1]$). For our purposes, the data that we want to discretize usually stems from the definition of the adjoint cross sections and distribution laws \cref{eq:adj_xs_flx_gen} that in turn depend on the direct cross section $\sigma_{i,j}(E)$; thus, in some regions of $\mathcal{O}$, $f_o$ will display resonances or other heterogeneous features that require a high density of points in $\mathcal{D}$ to be faithfully discretized. Some other regions in $\mathcal{O}$ might be `smoother', necessitating fewer discretization points, such as the cross sections outside of the resonance region. Using a uniform density of points, we would need an unacceptable number of data points to obtain a faithful reconstruction of $f_o$ everywhere.

To overcome these issues, we use two algorithms. The first algorithm finds `strategic' locations to place the discretization points in $\mathcal{O}$ so that the features of $f_o$ are correctly reconstructed in $f_d$ without using a prohibitive number of data points. To achieve this property, first it explores $\mathcal{O}$ by evaluating $f_o$ at test points, to ascertain whether there are heterogeneities, and then recursively constructs $f_a$, a temporary version of $f_d$, by adding points where the tolerance criterion is not respected. Once the function $f_o$ is correctly reconstructed by $f_a$, a second algorithm tries to find redundant points in $\mathcal{D}_a$ (the ensemble of discretization points of $f_a$), and remove them to reduce the discretized data size. The result of the successive application of these two algorithms is $f_d$: this discretized function respects the tolerance criterion with respect to $f_o$, and has a reduced data size $\text{Card}(\mathcal{D})\le \text{Card}(\mathcal{D}_a)$. The two algorithms will be detailed in the next sections.

\subsection{Additive algorithm}
\label{sec:disc_additive_algo}

We will start by describing the discretization algorithm responsible for `exploring' the function $f_o$ on the domain $\mathcal{O}$ and adding discretization points to obtain a first discretized function $f_a$. This strategy was inspired by the algorithm used by the data processing code NJOY to reconstruct linearly interpolated cross sections from ENDF files in the `RECONR' module \cite{macfarlane_njoy_2017}. This algorithm (and the one presented next) will be detailed in a unidimensional configuration. bidimensional generalization will be discussed in \cref{sec:disc_two_dim_gen}. 

\begin{figure}[hbt]
    \centering
    \includegraphics[width=.9\textwidth]{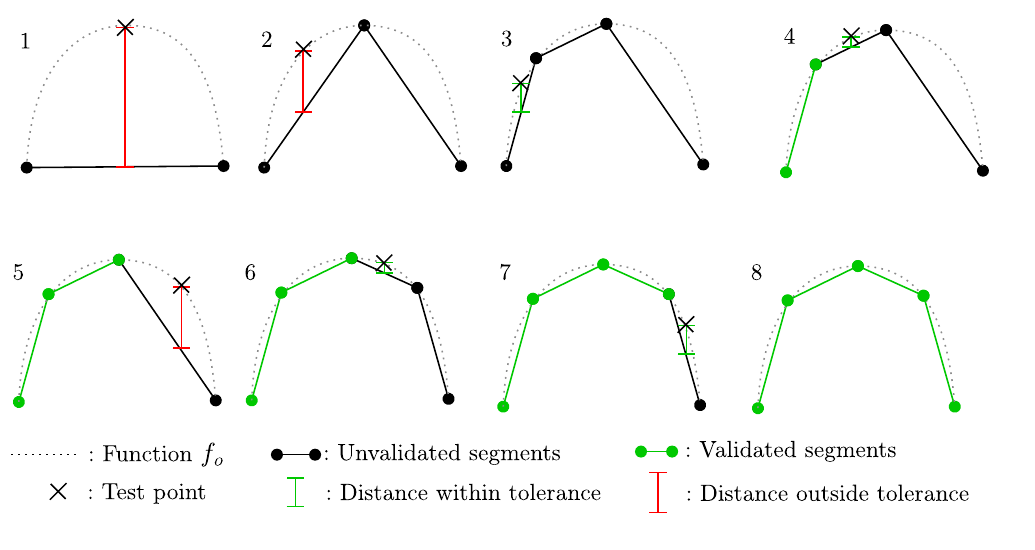}
    \caption{Discretization procedure using the additive algorithm. The dotted line represents the original function $f_o$, the segments represent the successive iterations of the discretized function $f_a$, black segments are not validated, while green segments are validated. The black cross represents the test point (here $n_\text{test}=1$) placed in the segment to test and where $f_o$ is evaluated. Vertical segments represent the distance between the segment interpolation and the true position of the test point. If it is green, then the interpolation is within tolerance; if it is red, then the interpolation needs to be refined. \textbf{Different steps, left to right, top to bottom:} 1: The segment is tested and the test point is outside tolerance; the test point is added. 2: The leftmost segment is tested and not validated; the test point is added. 3: The first segment is validated. 4: The second segment is validated. 5: The third segment is not validated; the test point is added. 6, 7: Both remaining segments are validated. 8: All segments are validated; the algorithm ends.
    }
    \label{fig:rec_add_disc}
\end{figure}

The additive algorithm takes as an argument a preliminary mesh spanning the entire interval $I$, and then tests each `segment' of the corresponding reconstructed function $f_a$. To ascertain whether a given segment correctly represents the actual function $f_o$, it evaluates $f_o$ at intermediary test points within said segment. If needed, the algorithm refines the segment by adding data points to the mesh until all the segments are validated. The specific procedure is as follows:
\begin{enumerate}
    \item The segments of the initial discretized function $f_a$ are stored on a stack of not yet validated segments. For this purpose, the ideal method consists in storing the leftmost unverified segment on the top of the stack.
    \item The validated segments are transferred into another stack representing validated segments.
    \item The leftmost non-validated segment is retrieved from the non-validated stack.
    \item A number $n_\text{test}$ of points $\tilde{x}_m$ are created in between the extremities $x_l$ and $x_{l+1}$ of the current segment.
    \item For all $\tilde{x}_m$, the function $f_0$ is evaluated, and the value $f_0(\tilde{x}_m)$ is compared with the interpolated value $f_a(\tilde{x}_m)$.
    \item If all the points are within the tolerance, then the current segment is transferred to the validated stack.
    \item If at least one of the points is outside the tolerance, the segments needs to be refined. All the test positions $\tilde{x}_m$ are added to the discretized function mesh. The current segment turns into $n_\text{test}+1$ new segments that are added to the stack of non-validated segments. 
    \item The algorithm goes back to step 3.
    \item When the non-validated segment stack is empty, all segments have been verified, and the algorithm ends.
\end{enumerate}
This discretization procedure is illustrated in \cref{fig:rec_add_disc}.

This additive algorithm can sometimes fail to properly discretize $f_o$: this happens when a segment is validated although the function $f_o$ is not faithfully represented by it. The recursive nature of the algorithm will amplify any validation error, since a validated segment is never tested again, potentially `missing' a whole tree of `offspring' segments that would have been added if the segment had not been incorrectly validated.

\begin{figure}[hbt]
    \centering
    \includegraphics[width=.7\textwidth]{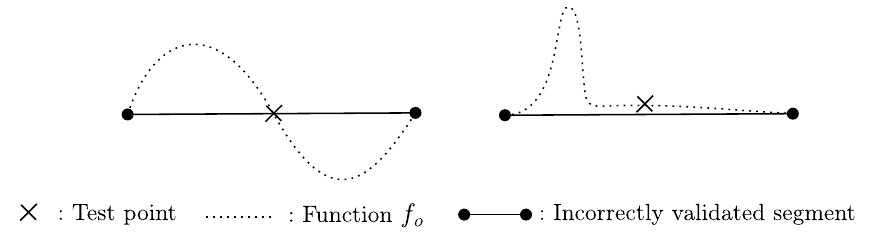}
    \caption{ Fail cases of the additive discretization algorithm where a segment is incorrectly validated. \textbf{On the left}, the function happens to be close to the segment at the test point. This will happen randomly as many segments are tested. \textbf{On the right}, the function features a heterogeneity, but in a region whose spatial extent is orders of magnitude smaller than the segment size. The test points will then likely land on a non-heterogeneous part of the function and incorrectly validate the segment. See text for a more thorough interpretation of both modes of failure.
    }
    \label{fig:rec_add_disc_fail}
\end{figure}

The incorrect validation of a segment can usually happen for two reasons (for an illustration, see \cref{fig:rec_add_disc_fail}). 

The first case corresponds to simple bad luck: if the function $f_o$ has important variations compared to the considered segment but happens to be close to the interpolation at all of the $n_\text{test}$ test points, then the segment will be incorrectly validated. In order to overcome this issue, we might use a bigger number of test points. If all of the test points have a given (independent) probability to lie within the tolerance of the segment, as $n_\text{test}$ increases, the probability of all test points being validated decreases exponentially. In some cases, the number of segments to be tested can be very large: since any error of validation can have a massive impact, it is safer to take some margins over $n_\text{test}$. When discretizing data affected by strong heterogeneities (for example, containing resonant cross sections), in our code we typically use $n_\text{test}=7$.

The second case can occur if the heterogeneities in the function $f_o$ have an extent that is very small compared to the segment size. In this case, statistically the test points will not land in the heterogeneity region, and the segment will be incorrectly validated. This problem could be alleviated by increasing $n_\text{test}$, but it would require a number of test positions proportional to the relative size difference between the tested segment and the heterogeneity region. A better strategy is to ensure that all heterogeneity regions will start with at least one point of the initial mesh in it. Instead of starting the algorithm with only the two points at the bounds of $\mathcal{O}$, we start with a linearly (or logarithmically) spaced mesh dense enough to ensure the aforementioned condition. In our code, when discretizing unidimensional functions, if $x$ is on a logarithmic scale we initially place 300 points per decade; otherwise, we initially place 100 linearly spaced points.

This algorithm has the inconvenience that it can only add points. In the end, the preliminary discretized function $f_a$ usually contains many redundant points.

\subsection{Douglas-Peucker algorithm}
\label{sec:disc_rdp}

The Douglas-Peucker algorithm (also known as the Ramer-Douglas-Peucker algorithm), is a procedure initially introduced in the field of cartography to compress segmented curves in maps \cite{douglas_algorithms_1973}; a recent summary of the procedure can be found e.g.~in Ref.~\citenum{hershberger_speeding_1992}.

\begin{figure}[hbt]
    \centering
    \includegraphics[width=.9\textwidth]{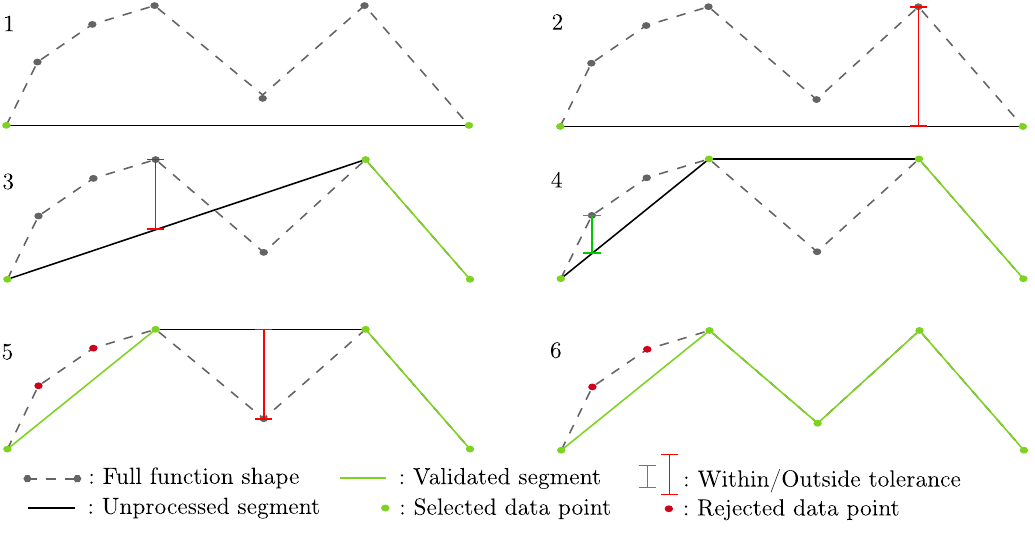}
    \caption{Discretization procedure using the Douglas-Peucker algorithm. The dotted line represents the function discretized by the additive algorithm $f_a$, the segments represent the successive iterations of the simplified discretized function $f_d$, black segments are not validated, while green segments are validated. Vertical segments represent the distance between the segment interpolation and the point on $f_a$. If it is green, then the interpolation is within tolerance; if it is red, then the interpolation needs to be refined. \textbf{Different steps, left to right, top to bottom:} 1: The two extremities of $f_a$ are added to $f_d$; the segment that they form is unprocessed. 2: The point farthest away from the segment is tested to be outside of tolerance. The point is added to $f_d$. Of the two sub-segments newly formed, the first is added to the not-validated segment stack; the second has no intermediary point in $f_a$ and can be validated straight away. 3: The farthest point from the not-validated segment is outside tolerance; the point is added to $f_d$ and the two sub-segments are not-validated. 4: The farthest point from the first not-validated segment is within tolerance; that segment is validated and all the points within it are discarded. 5: The farthest point from the last not-validated segment is outside tolerance; the point is added; both subsegments are validated by default as they contain no intermediary point in $f_a$. 6: No remaining not-validated segments; the algorithm ends
    }
    \label{fig:RDP_disc}
\end{figure}

The principle of this algorithm is to start with the set of initial points composing the original discretized function $f_a$, and recursively select data points that we want to keep, until the obtained function $f_d$ correctly represents $f_a$. Any points left over by this procedure are redundant, and are then discarded. The algorithm is the following:
\begin{enumerate}
    \item First select the two extremities of $f_a$, and put the segment formed by these points into a stack of unprocessed segments.
    \item Retrieve an unprocessed segment from the stack.
    \item Among all the points $x_l$ from $f_a$ that are within that segment, find the farthest away from the segment. 
    \item If this point is within the tolerance of the segment, then all the points are within the tolerance, the segment is validated, and all the points are discarded.
    \item If the farthest away point is not within the tolerance of the segment, the point is selected. The two newly formed sub-segments are added to the stack of unprocessed segments.
    \item The algorithm returns to step 2.
    \item When the stack of unprocessed segments is empty, the selected points form the discretized (and `reduced') function $f_d$, and the rest of the points are discarded.
\end{enumerate}
This discretization procedure is illustrated in \cref{fig:RDP_disc}.

The Douglas-Peucker algorithm is much more resilient than the additive algorithm presented above, in that it surely yields a correct approximation (within the given tolerance). This is because this algorithm works with a finite known set of points and can check all of them, contrary to the additive algorithm, which has to explore the continuous domain $\mathcal{O}$.

\subsection{Bidimensional generalization}
\label{sec:disc_two_dim_gen}

Both the additive algorithm presented in \cref{sec:disc_additive_algo} and the Douglas-Peucker algorithm presented in \cref{sec:disc_rdp} are unidimensional. There exist many algorithms to discretize bidimensional surfaces; however, they usually produce either triangle-based meshes or regular meshes. On the contrary, our Monte Carlo sampling strategy imposes a specific mesh type: a bidimensional surface composed of a set of unidimensional functions along the coordinate $x$, at fixed values of $w$ (see \cref{sec:bidim_interp}).

We generalized both algorithms presented above so they can discretize such bidimensional functions. The main idea is that our bidimensional discretized functions can be seen as a unidimensional discretized function along $w$, except that every discretized point of this function is itself a discretized function along $x$. Individual functions along $x$ can be discretized using the unidimensional algorithms presented above. The functions of functions (along $w$) can be also discretized using the same algorithms, by suitably generalizing the different operations previously done on points and segments so that they apply to discretized functions.

The notion of segments between two points can be generalized using the notion of linear interpolation (along $w$) of two functions (of variable $x$). For this purpose, we will use the interpolation scheme presented in \cref{sec:bidim_interp}, taking into account the edges of the distribution with unit-based interpolation.

The verification of the distance between two functions (and the test with respect to the tolerance) can be achieved in several ways. Indeed, there exist many mathematical norms in the space of functions (e.g.~1-norm, 2-norm, or $\infty$-norm). In our case, since we want to ensure that all the points remain within a given tolerance, we will use the $\infty$-norm, which effectively measures the maximal distance between two points of two functions. Finding the maximal distance numerically requires both functions to be discretized (and linearly interpolated between the data points); then we can simply assess the distance at all the discretization points from both functions. Since interpolated functions (that generalize the segment) following the method in \cref{sec:bidim_interp} are effectively linearly interpolated functions, this poses no problem. Care should be taken in the additive algorithm, when one creates test points from $f_o$: in the bidimensional generalization, these test points are functions of $x$. In order to be able to test the tolerance, these functions need to be discretized. The additive bidimensional algorithm needs to call the additive unidimensional algorithm to produce a discretized `slice' of the function $f_o$ at a given $\tilde{w}$. This requires the bidimensional additive algorithm to have access to the boundary functions $x_\text{min}(w)$ and $x_\text{max}(w)$ at any arbitrary $w$, in order to produce the initial regular mesh given to the unidimensional additive algorithm to obtain the `slice' of $f_o$ at the test point $\tilde{w}$.

Overall, the bidimensional discretization scheme first performs an additive discretization of $f_o$. The algorithm discretizes along $w$, using the unidimensional additive discretization algorithm along $x$ to produce the test points. The bidimensional discretized function $f_a$ can be then simplified in two ways using the Douglas-Peucker algorithm. Either one first simplifies all the individual unidimensional functions along $x$, and then proceeds to simplify the `main' function along $w$, or the other way around. The second method seems to be more efficient: since simplifying along $w$ requires removing entire functions along $x$, it is easier to do so with not-yet-simplified functions. The final tolerance obtained with this `two-stage' Douglas-Peucker application is twice as large as the individual tolerance.

\subsection{Some comments on the tolerance}
\label{sec:notion_of_tolerance}

An important part of the discretization is how the tolerance between the discretized function $f_d$ and the original function $f_o$ is determined. The first way of defining a tolerance would be to set a maximum absolute distance between $f_d$ and $f_o$. However, since the discretized data tends to span orders of magnitude (this is mainly due to important variations of $g(E)$ and of $\sigma(E)$ over the energy range), using an absolute tolerance seems ill-advised. We could instead define a relative tolerance, where the distance at point $p$ is computed as:
\begin{equation}
    d_\text{rel}(p)= \left|\frac{f_d(p) -f_o(p)}{f_o(p)}\right|,
\end{equation}
and $d_\text{rel}$ needs to stay below a given value. We could also use several other kinds of distance functions with a limit value as a tolerance.

Properly choosing a tolerance requires knowing the impact of using discretized adjoint data different from the reference definitions in \cref{eq:adj_xs_flx_gen}. As stated in \cref{sec:def_adj_data}, any adjoint data will be unbiased, as long as the correct weight correction is used (see \cref{eq:weight_corr}). However, the definition in \cref{eq:adj_xs_flx_gen} was devised to reduce variance. Since the function $g(E)$ requires a priori knowing the shape of the direct flux to maximize the performance of the Monte Carlo game (and the flux is generally not known exactly in advance), we should not worry too much about venturing away from the value given in \cref{eq:adj_xs_flx_gen}, since even this reference is typically imprecise.

In our code, we aim at staying within 10\% of the reference definition for unidimensional data, and 20\% for bidimensional data. The discrepancy between $f_o$ and $f_d$ will impact the variance of the adjoint game mainly through the weight correction factor \cref{eq:weight_corr}. If the value used for the adjoint distribution $f^\dagger_{i,j}$ or the adjoint cross section $\sigma^\dagger_{i,j}$ differs from the reference by a factor $r_{d/o} := f_d/f_o$, then the weight correction will differ by a factor $1/r_{d/o}$. The multiplicative nature of the weight correction justifies the use of a tolerance based on a relative distance.

A small improvement can be made to the relative-distance tolerance by noting that the $1/r_{d/o}$ factor will have more impact if the discretized version is smaller than the original function, since this can potentially lead to particles carrying huge weights, with a possibly catastrophic impact on the variance. On the contrary, even if $f_d$ is non-zero somewhere in $\mathcal{O}$ where $f_o$ is zero, leading to $r_{d/o}=\infty$, these particles will die (their weight being corrected by a factor $0$); furthermore their impact on the overall calculation might be limited if the fraction of particles that go through this region of $\mathcal{O}$ remains small. Therefore, we can tolerate large relative discrepancies if $f_o<f_d$ and if these events concern a small fraction of particles. In our processing code, this consideration has led to the implementation of a modified tolerance that becomes very permissive when the density $f_o$ is very small. To avoid discretizing phase-space regions that are several orders of magnitude less likely than others, we even use a `leveler' that assigns a minimum value for $f_d$ in regions representing, in total, fewer than $0.2\%$ of the cumulative distribution (hence, fewer than 0.2\% of particles going through this distribution). This improvement is not used in unidimensional discretizations that can be truncated, such as the function $\tilde{h}(E)$ used for isotropic elastic scattering \cref{sec:el_scat}.

The tolerance used at the different steps of discretization can be different. We usually use a tolerance 1.5 times smaller for the first step of additive discretization, to increase the chance of correctly exploring $f_o$ on $\mathcal{O}$; the supplementary data points are then deleted by the Douglas-Peucker algorithm.

Finally, in some situations the tolerance is not respected, to ensure the convergence of the additive algorithm. Because of its recursive nature, the additive algorithm can enter a situation where adding more data points will not help improve the precision of the discretization. Such a situation can occur when a discontinuity is present in $f_o$, especially in a bidimensional case when the discontinuity is neither in the $w$ nor the $x$ direction. In such cases, a maximum density of points is used to prevent infinite recursion. In our code, we use a maximum density of around $100~000\text{ points/decade}$ in the $x$ direction, and a maximum of $200\text{ functions/decade}$ in the $w$ direction.

\begin{figure}[hbtp]
    \centering
    \includegraphics[width=.9\textwidth]{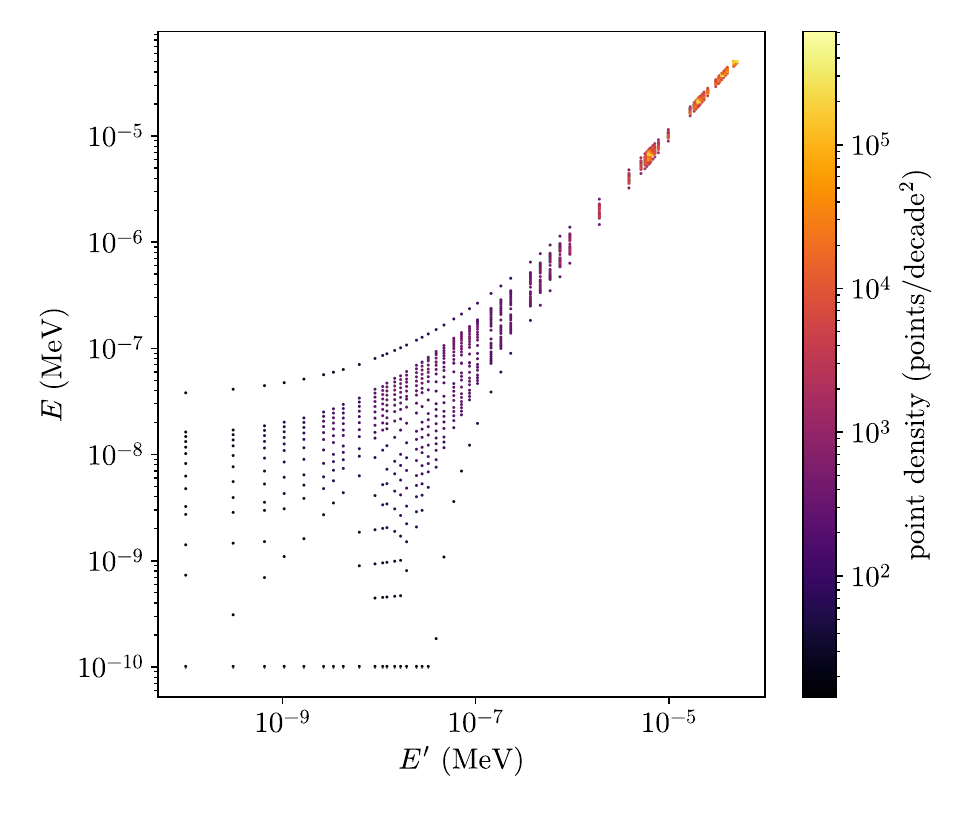}
    \caption{2D discretization mesh $\mathcal{D}$, for the energy-to-energy adjoint law $\tilde{f}^\dagger$ prepared for the adjoint SVT reaction of \textsuperscript{238}U. The discretized function is presented in the right part of \cref{fig:adj_U_svt}. Each dot on the scatter plot represents a discretization point $p_{k,l}$ in the $(E',E)$ space (corresponding, respectively, to $w$ and $x$). The color of the dots represents the local density of discretization points measured using a nearest neighbor technique. The discretization is able to adapt the density of functions along $E'$ and, within each function, the density of points along $E$. The regions of high discretization point density at $E\approx E' \in \{6.3,20,48\}\text{ eV}$ correspond to energies of the first resonances of \textsuperscript{238}U, impacting the adjoint distribution law produced. At these difficult points, the discretization briefly reaches the limit of 200 functions/decade. The effect of the resonances on the adjoint distribution law can also be observed in \cref{fig:adj_U_svt}. The first and last point of each function along $E$ are placed at the boundaries $E_\text{min}(E')$ and $E_\text{max}(E')$ of the distribution (defined in \cref{eq:svt_boundaries}). As the original function $f_o$ is orders of magnitude smaller close to the boundaries compared to along the $E=E'$ line, the `leveler' fills the regions close to the boundaries with a minimum value, avoiding expensive discretization in these regions. The `leveled' region, can also be seen in \cref{fig:adj_U_svt}. }
    \label{fig:2D_Mesh}
\end{figure}

For illustration, an example of discretization mesh $\mathcal{D}$ obtained using our code is presented in \cref{fig:2D_Mesh}.

%% file: Results.tex
\section{Numerical Results}
\label{sec:results}
\subsection{Production of the adjoint nuclear data library}
\label{sec:res_adj_data}

The procedures described in \cref{sec:prep_adj_data} to prepare the adjoint cross sections and distribution laws (including the discretization algorithms presented in \cref{sec:disc_algo}) have been developed in the Monte Carlo mini-app presented in Ref.~\citenum{rovel_general_2025}. This code implements both direct and adjoint transport simulations using the method described in \cref{sec:theory} (see Ref.~\citenum{rovel_general_2025} for a thorough discussion), and will be used in \cref{sec:res_reciprocity} to verify the adjoint Monte Carlo simulation procedure. 

The nuclear data was retrieved from the ENDF-B/VIII.0 library in the form of ACE files, processed with NJOY~\cite{macfarlane_njoy_2017}. Using the ALEXANDRIA module of \tripolir{} \cite{mancusi_overview_2024} to parse the ACE files, the data was then converted into files in JSON format. These files are used as input for our Monte Carlo mini-app. Taking human-readable files as an input allows us to easily modify existing nuclear data, or even create entirely synthetic nuclear data to explore the properties of adjoint transport. Our mini-app reads the input data file and produces the adjoint cross sections and adjoint distribution laws that will be needed to perform the adjoint transport. Adjoint nuclear data can also be exported in JSON format. This allows for easy visualization and manipulation of the produced discretized adjoint data.

Exporting the ENDF-B/VIII.0 library at 0~K (555 isotopes) into our JSON input format takes around 4~GB of disk space. This storage is not equally distributed, the largest nuclide being \textsuperscript{235}U, which represents at 10\% of the total storage space (400~MB); for comparison, the median nuclide takes 0.1\% (5~MB). The adjoint nuclear data produced in the same JSON format is smaller, and takes around 2~GB of disk space. The distribution of data sizes is less skewed, with the largest nuclide being again \textsuperscript{235}U, which represents 1\% of the total storage space (20~MB);  the median nuclide takes 0.15\% (3~MB). The computing time required to produce the aforementioned adjoint data is around 20 minutes on a single CPU. These values were obtained with the default setting of $\pm 10\%$ precision for unidimensional data, and $\pm 20\%$ precision for bidimensional data. Doubling the tolerance settings leads to a comparatively small reduction of -10\% in disk space size, while dividing it by a factor of 10 leads to an increase in the disk space size by a factor of 2.5. The central settings have been found to work well on several test cases. However, a more thorough analysis should be performed using real-life benchmark configurations for adjoint transport problems: this would allow assessing the impact of discretization precision on the size of the produced adjoint data and on the variance of the game. 

Since the adjoint transport simulation procedure described in \cref{sec:theory} demands both the adjoint and the direct nuclear data library (this latter being needed to perform the weight correction according to \cref{eq:weight_corr}), adjoint simulations will still require more memory that an equivalent direct simulation. 

\subsection{Adjoint transport simulations}
\label{sec:res_adj_calc}

\begin{figure}[hbtp]
    \centering
    \includegraphics[width=.9\textwidth]{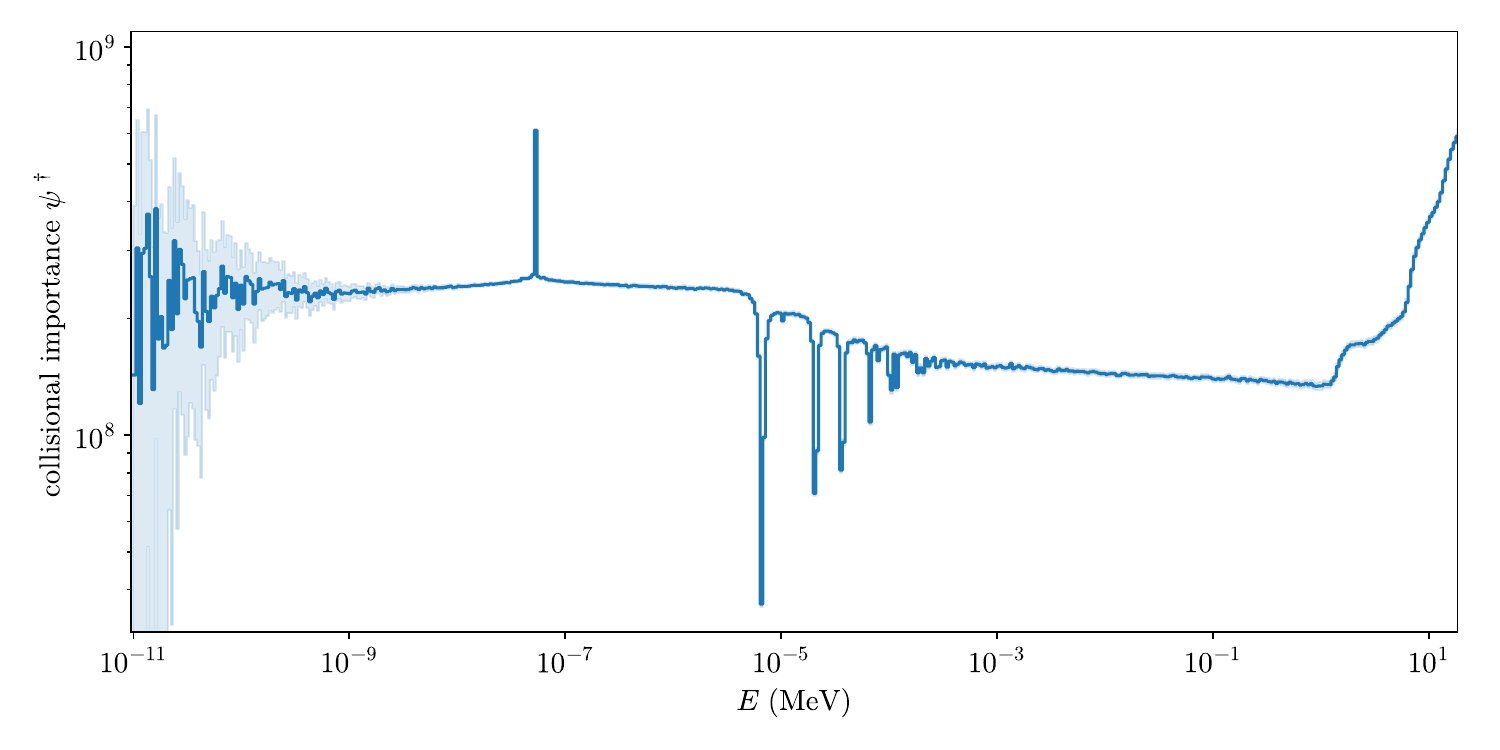}
    \caption{Collided importance $\psi^\dagger$ relative to a Dirac detector at $E^*=0.052\text{ eV}$ for a test problem consisting of \textsuperscript{238}U and \textsuperscript{1}H. See text for precise problem definition. The collision importance $\psi^\dagger$, which is proportional to the probability of a neutron scoring in the detector, has a maximum at the detector energy. The impact of the absorption resonances of \textsuperscript{238}U are clearly apparent in the dips of the importance at the corresponding energies. For each supplementary resonance that the neutron has to cross to get to the detector, the importance decreases, as the probability of the neutron surviving until reaching the detector decreases. At high energies ($E>1\text{ MeV}$), the possible fission of \textsuperscript{238}U increases the importance. }
    \label{fig:coll_import_U8H1}
\end{figure}

In order to test the validity of the adjoint sampling procedure, a simple benchmark was used. The simulation consists in a mixture of equal concentrations of \textsuperscript{238}U and \textsuperscript{1}H at a temperature of $600\text{ K}$ ($k_BT\approx 0.052\text{ eV}$). The medium has an infinite spatial extension, so that the problem only depends on the energy variable. Nuclear data is extracted from the ENDF-B/VIII.0 library, the Doppler broadening of the scattering kernel is activated under $400 k_BT$, we used the TSL of UO\textsubscript{2} for the elastic scattering of \textsuperscript{238}U and the TSL of H\textsubscript{2}O for the elastic scattering of \textsuperscript{1}H. The adjoint source is a delta distribution at $E^*=0.052\text{ eV}$ (corresponding to a point detector at the same energy), and the simulated energy range spans from $E_\text{min}= 10^{-5}\text{ eV}$ to $E_\text{max} = 20 \text{ MeV}$. During the game, the collision importance $\psi^\dagger$ is tallied over 500 logarithmically-spaced bins that span the entire energy range. The simulation uses $10^6$ particles grouped into $10^3$ batches, and enforces population control (splitting and Russian roulette). The simulation results are provided in \cref{fig:coll_import_U8H1}, to illustrate the capabilities of the simulation method described in this work.

\subsection{Reciprocity test}
\label{sec:res_reciprocity}

To assess the validity of the adjoint sampling procedure and of the produced adjoint nuclear data, we use a statistical test based on the reciprocity theorem, as recently proposed in Ref.~\citenum{rovel_preparing_2026}. For this purpose, we evaluate a series of responses using ten different sources and ten different detectors (yielding 100 different responses), with both direct and adjoint Monte Carlo games. The sources and detectors used for the test are described in detail in \cref{tab:sources_and_detectors}. This requires 10 direct simulations and 10 adjoint simulations: we adapt the number of particles for each simulation to attain a precision of about $0.1\%$ on all detectors.

\begin{table}
    \begin{subtable}[t]{0.45\textwidth}
        \centering
        \begin{tabular}{c|cc}
            \toprule
             Source number & $E_\text{min}$ (MeV) & $E_\text{max}$ (MeV)  \\
             \midrule
             1 & 0.100 & 0.158 \\
             2 & 0.158 & 0.251 \\
             3 & 0.251 & 0.398 \\
             4 & 0.398 & 0.631 \\
             5 & 0.631 & 1.00 \\
             6 & 1.00 & 1.59 \\
             7 & 1.59 & 2.51 \\
             8 & 2.51 & 3.98 \\
             9 & 3.98 & 6.31 \\
             10 & 6.31 & 10 
             
        \end{tabular}
    \end{subtable}
    \hspace{\fill}
    \begin{subtable}[t]{0.45\textwidth}
        \centering
        \begin{tabular}{c|cc}
            \toprule
             Detector number & $E_\text{min}$ (eV) & $E_\text{max}$ (eV)  \\
             \midrule
             1 & 0.01 & 0.02 \\
             2 & 0.02 & 0.0398 \\
             3 & 0.0398 & 0.0794 \\
             4 & 0.0794 & 0.158 \\
             5 & 0.158 & 0.316 \\
             6 & 0.316 & 0.631 \\
             7 & 0.631 & 1.26 \\
             8 & 1.26 & 2.51 \\
             9 & 2.51 & 5.01 \\
             10 & 5.01 & 10 
             
        \end{tabular}
    \end{subtable}
    
    \caption{Energy intervals for the sources (\textbf{on the left}) and detectors (\textbf{on the right}) used in the reciprocity test. The ten sources are uniformly distributed in ten logarithmically-spaced bins in the energy range $[0.1,10]\text{ MeV}$. The ten detectors have uniform response functions in ten logarithmically-spaced bins in the energy range $[0.1,10]\text{ eV}$. Sources are normalized to one source particle and detectors are normalized so that the integral of their response function over their energy range is equal to $1\text{ MeV}$. }
    \label{tab:sources_and_detectors}
\end{table}

According to the reciprocity theorem, for a given source-detector pair the direct and adjoint calculations must yield the same result (up to the statistical uncertainty). The discrepancies between the adjoint and the direct estimates (normalized by the combined standard deviation) should follow a standard normal distribution, and were analyzed using the Student $t$-test. Out of the 100 responses compared, only 3 have discrepancies above two standard deviations, and none exceed three. The precise results collected, the corresponding uncertainties and the Student $t$-variable are available in \cref{tab:direct_adj_results}. The adjoint simulations took, on average, 20 times longer than the direct simulations to obtain comparable statistical uncertainties. This result, however, is very problem-dependent and is only given as a rough order of magnitude of the relative computation time of adjoint calculations compared with direct calculations.

\begin{longtable}{ccccc}
\caption{Direct and adjoint results with their associated standard deviation in the reciprocity test. the $t$-value corresponds to the discrepancy between adjoint and direct results, expressed as a multiple of the combined standard deviation.} \\
\toprule
\textbf{Source} & \textbf{Detector} & \textbf{Direct Result ($\pm$ \%)} & \textbf{Adjoint Result ($\pm$ \%)} & \textbf{$t$-value} \\
\midrule
\endfirsthead
\toprule
\textbf{Source} & \textbf{Detector} & \textbf{Direct Result ($\pm$ \%)} & \textbf{Adjoint Result ($\pm$ \%)} & \textbf{$t$-value} \\
\midrule
\endhead
\bottomrule
\endfoot
\bottomrule
\endlastfoot
1 & 1 & 1.2391$\times 10^8$ ($\pm$ 0.11\%) & 1.2392$\times 10^8$ ($\pm$ 0.09\%) & -0.08 \\
1 & 2 & 1.4238$\times 10^8$ ($\pm$ 0.10\%) & 1.4231$\times 10^8$ ($\pm$ 0.10\%) & 0.39 \\
1 & 3 & 1.3126$\times 10^8$ ($\pm$ 0.09\%) & 1.3110$\times 10^8$ ($\pm$ 0.08\%) & 1.01 \\
1 & 4 & 7.6887$\times 10^7$ ($\pm$ 0.09\%) & 7.6768$\times 10^7$ ($\pm$ 0.10\%) & 1.16 \\
1 & 5 & 2.0167$\times 10^7$ ($\pm$ 0.09\%) & 2.0153$\times 10^7$ ($\pm$ 0.10\%) & 0.53 \\
1 & 6 & 2.6994$\times 10^6$ ($\pm$ 0.10\%) & 2.7012$\times 10^6$ ($\pm$ 0.06\%) & -0.55 \\
1 & 7 & 8.7648$\times 10^5$ ($\pm$ 0.11\%) & 8.7640$\times 10^5$ ($\pm$ 0.05\%) & 0.07 \\
1 & 8 & 4.0857$\times 10^5$ ($\pm$ 0.11\%) & 4.0826$\times 10^5$ ($\pm$ 0.06\%) & 0.61 \\
1 & 9 & 1.9675$\times 10^5$ ($\pm$ 0.11\%) & 1.9695$\times 10^5$ ($\pm$ 0.05\%) & -0.82 \\
1 & 10 & 1.0534$\times 10^5$ ($\pm$ 0.09\%) & 1.0540$\times 10^5$ ($\pm$ 0.05\%) & -0.54 \\
2 & 1 & 1.2311$\times 10^8$ ($\pm$ 0.11\%) & 1.2328$\times 10^8$ ($\pm$ 0.09\%) & -0.91 \\
2 & 2 & 1.4147$\times 10^8$ ($\pm$ 0.10\%) & 1.4162$\times 10^8$ ($\pm$ 0.11\%) & -0.71 \\
2 & 3 & 1.3052$\times 10^8$ ($\pm$ 0.10\%) & 1.3040$\times 10^8$ ($\pm$ 0.08\%) & 0.68 \\
2 & 4 & 7.6426$\times 10^7$ ($\pm$ 0.09\%) & 7.6409$\times 10^7$ ($\pm$ 0.10\%) & 0.17 \\
2 & 5 & 2.0064$\times 10^7$ ($\pm$ 0.10\%) & 2.0027$\times 10^7$ ($\pm$ 0.10\%) & 1.32 \\
2 & 6 & 2.6884$\times 10^6$ ($\pm$ 0.11\%) & 2.6870$\times 10^6$ ($\pm$ 0.06\%) & 0.43 \\
2 & 7 & 8.7311$\times 10^5$ ($\pm$ 0.11\%) & 8.7200$\times 10^5$ ($\pm$ 0.06\%) & 1.02 \\
2 & 8 & 4.0584$\times 10^5$ ($\pm$ 0.10\%) & 4.0579$\times 10^5$ ($\pm$ 0.05\%) & 0.11 \\
2 & 9 & 1.9562$\times 10^5$ ($\pm$ 0.10\%) & 1.9595$\times 10^5$ ($\pm$ 0.05\%) & -1.49 \\
2 & 10 & 1.0473$\times 10^5$ ($\pm$ 0.11\%) & 1.0485$\times 10^5$ ($\pm$ 0.06\%) & -0.92 \\
3 & 1 & 1.2225$\times 10^8$ ($\pm$ 0.10\%) & 1.2242$\times 10^8$ ($\pm$ 0.10\%) & -0.94 \\
3 & 2 & 1.4032$\times 10^8$ ($\pm$ 0.10\%) & 1.4050$\times 10^8$ ($\pm$ 0.13\%) & -0.76 \\
3 & 3 & 1.2936$\times 10^8$ ($\pm$ 0.09\%) & 1.2959$\times 10^8$ ($\pm$ 0.09\%) & -1.43 \\
3 & 4 & 7.5836$\times 10^7$ ($\pm$ 0.09\%) & 7.5908$\times 10^7$ ($\pm$ 0.11\%) & -0.66 \\
3 & 5 & 1.9902$\times 10^7$ ($\pm$ 0.09\%) & 1.9906$\times 10^7$ ($\pm$ 0.09\%) & -0.15 \\
3 & 6 & 2.6676$\times 10^6$ ($\pm$ 0.11\%) & 2.6671$\times 10^6$ ($\pm$ 0.08\%) & 0.14 \\
3 & 7 & 8.6672$\times 10^5$ ($\pm$ 0.11\%) & 8.6608$\times 10^5$ ($\pm$ 0.06\%) & 0.60 \\
3 & 8 & 4.0346$\times 10^5$ ($\pm$ 0.11\%) & 4.0321$\times 10^5$ ($\pm$ 0.06\%) & 0.52 \\
3 & 9 & 1.9461$\times 10^5$ ($\pm$ 0.11\%) & 1.9442$\times 10^5$ ($\pm$ 0.07\%) & 0.75 \\
3 & 10 & 1.0391$\times 10^5$ ($\pm$ 0.09\%) & 1.0405$\times 10^5$ ($\pm$ 0.06\%) & -1.26 \\
4 & 1 & 1.2112$\times 10^8$ ($\pm$ 0.11\%) & 1.2124$\times 10^8$ ($\pm$ 0.12\%) & -0.60 \\
4 & 2 & 1.3922$\times 10^8$ ($\pm$ 0.10\%) & 1.3882$\times 10^8$ ($\pm$ 0.13\%) & 1.75 \\
4 & 3 & 1.2835$\times 10^8$ ($\pm$ 0.09\%) & 1.2806$\times 10^8$ ($\pm$ 0.12\%) & 1.54 \\
4 & 4 & 7.5239$\times 10^7$ ($\pm$ 0.09\%) & 7.5072$\times 10^7$ ($\pm$ 0.12\%) & 1.46 \\
4 & 5 & 1.9727$\times 10^7$ ($\pm$ 0.09\%) & 1.9712$\times 10^7$ ($\pm$ 0.13\%) & 0.48 \\
4 & 6 & 2.6385$\times 10^6$ ($\pm$ 0.10\%) & 2.6414$\times 10^6$ ($\pm$ 0.08\%) & -0.85 \\
4 & 7 & 8.5760$\times 10^5$ ($\pm$ 0.11\%) & 8.5729$\times 10^5$ ($\pm$ 0.07\%) & 0.27 \\
4 & 8 & 3.9966$\times 10^5$ ($\pm$ 0.10\%) & 3.9876$\times 10^5$ ($\pm$ 0.09\%) & 1.64 \\
4 & 9 & 1.9268$\times 10^5$ ($\pm$ 0.12\%) & 1.9260$\times 10^5$ ($\pm$ 0.08\%) & 0.30 \\
4 & 10 & 1.0287$\times 10^5$ ($\pm$ 0.09\%) & 1.0311$\times 10^5$ ($\pm$ 0.09\%) & -1.82 \\
5 & 1 & 1.1930$\times 10^8$ ($\pm$ 0.10\%) & 1.1961$\times 10^8$ ($\pm$ 0.18\%) & -1.28 \\
5 & 2 & 1.3725$\times 10^8$ ($\pm$ 0.11\%) & 1.3732$\times 10^8$ ($\pm$ 0.18\%) & -0.24 \\
5 & 3 & 1.2652$\times 10^8$ ($\pm$ 0.10\%) & 1.2620$\times 10^8$ ($\pm$ 0.15\%) & 1.44 \\
5 & 4 & 7.4133$\times 10^7$ ($\pm$ 0.10\%) & 7.4286$\times 10^7$ ($\pm$ 0.14\%) & -1.18 \\
5 & 5 & 1.9442$\times 10^7$ ($\pm$ 0.09\%) & 1.9456$\times 10^7$ ($\pm$ 0.15\%) & -0.41 \\
5 & 6 & 2.6077$\times 10^6$ ($\pm$ 0.10\%) & 2.6020$\times 10^6$ ($\pm$ 0.11\%) & 1.42 \\
5 & 7 & 8.4787$\times 10^5$ ($\pm$ 0.11\%) & 8.4773$\times 10^5$ ($\pm$ 0.11\%) & 0.11 \\
5 & 8 & 3.9398$\times 10^5$ ($\pm$ 0.11\%) & 3.9430$\times 10^5$ ($\pm$ 0.12\%) & -0.51 \\
5 & 9 & 1.8990$\times 10^5$ ($\pm$ 0.12\%) & 1.8993$\times 10^5$ ($\pm$ 0.11\%) & -0.12 \\
5 & 10 & 1.0182$\times 10^5$ ($\pm$ 0.11\%) & 1.0184$\times 10^5$ ($\pm$ 0.12\%) & -0.13 \\
6 & 1 & 1.2647$\times 10^8$ ($\pm$ 0.12\%) & 1.2631$\times 10^8$ ($\pm$ 0.23\%) & 0.49 \\
6 & 2 & 1.4518$\times 10^8$ ($\pm$ 0.12\%) & 1.4527$\times 10^8$ ($\pm$ 0.23\%) & -0.23 \\
6 & 3 & 1.3389$\times 10^8$ ($\pm$ 0.11\%) & 1.3385$\times 10^8$ ($\pm$ 0.21\%) & 0.12 \\
6 & 4 & 7.8490$\times 10^7$ ($\pm$ 0.10\%) & 7.8501$\times 10^7$ ($\pm$ 0.19\%) & -0.07 \\
6 & 5 & 2.0561$\times 10^7$ ($\pm$ 0.10\%) & 2.0595$\times 10^7$ ($\pm$ 0.21\%) & -0.70 \\
6 & 6 & 2.7562$\times 10^6$ ($\pm$ 0.10\%) & 2.7556$\times 10^6$ ($\pm$ 0.16\%) & 0.11 \\
6 & 7 & 8.9509$\times 10^5$ ($\pm$ 0.11\%) & 8.9634$\times 10^5$ ($\pm$ 0.15\%) & -0.77 \\
6 & 8 & 4.1688$\times 10^5$ ($\pm$ 0.12\%) & 4.1725$\times 10^5$ ($\pm$ 0.15\%) & -0.44 \\
6 & 9 & 2.0129$\times 10^5$ ($\pm$ 0.11\%) & 2.0169$\times 10^5$ ($\pm$ 0.14\%) & -1.10 \\
6 & 10 & 1.0763$\times 10^5$ ($\pm$ 0.09\%) & 1.0754$\times 10^5$ ($\pm$ 0.15\%) & 0.46 \\
7 & 1 & 1.5088$\times 10^8$ ($\pm$ 0.13\%) & 1.5114$\times 10^8$ ($\pm$ 0.21\%) & -0.70 \\
7 & 2 & 1.7349$\times 10^8$ ($\pm$ 0.11\%) & 1.7352$\times 10^8$ ($\pm$ 0.20\%) & -0.08 \\
7 & 3 & 1.5995$\times 10^8$ ($\pm$ 0.11\%) & 1.5971$\times 10^8$ ($\pm$ 0.17\%) & 0.74 \\
7 & 4 & 9.3708$\times 10^7$ ($\pm$ 0.10\%) & 9.3855$\times 10^7$ ($\pm$ 0.18\%) & -0.75 \\
7 & 5 & 2.4564$\times 10^7$ ($\pm$ 0.11\%) & 2.4538$\times 10^7$ ($\pm$ 0.20\%) & 0.46 \\
7 & 6 & 3.2937$\times 10^6$ ($\pm$ 0.12\%) & 3.2872$\times 10^6$ ($\pm$ 0.13\%) & 1.13 \\
7 & 7 & 1.0691$\times 10^6$ ($\pm$ 0.13\%) & 1.0696$\times 10^6$ ($\pm$ 0.12\%) & -0.28 \\
7 & 8 & 4.9766$\times 10^5$ ($\pm$ 0.13\%) & 4.9752$\times 10^5$ ($\pm$ 0.12\%) & 0.16 \\
7 & 9 & 2.3972$\times 10^5$ ($\pm$ 0.11\%) & 2.4005$\times 10^5$ ($\pm$ 0.13\%) & -0.81 \\
7 & 10 & 1.2824$\times 10^5$ ($\pm$ 0.10\%) & 1.2881$\times 10^5$ ($\pm$ 0.13\%) & -2.63 \\
8 & 1 & 1.5822$\times 10^8$ ($\pm$ 0.11\%) & 1.5851$\times 10^8$ ($\pm$ 0.17\%) & -0.89 \\
8 & 2 & 1.8199$\times 10^8$ ($\pm$ 0.12\%) & 1.8163$\times 10^8$ ($\pm$ 0.19\%) & 0.92 \\
8 & 3 & 1.6779$\times 10^8$ ($\pm$ 0.11\%) & 1.6779$\times 10^8$ ($\pm$ 0.18\%) & -0.01 \\
8 & 4 & 9.8284$\times 10^7$ ($\pm$ 0.11\%) & 9.8445$\times 10^7$ ($\pm$ 0.18\%) & -0.78 \\
8 & 5 & 2.5780$\times 10^7$ ($\pm$ 0.11\%) & 2.5732$\times 10^7$ ($\pm$ 0.19\%) & 0.87 \\
8 & 6 & 3.4584$\times 10^6$ ($\pm$ 0.12\%) & 3.4524$\times 10^6$ ($\pm$ 0.14\%) & 0.98 \\
8 & 7 & 1.1223$\times 10^6$ ($\pm$ 0.13\%) & 1.1238$\times 10^6$ ($\pm$ 0.11\%) & -0.79 \\
8 & 8 & 5.2311$\times 10^5$ ($\pm$ 0.12\%) & 5.2257$\times 10^5$ ($\pm$ 0.12\%) & 0.60 \\
8 & 9 & 2.5225$\times 10^5$ ($\pm$ 0.12\%) & 2.5172$\times 10^5$ ($\pm$ 0.11\%) & 1.30 \\
8 & 10 & 1.3485$\times 10^5$ ($\pm$ 0.11\%) & 1.3496$\times 10^5$ ($\pm$ 0.13\%) & -0.50 \\
9 & 1 & 1.7688$\times 10^8$ ($\pm$ 0.14\%) & 1.7692$\times 10^8$ ($\pm$ 0.17\%) & -0.10 \\
9 & 2 & 2.0311$\times 10^8$ ($\pm$ 0.13\%) & 2.0260$\times 10^8$ ($\pm$ 0.19\%) & 1.12 \\
9 & 3 & 1.8739$\times 10^8$ ($\pm$ 0.11\%) & 1.8668$\times 10^8$ ($\pm$ 0.17\%) & 1.87 \\
9 & 4 & 1.0974$\times 10^8$ ($\pm$ 0.12\%) & 1.0979$\times 10^8$ ($\pm$ 0.17\%) & -0.24 \\
9 & 5 & 2.8783$\times 10^7$ ($\pm$ 0.12\%) & 2.8695$\times 10^7$ ($\pm$ 0.18\%) & 1.39 \\
9 & 6 & 3.8597$\times 10^6$ ($\pm$ 0.13\%) & 3.8514$\times 10^6$ ($\pm$ 0.13\%) & 1.20 \\
9 & 7 & 1.2507$\times 10^6$ ($\pm$ 0.15\%) & 1.2514$\times 10^6$ ($\pm$ 0.11\%) & -0.30 \\
9 & 8 & 5.8299$\times 10^5$ ($\pm$ 0.13\%) & 5.8284$\times 10^5$ ($\pm$ 0.13\%) & 0.14 \\
9 & 9 & 2.8162$\times 10^5$ ($\pm$ 0.13\%) & 2.8097$\times 10^5$ ($\pm$ 0.10\%) & 1.42 \\
9 & 10 & 1.5029$\times 10^5$ ($\pm$ 0.13\%) & 1.5052$\times 10^5$ ($\pm$ 0.11\%) & -0.90 \\
10 & 1 & 2.7233$\times 10^8$ ($\pm$ 0.13\%) & 2.7218$\times 10^8$ ($\pm$ 0.14\%) & 0.29 \\
10 & 2 & 3.1269$\times 10^8$ ($\pm$ 0.12\%) & 3.1251$\times 10^8$ ($\pm$ 0.15\%) & 0.29 \\
10 & 3 & 2.8846$\times 10^8$ ($\pm$ 0.12\%) & 2.8718$\times 10^8$ ($\pm$ 0.13\%) & 2.57 \\
10 & 4 & 1.6884$\times 10^8$ ($\pm$ 0.12\%) & 1.6875$\times 10^8$ ($\pm$ 0.12\%) & 0.31 \\
10 & 5 & 4.4281$\times 10^7$ ($\pm$ 0.11\%) & 4.4256$\times 10^7$ ($\pm$ 0.15\%) & 0.31 \\
10 & 6 & 5.9313$\times 10^6$ ($\pm$ 0.12\%) & 5.9211$\times 10^6$ ($\pm$ 0.11\%) & 1.06 \\
10 & 7 & 1.9227$\times 10^6$ ($\pm$ 0.14\%) & 1.9266$\times 10^6$ ($\pm$ 0.09\%) & -1.21 \\
10 & 8 & 8.9536$\times 10^5$ ($\pm$ 0.13\%) & 8.9536$\times 10^5$ ($\pm$ 0.10\%) & -0.00 \\
10 & 9 & 4.3145$\times 10^5$ ($\pm$ 0.14\%) & 4.3303$\times 10^5$ ($\pm$ 0.08\%) & -2.29 \\
10 & 10 & 2.3110$\times 10^5$ ($\pm$ 0.12\%) & 2.3122$\times 10^5$ ($\pm$ 0.09\%) & -0.33
\label{tab:direct_adj_results}
\end{longtable}

%% file: Conclusion.tex
\section{Conclusions}

In this work we have investigated the production of adjoint cross sections and adjoint distribution laws necessary to sample the adjoint collision event, a key ingredient for adjoint particle-transport Monte Carlo simulations. After succinctly recalling the general framework presented in our previous work \cite{rovel_general_2025} for adjoint Monte Carlo games,  we have built upon this formalism and detailed the procedure used to prepare adjoint nuclear data.

Our strategy has been as exhaustive as possible, addressing all the nuclear reactions required to be able to process a full modern nuclear data library. We have notably taken into account the treatment of the Doppler broadening of the elastic scattering kernel, the Thermal Scattering Laws (TSL), and the Unresolved Resonance Range (URR). We have then presented the discretization algorithms used to adequately select the positions of the discretization points where the adjoint data is stored in memory.

These techniques were used to produce an entire adjoint nuclear data library based on the ENDF-B/VIII.0 evaluation. Finally, we briefly tested an adjoint simulation using this data on a simple infinite-medium benchmark configuration. We verified the results produced by the adjoint games by relying on the reciprocity theorem.

This work is a further step towards the ambitious goal of establishing a full Monte Carlo sampling strategy for the adjoint Boltzmann equation. Attaining this objective requires further work on several directions. A detailed study of the behavior of adjoint Monte Carlo games in the case involving flights and spatially heterogeneous media is needed. The study of variance reduction applied to adjoint calculations, following the preliminary findings presented in our previous paper \cite{rovel_general_2025}, is also paramount to ensure the numerical stability of adjoint simulations. The final step will be to implement adjoint transport routines in the production Monte Carlo code \tripolir{}, and probe these algorithms on real-world radiation shielding applications.

%% file: Appendices.tex
\setcounter{equation}{0}
\section{Kinematics of Doppler-broadened elastic scattering}
\label{sec:proof_svt}

In this section we will investigate the elastic scattering on a target nucleus of relative mass $A$ whose velocity obeys a Maxwellian distribution. This will allow establishing \cref{eq:kernel_dbrc,eq:kernel_svt_iso,eq:svt_energy_to_energy_kernel}.

The incoming neutron speed is denoted $\mathbf{v}$, the incoming target speed is denoted $\mathbf{V}$, the outgoing neutron speed is $\mathbf{v}'$. The center of mass speed is then
\begin{equation}
    \boldsymbol{\mathcal{V}}= \frac{\mathbf{v}+A\mathbf{V}}{A+1}.
\end{equation}
We denote $\mathbf{q}=\mathbf{v}'-\mathbf{v}$ the impulse transfer, and $\mathbf{v}_r=\mathbf{v}-\mathbf{V}$ the relative speed between the neutron and the target. In the center of mass frame, the neutron speed is $\mathbf{v}_{cm}=\mathbf{v}-\boldsymbol{\mathcal{V}}=\mathbf{v}_r\frac{A}{A+1}$, the direction is  $\mathbf{\Omega}_{cm}=\mathbf{v}_{cm}/v_{cm}$, the outgoing speed is $\mathbf{v}'_{cm}= \mathbf{v'}-\boldsymbol{\mathcal{V}}$, the outgoing direction is $\mathbf{\Omega}'_{cm}=\mathbf{v}'_{cm}/v'_{cm}$, and the deflection cosine is $\mu_{cm}= \mathbf{\Omega}_{cm}\cdot \mathbf{\Omega}'_{cm}$.

Let us study the distribution of outgoing speeds $\mathbf{v}'$, at fixed $\mathbf{v}$ and $\mathbf{V}$:
\begin{align}
    \mathbb{P}(\mathbf{v}\to \mathbf{v}' | \mathbf{V})d\mathbf{v}' &= \mathbb{P}(v_{cm},\mathbf{\Omega}\to v_{cm}',\mathbf{\Omega}'_{cm}) d\mathbf{\Omega}'_{cm}dv'_{cm} \nonumber\\
    &= \frac{g(\mu_{cm})}{2\pi}\delta(v'_{cm}-v_{cm}) d\mathbf{\Omega}'_{cm}dv_{cm}'\nonumber\\
    &= \frac{g(\mu_{cm})}{2\pi(v_{cm})^2}\delta(v'_{cm}-v_{cm}) d\mathbf{v}'_{cm}\nonumber\\
    &=  \frac{g(\mu_{cm})}{2\pi(v_{cm})^2} \delta(||\mathbf{v}'-\boldsymbol{\mathcal{V}}||-||\mathbf{v}-\boldsymbol{\mathcal{V}}||)d\mathbf{v}'.
    \label{eq:apxsvt_ker1}
\end{align}
\Cref{eq:apxsvt_ker1} yields a density for going from $\mathbf{v}$ to $\mathbf{v'}$ at fixed $\mathbf{V}$, where $g(\mu_{cm})$ is the angular scattering distribution in the center of mass framework; we assumed the azimuthal distribution in the center of mass frame to be uniform, whence the $1/(2\pi)$ term. Now, we want to integrate \cref{eq:apxsvt_ker1} over the distribution of possible speeds $\mathbf{V}$; however, to perform the integration we need the Dirac-delta term within \cref{eq:apxsvt_ker1} to be expressed in terms of $\mathbf{V}$. We obtain then
\begin{align}
    \mathbb{P}(\mathbf{v}\to \mathbf{v}' | \mathbf{V})d\mathbf{v}' 
    &=  \frac{g(\mu_{cm})}{2\pi(v_{cm})^2} 2v_{cm} \delta(||\mathbf{v}'-\boldsymbol{\mathcal{V}}||^2-||\mathbf{v}-\boldsymbol{\mathcal{V}}||^2)d\mathbf{v}'\nonumber\\
    &= \frac{g(\mu_{cm})}{\pi v_{cm}}\delta\big(2\mathbf{q} \cdot\left( \frac{\mathbf{v}+A\mathbf{V}}{A+1}\right) - v'^2 + v^2 \big)d\mathbf{v}'\nonumber\\
    &=\frac{g(\mu_{cm})}{\pi v_{cm}}\frac{(A+1)}{2A} \delta\big(\mathbf{q}\cdot\mathbf{V} - \frac{(A+1)}{2A}q^2+\mathbf{q}\cdot\mathbf{v}\big)d\mathbf{v}'\nonumber\\
    &=\frac{g(\mu_{cm})}{2\pi v_{cm}q}\frac{(A+1)}{A} \delta\big(\mathbf{V}\cdot\mathbf{n}- \frac{A+1}{2A}q-v\mathbf{n} \big)d\mathbf{v}',
    \label{eq:apxsvt_ker2}
\end{align}
where we introduced $\mathbf{n}=\mathbf{q}/q$. From \cref{eq:apxsvt_ker2}, in order for a transition from $\mathbf{v}$ to $\mathbf{v}'$ to be possible, $\mathbf{V}$ needs to lie on a plane orthogonal to $\mathbf{n}$. We can then write :
\begin{equation}
    \label{eq:apxsvt_decomp_V}
    \mathbf{V} = \left( \frac{A+1}{2A}q+\mathbf{v}\cdot\mathbf{n}\right)\mathbf{n}+x\mathbf{e}_1+ y\mathbf{e}_2,
\end{equation}
with $\mathbf{e}_1,\mathbf{e}_2,\mathbf{n}$ forming an orthonormal basis.

After properly introducing the density, we can integrate over the different target velocities $\mathbf{V}$. From \cref{eq:full_target_speed_distrib} in vector form, the distribution of target velocities is:
\begin{equation}
    \mathbb{P}(\mathbf{V}|\mathbf{v})d\mathbf{V}= C^{-1}(v) \frac{1}{v} v_r \sigma_{0}(v_r)\frac{\beta^{3/2}}{\pi^{3/2}}e^{-\beta V^2}d\mathbf{V}.
\end{equation}
The total probability of going from $\mathbf{v}$ to $\mathbf{v'}$ is then 
\begin{align}
\mathbb{P}(\mathbf{v}\to \mathbf{v}') 
=& \int \mathbb{P}(\mathbf{v}\to \mathbf{v}' | \mathbf{V})\mathbb{P}(\mathbf{V}|\mathbf{v})d\mathbf{V} \nonumber\\
=& \int \frac{g(\mu_{cm})}{2\pi v_{cm} q}\frac{(A+1)}{A} \delta\big(\mathbf{V}\cdot\mathbf{n}- \frac{A+1}{2A}q-\mathbf{v}\cdot\mathbf{n} \big)C^{-1}(v) \frac{1}{v} v_r \sigma_{0}(v_r)\frac{\beta^{3/2}}{\pi^{3/2}}e^{-\beta V^2}d\mathbf{V}\nonumber\\
=& \int_{x=-\infty}^\infty  \int_{y=-\infty}^\infty \frac{g(\mu_{cm})}{2\pi v_{cm} q}\frac{(A+1)}{A} C^{-1}(v) \frac{1}{v} v_r \sigma_{0}(v_r)\frac{\beta^{3/2}}{\pi^{3/2}}\nonumber\\
&\qquad\qquad\qquad\times\text{exp}\left(-\beta(x^2+y^2 +(\frac{A+1}{2A}q+\mathbf{v}\cdot\mathbf{n}))^2 )\right)dxdy\nonumber\\
=& \left(\frac{A+1}{A}\right)^2\frac{\beta^{3/2}C^{-1}(v)}{2\pi^{5/2}vq}\text{exp}\left(-\beta(\frac{A+1}{2A}q+\mathbf{v}\cdot\mathbf{n})^2\right)\nonumber\\
&\times \int_{x=-\infty}^\infty  \int_{y=-\infty}^\infty g(\mu_{cm})  \sigma_{0}(v_r)e^{-\beta(x^2+y^2) }dxdy
\label{eq:apxsvt_ker3}
\end{align}
Converting \cref{eq:apxsvt_ker3} from a speed density into an energy-direction density, and integrating along the azimuthal angle $\varphi_l$, allows retrieving the general kernel from \cref{eq:kernel_dbrc}: 
\begin{equation*}
    \begin{aligned}
        \mathbb{P}(E\to E', \mu_l) 
        =&  \left(\frac{A+1}{A}\right)^2\frac{\beta^{3/2}C^{-1}(v)}{2\pi^{5/2}qv}\text{exp}\left(-\beta(\frac{A+1}{2A}q+\mathbf{v}\cdot\mathbf{n})^2\right)\times 2\pi v'\\
        &\times \frac{\pi}{\beta} \int_{x=-\infty}^\infty  \int_{y=-\infty}^\infty g(\mu_{cm})  \sigma_{0}(v_r)\frac{\beta}{\pi}e^{-\beta(x^2+y^2) }dxdy \\
        =&  \left(\frac{A+1}{A}\right)^2\frac{\beta^{1/2}C^{-1}(v)}{\pi^{1/2}q}\sqrt{\frac{E'}{E}}\text{exp}\left(-\frac{\beta}{q^2}(E'-E+\frac{q^2}{2A})^2\right)\\
        &\times \int_{x=-\infty}^\infty  \int_{y=-\infty}^\infty g(\mu_{cm})  \sigma_{0}(v_r)\frac{\beta}{\pi}e^{-\beta(x^2+y^2) }dxdy,
    \end{aligned}
\end{equation*}
where the argument in the exponential was transformed as follows: 
\begin{align}
    &-\beta(\frac{A+1}{2A}q+\mathbf{v}\cdot\mathbf{n})^2 \nonumber\\
    =& -\frac{\beta}{q^2}(\frac{A+1}{2A}q^2+\mathbf{v}\cdot\mathbf{q})^2 \nonumber\\
    =& -\frac{\beta}{q^2}(\frac{A+1}{2A}q^2-v^2+\frac{v^2+v'^2-q^2}{2})^2 \nonumber\\
    =& -\frac{\beta}{q^2}(\frac{v'^2-v^2}{2}+\frac{q^2}{2A})^2 \nonumber\\
    =& -\frac{\beta}{q^2}(E'-E+\frac{q^2}{2A})^2.
\end{align}

Now, we will make use of the isotropic and CXS hypotheses; we obtain then $\sigma_0(v_r)=\sigma_0$, $g(\mu_{cm})=1/2$ and $C(v)=\sigma_0 \tilde{C}(v)$, which yields
\begin{align}
&\mathbb{P}(E\to E',\mu_l)\nonumber\\
&=\left(\frac{A+1}{A}\right)^2\frac{\beta^{1/2}C^{-1}(v)}{\pi^{1/2}q}\sqrt{\frac{E'}{E}}\text{exp}\left(-\frac{\beta}{q^2}(E'-E+\frac{q^2}{2A})^2\right) \int_{x=-\infty}^\infty  \int_{y=-\infty}^\infty \frac{\sigma_0}{2}\frac{\beta}{\pi}  e^{-\beta(x^2+y^2) }dxdy \nonumber\\
&= \left(\frac{A+1}{A}\right)^2\frac{\beta^{1/2}\tilde{C}^{-1}(v)}{2\pi^{1/2}q}\sqrt{\frac{E'}{E}}\text{exp}\left(-\frac{\beta}{q^2}(E'-E+\frac{q^2}{2A})^2\right).
\label{eq:apxsvt_ker4}
\end{align}
Note that \cref{eq:apxsvt_ker4} corresponds to \cref{eq:kernel_svt_iso}. Now, let us integrate along $\mu_l$:
\begin{align}
    \mathbb{P}(E\to E') =& \int_{-1}^1 \mathbb{P}(E\to E',\mu_l)d\mu_l\nonumber\\
    =& \int_{-1}^1 \left(\frac{A+1}{A}\right)^2\frac{\beta^{1/2}\tilde{C}^{-1}(v)}{2\pi^{1/2}q}\sqrt{\frac{E'}{E}}\text{exp}\left(-\frac{\beta}{q^2}(E'-E+\frac{q^2}{2A})^2\right)d\mu_l.
    \label{eq:apxsvt_ker5}
\end{align}
In \cref{eq:apxsvt_ker5}, the only term that depends on $\mu_l$ is $q$, via the relation
\begin{equation}
    q := ||\mathbf{v}'-\mathbf{v}||= \sqrt{2\left(E'+E-2\mu_l\sqrt{EE'}\right)}.
    \label{eq:apxsvt_defq}
\end{equation}
Let us change variables to integrate over $q$. We have:
\begin{equation}
    dq = - \frac{2\sqrt{EE'}}{q}d\mu_l \implies d\mu_l = -\frac{q}{2\sqrt{EE'}}dq,
\end{equation}
and the new integration bounds are
\begin{equation}
    \begin{aligned}
        q_\text{min} &= |\sqrt{2E}-\sqrt{2E'}|\\
        q_\text{max} &= \sqrt{2E}+\sqrt{2E'},
    \end{aligned}
\end{equation}
which leads to
\begin{align}
    \mathbb{P}(E\to E') =& \int_{q_\text{min}}^{q_\text{max}} \left(\frac{A+1}{A}\right)^2\frac{\beta^{1/2}\tilde{C}^{-1}(v)}{2\pi^{1/2}q}\sqrt{\frac{E'}{E}}\text{exp}\left(-\frac{\beta}{q^2}(E'-E+\frac{q^2}{2A})^2\right)\frac{q}{2\sqrt{EE'}}dq \nonumber\\
    =& \left(\frac{A+1}{A}\right)^2\frac{\beta^{1/2}\tilde{C}^{-1}(v)}{4\pi^{1/2}E}\int_{q_\text{min}}^{q_\text{max}} \text{exp}\left(-\beta(\frac{\Delta E}{q}+\frac{q}{2A})^2\right)dq,
    \label{eq:apxsvt_ker6}
\end{align}
where we introduced $\Delta E= E'-E$. Let us denote $I$ the integral term in \cref{eq:apxsvt_ker6}. Furthermore, let us define $a = \sqrt{\beta}\Delta E$ and $b = \sqrt{\beta}/(2A)$. Then, the exponent in $I$ can be re-written as $-(bq+a/q)^2$. We will now introduce
\begin{equation}
\begin{aligned}
    x&= bq + \frac{a}{q}\\
    y&= bq - \frac{a}{q}
\end{aligned}
\label{eq:apxsvt_glasser_var}
\end{equation}
to help perform another change of integration variable. These quantities $x$ and $y$ satisfy
\begin{equation}
    \frac{dx+dy}{2b} = dq,
\end{equation}
and can be converted from one to another using
\begin{equation}
    x^2 = y^2+4ab.
\end{equation}
We obtain
\begin{align}
    I =& \int_{q_\text{min}}^{q_\text{max}} \text{exp}\left( -(bq+a/q)^2\right)dq\nonumber\\
    =&\frac{1}{2b}\left[ \int_{x_\text{min}}^{x_\text{max}} e^{-x^2}dx + \int_{y_\text{min}}^{y_\text{max}} e^{-x^2}dy \right]\nonumber\\
    =&\frac{1}{2b}\left[ \int_{x_\text{min}}^{x_\text{max}} e^{-x^2}dx + e^{-4ab}\int_{y_\text{min}}^{y_\text{max}} e^{-y^2}dy \right]\nonumber\\
    =& \frac{\sqrt{\pi}}{4b} \left[\text{erf}\left(x\right)\right]^{x_\text{max}}_{x_\text{min}}+ \frac{e^{-4ab}\sqrt{\pi}}{4b} \left[\text{erf}\left(y\right)\right]^{y_\text{max}}_{y_\text{min}}.
    \label{eq:apxsvt_integral}
\end{align}
Now let us replace $a$ and $b$ by their values:
\begin{equation}
    4ab=\frac{2\beta \Delta E}{A}=\frac{E'-E}{k_bT},
\end{equation}
and evaluate the bounds of $x$ and $y$. For $x_\text{max}$, we have:
\begin{align}
    x_\text{max}&=bq_\text{max}+a/q_\text{max}\\
    &=\sqrt{\frac{\beta}{2}}\left(\frac{\sqrt{E}+\sqrt{E'}}{A}+\frac{E'-E}{\sqrt{E}+\sqrt{E'}}\right) \nonumber\\
    &= \sqrt{\frac{\beta}{2}}\left(\frac{\sqrt{E}+\sqrt{E'}}{A}+\sqrt{E'}-\sqrt{E}\right)\nonumber\\
    &= \sqrt{\frac{\beta}{2}} \left( \frac{A+1}{A}\sqrt{E'}-\frac{A-1}{A}\sqrt{E}\right)\nonumber\\
    &= \frac{\eta \sqrt{E'}-\rho\sqrt{E}}{\sqrt{k_bT}},
\end{align}
where 
\begin{equation}
    \eta = \frac{A+1}{2\sqrt{A}}
\end{equation}
and
\begin{equation}
    \rho = \frac{A-1}{2\sqrt{A}}.
\end{equation}
For $y_\text{max}$, we have
\begin{align}
    y_\text{max}&=bq_\text{max}-a/q_\text{max}\nonumber\\
    &= \frac{\eta \sqrt{E}-\rho\sqrt{E'}}{\sqrt{k_bT}}.
\end{align}
The lower bounds of $x$ and $y$ need to distinguish the cases $E<E'$ and $E>E'$: we will use upper signs in the former case, and lower signs for the latter case, which yields
\begin{align}
     x_\text{min}&=bq_\text{min}+a/q_\text{min}\\
    &=\sqrt{\frac{\beta}{2}}\left(\frac{\mp\sqrt{E}\pm\sqrt{E'}}{A}\pm\frac{E'-E}{\sqrt{E'}-\sqrt{E}}\right) \nonumber\\
    &= \sqrt{\frac{\beta}{2}}\left(\frac{\mp\sqrt{E}\pm\sqrt{E'}}{A}\pm\sqrt{E'}\pm\sqrt{E}\right)\nonumber\\
    &= \sqrt{\frac{\beta}{2}} \left( \pm\frac{A+1}{A}\sqrt{E'}\pm\frac{A-1}{A}\sqrt{E}\right)\nonumber\\
    &= \pm\frac{\eta \sqrt{E'}+\rho\sqrt{E}}{\sqrt{k_bT}},
\end{align}
and finally
\begin{align}
    y_\text{min}&=bq_\text{min}-a/q_\text{min}\nonumber\\
    &= \mp\frac{\eta \sqrt{E}+\rho\sqrt{E'}}{\sqrt{k_bT}}.
\end{align}
Combining the value of $I$ from \cref{eq:apxsvt_integral} with the computed bounds and \cref{eq:apxsvt_ker6}, we get the final expression
\begin{equation}
    \begin{aligned}
        \mathbb{P}(E\to E') =& \tilde{C}^{-1}(E)\frac{\eta^2}{2E}\Bigg[ 
        \text{exp}\left(\frac{E-E'}{k_bT}\right)\times\left(\text{erf}\left(\frac{\eta\sqrt{E}-\rho\sqrt{E'}}{\sqrt{k_b T}}\right) \pm\text{erf}\left(\frac{\eta\sqrt{E}+\rho\sqrt{E'}}{\sqrt{k_b T}}\right)\right) \\
         &+ \text{erf}\left( \frac{\eta\sqrt{E'}-\rho\sqrt{E}}{\sqrt{k_b T}} \right)
        \mp \text{erf} \left( \frac{\eta\sqrt{E'}+\rho\sqrt{E}}{\sqrt{k_b T}} \right)
\Bigg],
    \end{aligned}
\end{equation}
which corresponds to \cref{eq:svt_energy_to_energy_kernel}. 